\documentclass[twocolumn]{aastex631}   
\usepackage{epsf}
\usepackage{color}
\usepackage{color}
\usepackage{enumitem}
\usepackage{mathtools}

\usepackage[mathscr]{euscript}
 \let\mathscr\relax
\usepackage[scr]{rsfso}

\shorttitle{Stellar Populations at $z\!\!=$9--11}
\shortauthors{Tacchella et al.}

\newcommand{\sol}{$_{\odot}$}

\begin{document}
\title{\vspace{-0.2cm}On the Stellar Populations of Galaxies at $z\!\!=$9--11: \\The Growth of Metals and Stellar Mass at Early Times}
\author[0000-0002-8224-4505]{Sandro Tacchella}
\affiliation{Department of Physics, Ulsan National Institute of Science and Technology (UNIST), Ulsan 44919, Republic of Korea}
\affiliation{Center for Astrophysics $\vert$ Harvard \& Smithsonian, Cambridge, MA, USA}
\email{tacchella@unist.ac.kr, stevenf@astro.as.utexas.edu}
\author[0000-0001-8519-1130]{Steven L. Finkelstein}
\affiliation{Department of Astronomy, The University of Texas at Austin, Austin, TX, USA}
\author[0000-0002-9921-9218]{Micaela Bagley}
\affiliation{Department of Astronomy, The University of Texas at Austin, Austin, TX, USA}
\author[0000-0001-5414-5131]{Mark Dickinson}
\affiliation{National Optical-Infrared Astronomy Research Laboratory, Tucson, AZ, USA}
\author[0000-0001-7113-2738]{Henry C. Ferguson}
\affiliation{Space Telescope Science Institute, 3700 San Martin Dr., Baltimore, MD 21218, USA}
\author[0000-0002-7831-8751]{Mauro Giavalisco}
\affiliation{University of Massachusetts, Amherst, MA, USA}
\author[0000-0002-9231-1505]{Luca Graziani}
\affiliation{Dipartimento di Fisica, Sapienza, Universit\`{a} di Roma, Piazzale Aldo Moro 5, 00185, Roma, IT}
\author{Norman A. Grogin}
\affiliation{Space Telescope Science Institute, 3700 San Martin Dr., Baltimore, MD 21218, USA}
\author[0000-0001-6145-5090]{Nimish Hathi}
\affiliation{Space Telescope Science Institute, 3700 San Martin Dr., Baltimore, MD 21218, USA}
\author[0000-0001-6251-4988]{Taylor A. Hutchison}
\affiliation{Department of Physics and Astronomy, Texas A\&M University, College Station, TX, 77843-4242 USA}
\affiliation{George P. and Cynthia Woods Mitchell Institute for Fundamental Physics and Astronomy,\\ Texas A\&M University, College Station, TX, 77843-4242 USA}
\author[0000-0003-1187-4240]{Intae Jung}
\affiliation{Astrophysics Science Division, NASA Goddard Space Flight Center, Greenbelt, MD 20771, USA}
\affiliation{Department of Physics, The Catholic University of America, Washington, DC 20064, USA}
\author[0000-0002-6610-2048]{Anton M. Koekemoer}
\affiliation{Space Telescope Science Institute, 3700 San Martin Dr., Baltimore, MD 21218, USA}
\author[0000-0003-2366-8858]{Rebecca L. Larson}
\altaffiliation{NSF Graduate Fellow}
\affiliation{Department of Astronomy, The University of Texas at Austin, Austin, TX, USA}
\author[0000-0001-7503-8482]{Casey Papovich}
\affiliation{Department of Physics and Astronomy, Texas A\&M University, College Station, TX, 77843-4242 USA}
\affiliation{George P. and Cynthia Woods Mitchell Institute for Fundamental Physics and Astronomy,\\ Texas A\&M University, College Station, TX, 77843-4242 USA}
\author{Norbert Pirzkal}
\affiliation{Space Telescope Science Institute, 3700 San Martin Dr., Baltimore, MD 21218, USA}
\author[0000-0003-2349-9310]{Sof\'ia Rojas-Ruiz}\altaffiliation{Fellow
  of the International Max Planck Research School for\\ Astronomy and
  Cosmic Physics at the University of \\ Heidelberg (IMPRS--HD)}
\affiliation{Max-Planck-Institut f\"{u}r Astronomie, K\"{o}nigstuhl 17, D-69117, Heidelberg, Germany}
\author{Mimi Song}
\affiliation{University of Massachusetts, Amherst, MA, USA}
\author{Raffaella Schneider}
\affiliation{Dipartimento di Fisica, Sapienza, Universit\`{a} di Roma, Piazzale Aldo Moro 5, 00185, Roma, IT}
\author{Rachel S. Somerville}
\affiliation{Center for Computational Astrophysics, Flatiron Institute, NY, USA}
\author[0000-0003-3903-6935]{Stephen M. Wilkins}
\affiliation{Astronomy Centre, University of Sussex, Falmer, Brighton BN1 9QH, UK}
\author[0000-0003-3466-035X]{L. Y. Aaron Yung}
\affiliation{Astrophysics Science Division, NASA Goddard Space Flight Center, Greenbelt, MD 20771, USA}

\begin{abstract}
We present a detailed stellar population analysis of 11 bright ($H<26.6$) galaxies at $z=9-11$ (three spectroscopically confirmed) to constrain the chemical enrichment and growth of stellar mass of early galaxies. We use the flexible Bayesian spectral energy distribution (SED) fitting code \texttt{Prospector} with a range of star-formation histories (SFHs), a flexible dust attenuation law, and a self-consistent model of emission lines. This approach allows us to assess how different priors affect our results and how well we can break degeneracies between dust attenuation, stellar ages, metallicity, and emission lines using data that probe only the rest-frame ultraviolet (UV) to optical wavelengths. We measure a median observed UV spectral slope $\beta=-1.87_{-0.43}^{+0.35}$ for relatively massive star-forming galaxies ($9<\log(M_{\star}/M_{\odot})<10$), consistent with no change from $z=4$ to $z=9–10$ at these stellar masses, implying rapid enrichment. Our SED-fitting results are consistent with a star-forming main sequence with sublinear slope ($0.7\pm0.2$) and specific star-formation rates of $3-10~\mathrm{Gyr}^{-1}$. However, the stellar ages and SFHs are less well constrained. Using different SFH priors, we cannot distinguish between median mass-weighted ages of $\sim50-150$ Myr, which corresponds to 50\% formation redshifts of $z_{50}\sim10-12$ at $z\sim9$ and is of the order of the dynamical timescales of these systems. Importantly, models with different SFH priors are able to fit the data equally well. We conclude that the current observational data cannot tightly constrain the mass-buildup timescales of these $z=9-11$ galaxies, with our results consistent with SFHs implying both a shallow and steep increase in the cosmic SFR density with time at $z>10$.
\end{abstract}

\vspace{-1.2cm}

\keywords{early universe --- galaxies: formation --- galaxies: evolution --- galaxies: high-redshift --- stars: formation}

\section{Introduction}\label{sec:intro}

The past decade has seen observational studies leap into the epoch of reionization, the time in the early universe when energetic photons (presumably from early star formation) ionized the gas in the intergalactic medium (IGM). Advances in near-IR imaging both in space with the {\it Hubble Space Telescope} ({\it HST}) and from the ground (with, e.g., Subaru and VISTA) have allowed the discovery of large samples of dropout galaxy candidates at $6<z<11$ \citep[e.g.,][]{oesch10, ellis13, oesch18,finkelstein15, finkelstein21, bouwens15, bouwens21_rebels, bouwens21, bowler15, bowler17, mcleod16, livermore17, atek18, harikane21}. Studying the properties of these galaxies, which exist at a time less than 1 Gyr after the Big Bang, can provide key constraints on the buildup of both stellar mass and heavy elements in early galaxies. In particular, the stellar population of galaxies at the earliest probed cosmic times ($z>8$) ought to supply crucial information on the formation of the first stars and galaxies. 

For instance, the rest-frame ultraviolet (UV) colors of these early galaxies can inform us about the earliest phases of chemical enrichment. The UV color is sensitive to dust attenuation, stellar ages, and stellar metallicities \citep[e.g.,][]{wilkins11}. At these early times, dust attenuation is believed to dominate, though very low metallicities can result in extremely blue colors. Early results at $z\sim7$ found some evidence that the faintest galaxies in the Hubble Ultra Deep Field ($m_{\rm AB}\sim29$; stellar mass of $\log(M_{\star}/M_{\odot})\sim7-8$) had rest-frame UV colors consistent with essentially no metals \citep[e.g.,][]{bouwens10b,finkelstein10}, though follow-up studies with larger samples accounting for selection biases \citep[e.g.,][]{dunlop12} were able to rule out Population III-dominated galaxies at this epoch \citep{finkelstein12a, dunlop13, bouwens14}. 

Looking at the full dynamic range of the galaxy population, correlations of the rest-UV color have been found with both the UV luminosity \citep{bouwens14, stefanon21_beta} and stellar mass \citep{finkelstein12a, bhatawdekar21}, where more luminous/massive systems have redder observed colors. In particular, \citet{finkelstein12a} found that the most massive galaxies in their sample ($\log(M_{\star}/M_{\odot})\sim9-10$) had similarly red rest-UV colors from $z=4-7$, indicating a roughly constant level of dust attenuation in these galaxies. Pushing these measurements to even earlier cosmic epochs can constrain exactly when dust began forming in the early universe, potentially constraining the respective efficiencies of different dust production mechanisms \citep[e.g.,][]{valiante11, valiante14, mancini15, mancini16, popping17_dust, aoyama18, graziani20}.

Additionally, the rest-frame UV emission of these high-redshift galaxies, which can be currently probed with {\it HST} in the near-IR, contains a wealth of information regarding the ages of the stars: relatively young stars ($\sim10^7~\mathrm{yr}$) will dominate the observed emission in both the far- and near-UV rest frame, while older stars ($\sim10^9~\mathrm{yr}$) will contribute more to the near-UV than they do to the far-UV \citep[e.g.,][]{conroy13_rev}. As at early cosmic times the stellar populations are younger (with an upper limit given by the age of the universe), the rest-UV light can be used to infer stellar ages and stellar masses. A major challenge in using the UV as a tracer for the stellar age and the star-formation history in general is the degeneracy with other galaxy properties, including the wavelength-dependent attenuation of the UV emission by dust and the metallicity content of the stars \citep[e.g.,][]{papovich01}. 

Extending the wavelength coverage further into the rest-frame optical is helpful to constrain the stellar populations and break some of these degeneracies. This can currently be done with {\it Spitzer}/IRAC for $z>8$ galaxies \citep[e.g.,][]{stefanon19}, though it remains challenging due to low signal-to-noise and deblending issues. Furthermore, emission lines such as H$\beta$, [\ion{O}{3}] and [\ion{O}{2}] can contaminate the IRAC bands and therefore can be confused with a strong Balmer/$4000~\mathrm{\AA}$ break \citep{labbe13, finkelstein13, smit15, faisst16, roberts-borsani16, de-barros19, endsley21_ew}. 

Several studies have made use of combining {\it HST} with {\it Spitzer}/IRAC data in order to constrain the stellar populations of $z>8$ galaxies. \citet{stefanon19} measured for 18 bright $z=8$ galaxies an average stellar mass of $M_{\star}=10^{9.1_{-0.4}^{+0.5}}~M_{\odot}$, star-formation rate (SFR) of $\mathrm{SFR}=32_{-32}^{+44}~M_{\odot}~\mathrm{yr}^{-1}$, and stellar age of $22_{-22}^{+69}~\mathrm{Myr}$. At higher redshifts, MACS1149-JD1 at $z_{\rm spec}=9.11$ gained a lot of attention due to its red IRAC color, which was attributed to old stellar populations \citep{Zheng12}. In particular, \citet{hashimoto18} inferred an age of $290\pm150~\mathrm{Myr}$ by fitting a young and old stellar population to the data \citep[see also][]{roberts-borsani20_spitzer}. \citet{laporte21} studied six $z\sim9$ galaxies selected to have 4.5$\mu$m flux excesses (out of which 3 have a spectroscopic redshift) and found stellar ages (here referring to the time since beginning of star formation) of $200-500$ Myr (age of MACS1149-JD1 is consistent with 500 Myr), with the best fit being always obtained for a delayed or constant star-formation history (SFH). 

Constraints on these ages provide our crucial first glimpse into the buildup of stellar mass at $z >$ 10. One of the major systematic uncertainties on the ages of these galaxies from previous studies is the choice of the SFH. The derived ages are crucially dependent on this assumption \citep[e.g.,][]{papovich11, curtis-lake13, schaerer13_sed, buat14, leja19, lower20, tacchella21_quench}. This choice is also directly apparent when deriving the evolution of the cosmic SFR density from these ages and inferred SFHs. The cosmic SFR density is usually inferred from the UV luminosity function, with some studies suggesting the cosmic SFR declines with redshift more steeply at $z >$ 8 than at 4 $< z <$ 8 \citep[e.g.,][]{oesch18,bouwens21}, while others suggest the evolution continues with a more shallow decline \citep[e.g.,][]{mcleod16, finkelstein16}. \citet{laporte21} investigated this by averaging the best-fit SFHs of their six galaxies, determining that these galaxies formed $\sim70\%$ of their mass by $z=10$, which favors a smooth increase in the cosmic SFR density with time. However, the extent to which the priors on the assumed SFH and stellar population parameters impact the results have not yet been deeply investigated.

We present a new analysis of the properties of $z\sim9$ galaxies using a more flexible treatment of the SFH. We use a newly published sample of moderately bright $z >$ 8.5 galaxies \citep[][hereafter F21]{finkelstein21}. These sources are selected in the {\it HST} CANDELS fields \citep{grogin11, koekemoer11} and thus have an array of deep {\it HST} imaging available; yet they are also selected to be bright ($m_{\rm F160W} <$ 26.6), allowing meaningful {\it Spitzer}/IRAC constraints on their rest-frame optical fluxes, which are crucial to constrain their stellar populations.

We perform a careful inference on the stellar populations by using \texttt{Prospector}, a flexible Bayesian spectral energy distribution (SED) fitting code \citep{johnson21}. In particular, we expand upon previous $z>6$ SED investigations by adopting a range of simple and flexible models for the SFHs, a flexible dust attenuation law, self-consistent modeling of emission lines, and variable IGM absorption. We explore the dust reddening in these galaxies and thoroughly investigate how our inferred stellar ages depend on the adopted SFH prior. We conclude that our data are unable to meaningfully constrain the SFHs of these high-$z$ galaxies, consistent with findings at lower redshifts \citep{strait21}. Specifically, the SFHs can be consistent with either a rapid or slow increase in the cosmic SFR density with time at $z>9$. 

The paper is structured as follows. Section~\ref{sec:sample} introduces the galaxy sample and its selection. Section~\ref{sec:prospector} describes in detail the assumptions in our SED modeling. Section~\ref{sec:enrichment} discusses our key results concerning the chemical enrichment, while Section~\ref{sec:mass_growth} focuses on the inferred growth of stellar mass and its implication on early cosmic star formation. We conclude in Section~\ref{sec:conclusion}.

Throughout this work, all magnitudes are presented in the AB system, and we assume for the cosmological parameters $H_0=67.74~\mathrm{km}~\mathrm{s}^{-1}~\mathrm{Mpc}^{-1}$, $\Omega_{\rm M}=0.309$ and $\Omega_{\Lambda}=0.691$, consistent with the recent \citet{planck-collaboration20} measurements.

\section{Galaxy Sample}
\label{sec:sample}

In this work, we study the sample of 11 bright ($H<26.6$) galaxy candidates selected in the CANDELS fields by F21. Many of these sources were also presented in \citet{oesch18} and \citet{bouwens19}. In F21 the authors created new photometric catalogs for each of the five CANDELS fields measuring accurate colors and total fluxes for all available {\it HST} imaging bands and obtained deblended photometry in the IRAC/{\it Spitzer} bands using \texttt{TPHOT}, following \citet{song16} and \citet{merlin16}. Photometric redshifts were measured using \texttt{EAZY} \citep{brammer08}, using a large set of templates including a very blue template to match the expected colors of some high-redshift galaxies. Candidates were initially selected using a combination of criteria designed to select well-detected objects with photometric redshifts robustly constrained to be at $z >$ 8. F21 noted that using {\it Spitzer}/IRAC in tandem with {\it HST} in the initial selection process more robustly removed potential contaminating systems but also resulted in tighter redshift constraints for likely high-redshift candidates.

This initial sample of galaxies was vetted in a variety of ways, including several screens against non-galactic sources (noise, persistence, stellar sources), the addition of ground-based photometric constraints, and follow-up {\it HST} imaging in additional filters. The final sample of 11 galaxies continued to satisfy all stringent criteria for a likely high-redshift nature. We list these 11 sources in Table 1, and make use of the photometry published in the tables in F21. We note that three of these sources are spectroscopically confirmed, as noted in Table 1. We refer the reader to F21 for further details on the photometric measurements, photometric redshifts, and sample validation.

\begin{deluxetable}{cccc}
\vspace{2mm}
\tabletypesize{\small}
\tablecaption{Bright ($H <$ 26.6) $z >$ 8.5 Galaxy Sample}
\tablewidth{\textwidth}
\tablehead{
\colhead{ID} & \colhead{R.A.} & \colhead{Decl.} & \colhead{m$_{F160W}$}\\
\colhead{$ $}  & \colhead{(J2000)} & \colhead{(J2000)} & \colhead{(mag)}}
\startdata
EGS-6811$^{\ast}$&215.035385&52.890666&25.2\\
EGS-44164$^{\ast}$&215.218737&53.069859&25.4\\
EGS-68560&214.809021&52.838405&25.8\\
EGS-20381&215.188415&53.033644&26.0\\
EGS-26890&214.967536&52.932966&26.1\\
EGS-26816&215.097775&53.025095&26.1\\
EGS-40898&214.882993&52.840414&26.5\\
COSMOS-20646&150.081846&2.262751&25.4\\
COSMOS-47074&150.126386&2.383777&26.3\\
UDS-18697&34.255636&-5.166606&25.3\\
GOODSN-35589$^{\ast}$&189.106061&62.242040&25.8
\enddata
\tablecomments{The sample of photometric redshift selected $z >$ 8.5 galaxies studied in this work are taken from F21. $^{\ast}$These objects have spectroscopic redshifts, as listed in Table 3.}
\label{tab:sample}
\end{deluxetable}

\begin{deluxetable*}{p{0.1\textwidth} p{0.4\textwidth} p{0.4\textwidth}}
\tablecaption{Summary of 14 parameters and priors for the fiducial physical model with a flexible star-formation history (SFH) within \texttt{Prospector}. \label{tab:parameters}}
\tablehead{
\colhead{Parameter} & \colhead{Description} & \colhead{Prior}
}
\startdata
\hline
$z_{\rm phot}$ & redshift & prior from \texttt{EAZY} or fixed to $z_{\rm spec}$ \\
$\log(\mathrm{Z}/\mathrm{Z}_{\odot})$ & stellar metallicity & uniform: $\mathrm{min}=-2.0$, $\mathrm{max}=0.19$ \\
$\log(\mathrm{M}_{\star}/\mathrm{M}_{\odot})$ & total stellar mass formed & uniform: $\mathrm{min}=6$, $\mathrm{max}=12$\\
SFH & flexible SFH: ratio of the SFRs in adjacent time bins of the $N_{\rm SFH}$-bin SFH ($N_{\rm SFH}-1$ parameters total, with default choice $N_{\rm SFH}=6$); parametric SFH: delayed-$\tau$ model with two free parameters & see Section~\ref{subsec:prior_sfh} for details \\
$n$ & power-law modifier to shape of the \citet{calzetti00} attenuation curve of the diffuse dust & uniform: $\mathrm{min}=-1.0$, $\mathrm{max}=0.4$ \\
$\hat{\tau}_{\rm dust, 2}$ & diffuse dust optical depth & clipped normal: $\mathrm{min}=0$, $\mathrm{max}=4$, $\mu=0.3$, $\sigma=1$\\
$\hat{\tau}_{\rm dust, 1}$ & birth-cloud dust optical depth & clipped normal in ($\tau_{\rm dust, 1}/\tau_{\rm dust, 2}$): $\mathrm{min}=0$, $\mathrm{max}=2$, $\mu=1$, $\sigma=0.3$ \\
$\log(\mathrm{Z}_{\mathrm{gas}}/\mathrm{Z}_{\odot})$ & gas-phase metallicity & uniform: $\mathrm{min}=-2.0$, $\mathrm{max}=0.5$ \\
$\log(\mathrm{U})$ & ionization parameter for the nebular emission & uniform: $\mathrm{min}=-4.0$, $\mathrm{max}=-1$ \\
$f_{\rm IGM}$ & scaling of the IGM attenuation curve & clipped normal: $\mathrm{min}=0$, $\mathrm{max}=2$, $\mu=1$, $\sigma=0.3$ \\
\hline
\enddata
\end{deluxetable*}

\section{Constraining Stellar Population Posteriors with \texttt{Prospector}}
\label{sec:prospector}

We constrain the stellar populations by using \texttt{Prospector} \citep{johnson21}, a fully Bayesian inference code to derive stellar population properties from photometric and/or spectroscopic data. \texttt{Prospector} has been mainly employed on galaxies at lower redshifts \citep[e.g.,][]{leja19, webb20, belli21, tacchella21_quench}. The \texttt{Prospector} fit for one high-$z$ galaxy (GOODSN-35589) has been presented in \citet{johnson21} as a demonstration. We adopt a similar physical model for the galaxy SED as in \citet{johnson21} with details given in Section~\ref{subsec:model_sed}, with  Section~\ref{subsec:prior_sfh} highlighting the prior on the SFH. Section~\ref{subsec:prospector_results} assesses the goodness of the SED fits and shows the dust attenuation curve posteriors. Finally, Section~\ref{subsec:photo_z} compares the photometric redshifts from \texttt{EAZY} to the ones obtained here with \texttt{Prospector}.

\subsection{Physical model for the galaxy SED}
\label{subsec:model_sed}

We use the Flexible Stellar Population Synthesis (\texttt{FSPS}) package \citep{conroy09a, conroy10} with the MIST stellar evolutionary tracks and isochrones \citep{choi16, dotter16}. The MIST isochrones include the effects of rotation that boost the ionizing flux production of massive stars in a manner similar to the effect of binaries \citep{choi17}. Furthermore, throughout this work, we assume a \citet{chabrier03} initial mass function.

Our fiducial physical model consists of 14 free parameters describing the
contribution of stars, gas and dust (Table~\ref{tab:parameters}). While not all 14 parameters are constrained by the photometric data, the use of a highly flexible model together with physically motivated priors prevents the results from being overinterpreted. Therefore, the choice of the priors is important. As we show in Section~\ref{sec:mass_growth}, a key conclusion of the paper is that the inferred early mass growth of galaxies heavily depends on the prior on the SFH.

In our fiducial runs, we adopt the \texttt{EAZY} posterior (Section~\ref{sec:sample}; F21) as a prior for the photometric redshift or fix the redshift to the spectroscopic redshift $z_{\rm spec}$ when available. Three galaxies have a spectroscopic redshift: EGS-6811 with $z_{\rm spec}=8.678$ \citep{zitrin15}, EGS-44164 with $z_{\rm spec}=8.665$ (Larson et al., submitted), GOODSN-35589 with $z_{\rm spec}=10.957$ \citep{oesch16,jiang20_lya}. The main motivation for us to assume the \texttt{EAZY} posterior as redshift prior is that it allows us to focus on posterior sampling on the stellar population part instead of the redshift space (\texttt{Prospector} is a rather expensive to run in terms of time) and also to propagate the redshift uncertainty into the \texttt{Prospector} modeling. We model the chemical enrichment histories of the galaxies with a delta function, i.e., assuming that all stars within the galaxy have the same metal content with scaled-Solar abundances.  This single metallicity is varied with a prior that is uniform in $\log(Z/Z_{\odot})$ between $-2.0$ and 0.19, where $Z_{\odot}=0.0142$ \citep{asplund09}. 

One of the key strengths of the SED fitting code \texttt{Prospector} is the possibility to adopt flexible SFHs. Specifically, we adopt SFHs in our fiducial model, which do \textit{not} assume a certain shape with time\footnote{Since our fiducial SFHs are not a parametric function of $t$, these SFHs are sometimes also called non-parametric.} and are simply partitioned into time bins. The SFHs are characterized by the ratios of the SFRs in adjacent time bins. There are 5 free parameters for 6 time bins in addition to the total stellar mass. Furthermore, we also explore a parametric, delayed-$\tau$ model (two free parameters). Details about the SFH prior are given in Section~\ref{subsec:prior_sfh}. For the total stellar mass $M_{\star}$, we assume a flat prior in log-space in the range of $6<\log(M_{\star}/M_{\odot})<12$. Throughout this work, the stellar mass $M_{\star}$ denotes the integral of the SFH, i.e. it is the mass of all stars ever formed. 

We  model  dust  attenuation  using  a  two-component dust attenuation  model with a flexible attenuation curve (see \citealt{charlot00}). The first component is a birth-cloud component in our model that attenuates nebular emission and stellar emission only from stars formed in the last 10 Myr (attenuation law is a power law with a slope of $-1$). The second component is a diffuse component that has a variable attenuation curve and attenuates all stellar and nebular emission from the galaxy. We use the prescription from \citet{noll09} with a \citet{kriek13} attenuation curve, where the slope $n$ of the curve (dust index) is a free parameter and is directly linked to the strength of the UV bump. The dust index $n$ is modeled as an offset from the slope of the UV attenuation curve from \citet{calzetti00}. In total, the attenuation prescription has three free parameters: ($i$) the slope $n$ (flat prior between $-1.0<n<0.4$); ($ii$) the normalization $\hat{\tau}_{\rm dust, 2}$ of the diffuse dust component (flat prior between $0<\hat{\tau}_{\rm dust, 2}<4.0$); and ($iii$) the normalization $\hat{\tau}_{\rm dust, 1}$ of the birth-cloud component, which we model as a ratio with respect to the diffuse component (prior is a clipped normal centered on 1 with width of 0.3 in the range of $0 < \hat{\tau}_{\rm dust, 1}/\hat{\tau}_{\rm dust, 2} < 2.0$, motivated by \citealt{calzetti94, price14}).

\begin{figure*}
\centering
\includegraphics[width=\textwidth]{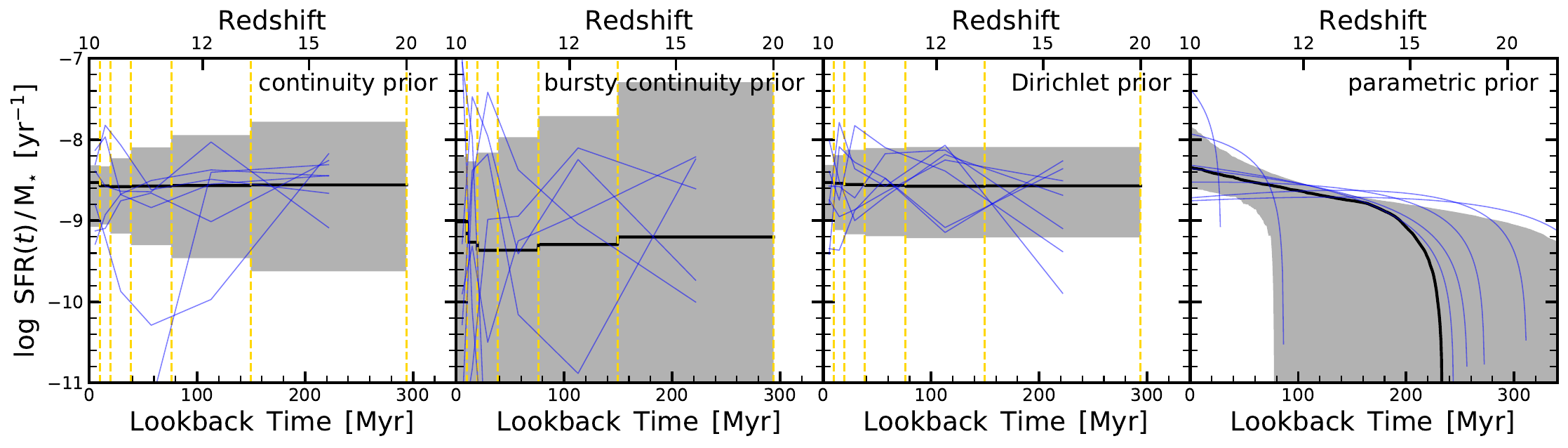}
\caption{Different choices for the SFH prior produce different behavior in $\mathrm{SFR}(t)$ (see also \citealt{leja19_nonparm} and \citealt{tacchella21_quench}). The panels from left to right show 100,000 random draws from the continuity prior (our fiducial SFH prior), the bursty continuity prior, the Dirichlet prior and the parametric prior (Section~\ref{subsec:prior_sfh}). The solid black lines mark the median in time for these draws, while the gray shaded regions indicate the 16-84th percentiles. The vertical yellow lines show the six time bins for the flexible SFHs. In each panel, seven individual draws (i.e., SFHs) are shown as blue lines to illustrate the behavior of different priors. The bursty continuity prior is weighted in order to produce multiple bursts and quenching episodes in comparison with the smoother continuity prior and Dirichlet prior. The parametric prior introduces a specific shape with an increasing SFH with time. All of these priors are able to fit the data equally well.}
\label{fig:prior_sfh}
\end{figure*}

The nebular emission (emission lines and continuum) is self-consistently modeled \citep{byler17}. We have two parameters: the gas-phase metallicity ($Z_{\rm gas}$) and the ionization parameter ($U$). We assume a flat prior in log-space for the metallicity ($-2.0<\log(Z_{\rm gas}/Z_{\odot})<0.5$) and ionization ($-4<\log(U)<-1$) parameters. Importantly, we do not link  the gas-phase metallicity $Z_{\rm gas}$ and the stellar metallicity $Z_{\star}$, i.e. $Z_{\rm gas}$ and $Z_{\star}$ are decoupled from each other. We choose to decouple the gas-phase from the stellar metallicity because it allows us to cover both cases where the $Z_{\rm gas}$ is smaller or larger than the $Z_{\star}$. Both cases are expected in the evolution of galaxies. Specifically, in the case of a closed-box chemical model, we expect the stellar metallicity to be always smaller than the gas-phase metallicity. This might be true in certain phases of the galaxy's lifetime. However, galaxies also accrete new gas, which typically has a lower metallicity than gas (and stars) already present in the galaxy, leading to a gas-phase metallicity that can be lower than the stellar metallicity. The main consequence of this assumption is an overall larger flexibility in the SED modeling, in particular in regards to the emission line strengths. 

As the photometry probes the rest-frame $\lambda<1216~\mathrm{\AA}$ spectrum at high redshifts, we include a $z$-dependent IGM attenuation following \citet{madau95}. This includes a free parameter that scales the total IGM opacity ($f_{\rm IGM}$), intended to account for line-of-sight variations in the total opacity. We adopt for $f_{\rm IGM}$ a clipped Gaussian prior distribution centered on 1, with a dispersion of 0.3 and clipped at 0 and 2.

\subsection{Priors for the SFH}
\label{subsec:prior_sfh}

In order to explore the robustness of our inferred mass assembly histories, we want to explore the dependence of our results on the assumed SFH prior. As noted above, the strength of \texttt{Prospector} is the possibility of adopting a flexible SFH (see also \citealt{iyer17, iyer19} for another SED fitting code with a flexible SFH approach). We assume four different priors for the SFH: three are flexible SFHs (``continuity prior'', ``bursty continuity prior'' and ``Dirichlet prior''), while the fourth is a parametric SFH with the shape of the delayed-$\tau$ model (``parametric prior''). The strength of the flexible SFH priors is that they are not parametric functions of time (in contrast to the parametric delayed-$\tau$ prior), which allows for a large flexibility regarding the shape of the SFH. Figure~\ref{fig:prior_sfh} illustrates the behavior of these four different priors by plotting the median trend of the SFH and individual draws. 

For the flexible SFHs, we assume that the SFH can be described by $N_{\rm SFH}$ time bins, where the SFR within each bin is constant. We fix $N_{\rm SFH}=6$ and specify the time bins in lookback time. The first bin is fixed at $0-10$ Myr to capture variation in the recent SFH of galaxies, while the other bins are spaced equally in logarithmic time between 10 Myr and a lookback time that corresponds to $z=20$, i.e., we assume the $\mathrm{SFR}=0~M_{\odot}/\mathrm{yr}$ at $z>20$ (a reasonable assumption given what observational constraints and theoretical predictions exist for this epoch; \citealt{maio10, bowman18, jaacks19}). These time bins are plotted as vertical dashed yellow lines in Figure~\ref{fig:prior_sfh}. We fit for the ratio between the time bins ($N_{\rm SFH}-1$ free parameters) and the total stellar mass formed, which has a flat prior in log-space in the range of $6<\log(M_{\star}/M_{\odot})<12$. 

Impact of the choice regarding the number of time bins has been extensively discussed in \citet[][see their Appendix A]{leja19_nonparm}. They explored varying the number of time bins between $N_{\rm SFH}=4-14$ and show that the results of the mock analysis are largely insensitive to the number of bins as long as $N_{\rm SFH}\gtrsim4$. Although their mock analysis cannot be translated to our analysis one-to-one, we think that this main conclusion still holds because they investigated lookback times of 10 Gyr, much longer than the age of the universe at the epochs of our objects. Therefore, our log-spaced bins with a width of typically less than 100 Myr should be enough to convey all of the necessary information in the data.

\begin{figure*}
\centering
\includegraphics[width=\textwidth]{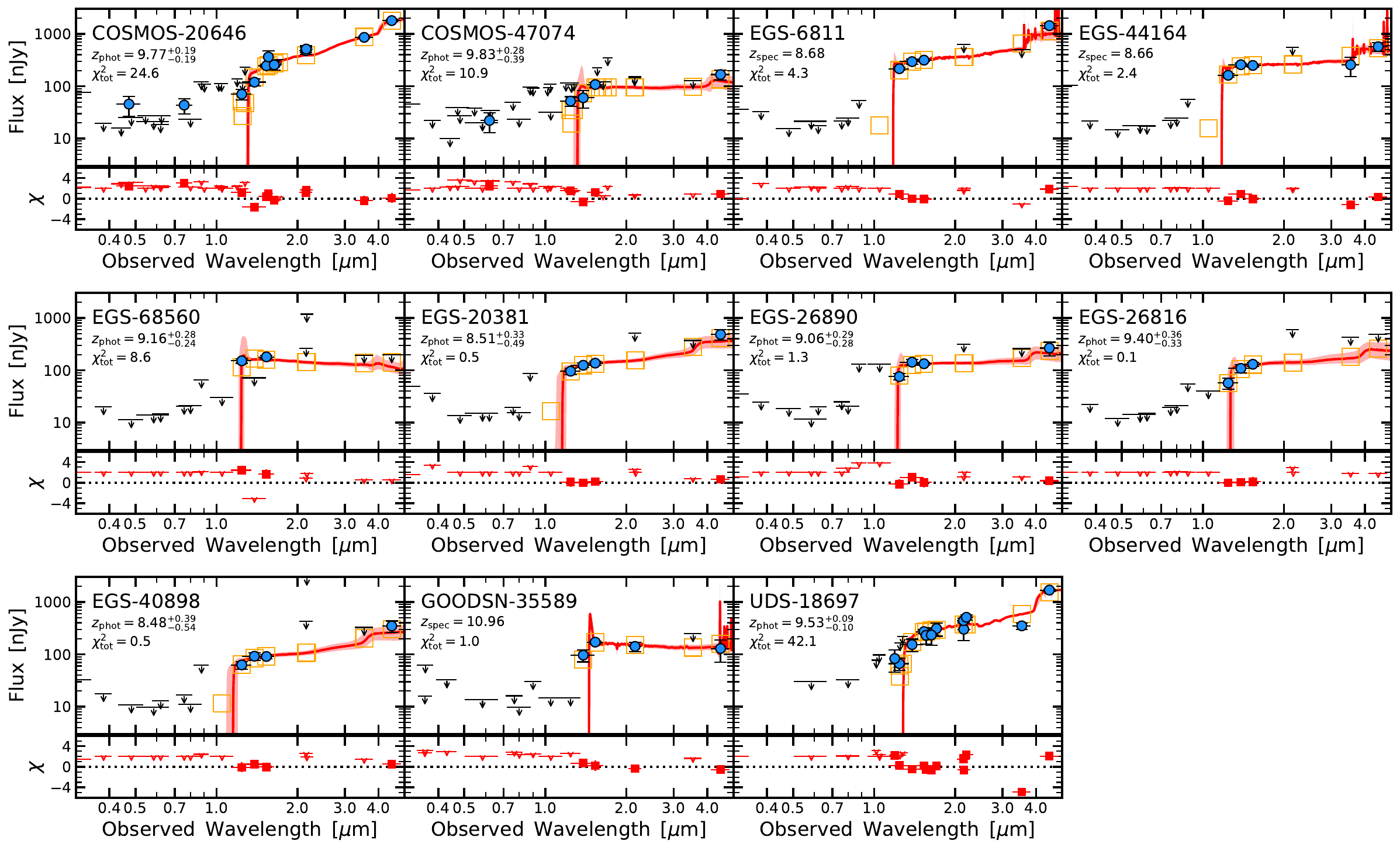}
\caption{The observed and model posterior spectral energy distributions (SEDs) for the 11 $z=9-11$ galaxy candidates in our sample. The blue symbols mark the fluxes of the detected bands, while the arrows show the upper limits (2$\sigma$ confidence). The red line and shaded region indicate the median and 16-84th percentile of the posterior SED. The orange squares plot the model fluxes for the median model. The lower panels show the $\chi$ distribution. The model is able to reproduce the data overall well. For three galaxies (EGS-6811, EGS-44164 and GOODSN-35589), the redshift has been fixed during the fitting to their spectroscopic redshift, which leads to the manifestation of the emission lines in the posterior model SEDs. For the remainder, the redshift prior was set to the F21 EAZY posterior, with the photometric redshift listed here measured from the resulting \texttt{Prospector} posterior. For each fit, we also give the total $\chi^2_{\rm tot}$.}
\label{fig:SEDs}
\end{figure*}

The priors for flexible SFHs are extensively discussed in \citet[][see also \citealt{tacchella21_quench}]{leja19_nonparm}, while parametric SFHs are explored in \citet{carnall19_sfh}. We adopt a continuity prior as well as the Dirichlet prior. For the continuity prior, we directly fit for the $\Delta\log(\mathrm{SFR})$ between adjacent time bins. We adopt the Student's t-distribution $\Delta\log(\mathrm{SFR})$. For the ``continuity prior'', we assume a Student's t-distribution with $\sigma=0.3$ and $\nu=2$, which weights against sharp transitions and is motivated by simulated SFHs at $z\sim1$ \citep{leja17}. This is our fiducial SFH prior. For the ``bursty continuity prior'', we adopt $\sigma=1.0$ and $\nu=2$, which leads to a more variable (i.e., bursty) SFH. In the case of the ``Dirichlet prior'', the fractional sSFR for each time bin follows a Dirichlet distribution \citep{leja17}. We assume a concentration parameter of 1, which weights toward smooth SFHs. As shown in Figure~\ref{fig:prior_sfh}, both the continuity and the Dirichlet prior include a symmetric prior in age and sSFR and an expectation value of constant $\mathrm{SFR}(t)$. The key difference from the Dirichlet prior is that the continuity prior explicitly weights against sharp changes in $\mathrm{SFR}(t)$.

Finally, for the parametric SFH, we assume a delayed-$\tau$ model of the form:

\begin{equation}
    \mathrm{SFR}(t) = (t-t_{\rm a})e^{-(t-t_{\rm a})/\tau}
    \label{eq:delayed_tau}
\end{equation}

\noindent
The parameter $\tau$ is varied as $\log(\tau)$ within a uniform prior in the range of $-1.0<\log(\tau)<10.0$, and the parameter $t_{\rm a}$ with a uniform prior between 1 Myr and the age of the universe at the galaxies' redshift $z_{\rm phot}$ ($t_{\rm H}(z_{\rm obs})$). Despite this large prior space for the parameters $\tau$ and $t_{\rm a}$, the resulting SFH from the parametric prior follows a specific shape of an increasing SFH with time, as shown in Figure~\ref{fig:prior_sfh}, consistent with constraints on SFHs in the epoch of reionization \citep[e.g.,][]{papovich11}.

\subsection{Resulting posteriors}
\label{subsec:prospector_results}

After setting up the physical galaxy SED model with 14 free parameters, we fit this model to the photometric data (Section~\ref{sec:sample}) within the \texttt{Prospector} framework using the dynamic nested sampling algorithm \texttt{dynesty} \citep{speagle20}, which allows us to perform an efficient sampling of the high-dimensional and complex parameter space. A strength of \texttt{Prospector} together with \texttt{dynesty} is its ability to infer full posterior distributions of the SED parameters and their degeneracies. We discuss these SED parameters and their inferred properties, such as the stellar mass, metallicity, and SFH, in the upcoming sections, but see Table~\ref{tab:posteriors} for a summary of the main physical parameters. Here, we focus on the resulting SEDs and compare them with the measured photometry in order to assess the goodness of the fits. Then we briefly discuss the resulting posteriors of the dust attenuation parameters. 

\begin{deluxetable*}{lcccccccc}\tabletypesize{\footnotesize}
\label{tab:posteriors}
\tablecaption{Results for the main physical parameters from our fiducial run assuming the continuity prior. The values are the median of the posterior, while the errors indicate the 16-84th percentiles. \label{tab:fiducial_results}}
\tablehead{
\colhead{ID} & \colhead{redshift} & \colhead{$M_{\rm UV, obs}$} & \colhead{UV slope $\beta$} & \colhead{$\log~M_{\star}$} & \colhead{$\log~\mathrm{SFR}_{50}$} & \colhead{$\log~\mathrm{sSFR}_{50}$} & \colhead{$A_{\rm V}$} & \colhead{$\log~Z$} \\
\colhead{} & \colhead{} & \colhead{$[\mathrm{mag}]$} & \colhead{} & \colhead{$[M_{\odot}]$} & \colhead{$[M_{\odot}~\mathrm{yr}^{-1}]$} & \colhead{$[\mathrm{Gyr}^{-1}]$} & \colhead{$[\mathrm{mag}]$} & \colhead{$[Z_{\odot}]$}
}
\startdata
\hline
COSMOS-20646 & $9.77_{-0.19}^{+0.19}$ & $-22.10_{-0.09}^{+0.09}$ & $-0.62^{+0.11}_{-0.12}$ & $10.9_{-0.2}^{+0.2}$ & $2.5_{-0.3}^{+0.3}$ & $0.7_{-0.4}^{+0.3}$ & $0.9_{-0.5}^{+0.6}$ & $-0.4_{-0.7}^{+0.3}$ \\
COSMOS-47074 & $9.83_{-0.39}^{+0.28}$ & $-21.01_{-0.12}^{+0.12}$ & $-2.11^{+0.20}_{-0.18}$ & $9.3_{-0.4}^{+0.3}$ & $1.0_{-0.2}^{+0.3}$ & $0.8_{-0.4}^{+0.3}$ & $0.1_{-0.1}^{+0.3}$ & $-1.2_{-0.6}^{+0.7}$ \\
EGS-6811 & $z_{\rm spec}=8.68$ & $-22.10_{-0.05}^{+0.05}$ & $-1.61^{+0.18}_{-0.12}$ & $10.6_{-0.3}^{+0.2}$ & $2.0_{-0.3}^{+0.4}$ & $0.5_{-0.4}^{+0.3}$ & $0.7_{-0.4}^{+0.6}$ & $-0.5_{-0.7}^{+0.4}$ \\
EGS-44164 & $z_{\rm spec}=8.66$ & $-21.87_{-0.05}^{+0.05}$ & $-1.87^{+0.11}_{-0.11}$ & $10.2_{-0.2}^{+0.2}$ & $1.6_{-0.3}^{+0.4}$ & $0.5_{-0.4}^{+0.4}$ & $0.4_{-0.2}^{+0.5}$ & $-1.3_{-0.4}^{+0.6}$ \\
EGS-68560 & $9.16_{-0.24}^{+0.28}$ & $-21.47_{-0.09}^{+0.09}$ & $-2.37^{+0.15}_{-0.11}$ & $9.1_{-0.3}^{+0.2}$ & $1.0_{-0.2}^{+0.2}$ & $1.0_{-0.2}^{+0.2}$ & $0.1_{-0.0}^{+0.1}$ & $-1.5_{-0.3}^{+0.5}$ \\
EGS-20381 & $8.51_{-0.49}^{+0.33}$ & $-21.12_{-0.14}^{+0.18}$ & $-1.60^{+0.22}_{-0.20}$ & $10.0_{-0.4}^{+0.3}$ & $1.5_{-0.3}^{+0.4}$ & $0.6_{-0.4}^{+0.4}$ & $0.6_{-0.3}^{+0.5}$ & $-1.0_{-0.6}^{+0.7}$ \\
EGS-26890 & $9.06_{-0.28}^{+0.29}$ & $-21.25_{-0.10}^{+0.10}$ & $-1.94^{+0.14}_{-0.21}$ & $9.6_{-0.3}^{+0.3}$ & $1.3_{-0.3}^{+0.3}$ & $0.7_{-0.4}^{+0.3}$ & $0.2_{-0.2}^{+0.4}$ & $-1.1_{-0.6}^{+0.7}$ \\
EGS-26816 & $9.40_{-0.33}^{+0.36}$ & $-21.28_{-0.13}^{+0.13}$ & $-1.80^{+0.33}_{-0.24}$ & $9.7_{-0.4}^{+0.3}$ & $1.4_{-0.3}^{+0.3}$ & $0.8_{-0.4}^{+0.3}$ & $0.3_{-0.2}^{+0.4}$ & $-1.1_{-0.6}^{+0.8}$ \\
EGS-40898 & $8.48_{-0.54}^{+0.39}$ & $-20.70_{-0.16}^{+0.23}$ & $-1.45^{+0.28}_{-0.28}$ & $9.9_{-0.4}^{+0.4}$ & $1.4_{-0.4}^{+0.4}$ & $0.6_{-0.4}^{+0.4}$ & $0.6_{-0.3}^{+0.6}$ & $-1.0_{-0.7}^{+0.6}$ \\
GOODSN-35589 & $z_{\rm spec}=10.96$ & $-21.71_{-0.09}^{+0.08}$ & $-2.34^{+0.13}_{-0.14}$ & $9.1_{-0.2}^{+0.3}$ & $1.1_{-0.2}^{+0.2}$ & $1.0_{-0.2}^{+0.2}$ & $0.1_{-0.0}^{+0.1}$ & $-1.1_{-0.6}^{+0.8}$ \\
UDS-18697 & $9.54_{-0.10}^{+0.09}$ & $-22.14_{-0.07}^{+0.07}$ & $-1.41^{+0.08}_{-0.06}$ & $11.0_{-0.2}^{+0.4}$ & $1.7_{-0.6}^{+0.6}$ & $-0.4_{-0.5}^{+0.4}$ & $0.5_{-0.4}^{+0.9}$ & $-0.0_{-0.5}^{+0.1}$ \\
\hline
\enddata
\tablecomments{We list the galaxy identifier (ID), the redshift (photometric if not $z_{\rm spec}$ specified), the absolute UV magnitude at rest-frame $1500~\mathrm{\AA}$ ($M_{\rm UV, obs}$), the UV spectral slope ($\beta$), the stellar mass ($M_{\star}$), the SFR averaged over past 50 Myr ($\mathrm{SFR}_{50}$), the specific SFR averaged over past 50 Myr ($\mathrm{sSFR}_{50}$), the dust attenuation at $5500~\mathrm{\AA}$ ($A_{\rm V}$), and the stellar metallicity ($Z$).}
\end{deluxetable*}

\subsubsection{SEDs and goodness of fit}

Figure~\ref{fig:SEDs} shows the observed and modeled posterior SEDs for our 11 $z=9-11$ galaxy candidates. The blue circles show the detected photometric bands, while the arrows mark the upper limits. The red solid lines and shaded regions indicate the median and the 16-84th percentile of the posterior SED. These are the results for our fiducial SED run, where we adopt the continuity prior for the SFH and the \texttt{EAZY} posterior as the redshift prior (if no spectroscopic redshift is available). Although we include emission lines in all the fits (Section~\ref{subsec:model_sed}), they are typically not visible in the posterior SED because they are ``smeared'' out when marginalizing over the redshift posterior distribution. Nominal exceptions are the galaxies for which we fix the redshift to the spectroscopic redshift (e.g., EGS-6811, EGS-68560, and GOODSN-35589): for those objects, the emission lines are clearly visible.

The detections below the Lyman-$\alpha$ break in COSMOS-20646 and COSMOS-47074 are discussed in F21 (see their Figure 15). Briefly, the detections are only marginally significant ($\mathrm{SNR}=2-3$) and are also slightly offset from the source position. Consistent with the Prospector analysis here (see Section~\ref{subsec:photo_z}), when including these fluxes in the photometric redshift modeling (along with all the non-detections/upper limits), F21 also found the preference is still for a high-redshift solution, although a low-redshift solution is possible for both sources at a low (10\%) probability level.

The lower panels of Figure~\ref{fig:SEDs} show the $\chi$ values for the individual passbands, which is the difference between observed to model fluxes normalized by the observed errors. The individual $\chi$ values are typically around 1. We also quote the total $\chi^2_{\rm tot}$, which we estimate by summing the individual $\chi$ values of the detected bands. We do not quote the reduced $\chi^2$ values as the number of degrees of freedom is not well defined in a non-linear model such as considered here \citep{andrae10}. In summary, the model is able to reproduce the observational data well within the observational uncertainties.

These results are for our fiducial, continuity SFH prior, but the other SFH priors are able to reproduce the observational data equally well. Specifically, we find very similar (differences amount to less than 20\%) $\chi$ values for all the four SFH priors (Section~\ref{subsec:prior_sfh}). Furthermore, none of these priors is preferred by the data: the Bayes factor, i.e. the ratio of the evidences between the different models, is around 1. Specifically, the median and 16-84th percentile of Bayes factor of the bursty continuity prior, of the Dirichlet prior and of the parametric prior (all relative to the fiducial continuity prior) are $0.7_{-0.1}^{+0.7}$, $0.9_{-0.3}^{+0.1}$, and $1.1_{-0.3}^{+0.3}$, respectively. This inability to identify a preferred model can be attributed to both the small sample size and the limited information content of the observational data.

\subsubsection{Attenuation curve}
\label{subsec:attenuation_curve}

As highlighted in the previous section, the rest-frame wavelength coverage is limited due to the high-redshift nature of these sources. Specifically, we only cover the rest-frame UV and the Balmer/$4000~\mathrm{\AA}$-break, though the latter is only constrained by the IRAC photometry, which suffers from systematic uncertainties related to deblending (see also F21 and Appendix~\ref{appsec:galfit_phot}) and can also be contaminated by strong emission lines. Therefore, in order to constrain the buildup of stellar mass (i.e., SFHs), we need to properly interpret the rest-UV emission. 

\begin{figure}
\includegraphics[width=\linewidth]{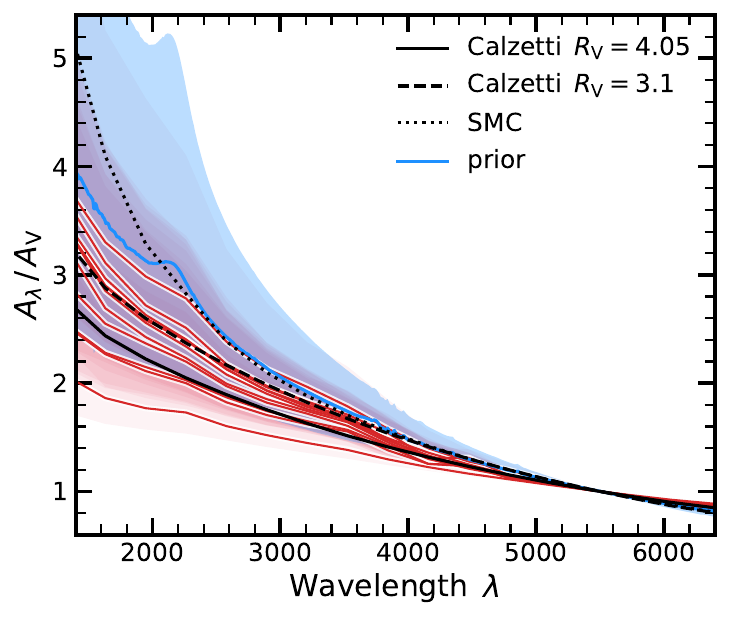}
\caption{Our fitted attenuation law for all 11 galaxies in our sample in comparison with the \citet{calzetti00} attenuation curve and an SMC-like \citep{pei92} attenuation curve. The SMC curve is shown as a dotted line, while the \citet{calzetti00} curves with $R_{\rm V}=3.1$ and $R_{\rm V}=4.05$ are plotted as dashed and solid lines, respectively. The blue solid line marks the median of the prior, while the associated shaded region indicates the 16-84th percentile of the prior. The individual solid lines and shaded regions in red show the median and 16-84th percentiles of our inferred posteriors. We find significant variations from galaxy to galaxy, and the posteriors are rather broad.}
\label{fig:attenuation_law}
\end{figure}

Flexibility in the attenuation law is motivated from observations \citep[e.g.,][]{johnson07, kriek13, battisti16, salmon16, salim18_curves} and theory \citep[e.g.,][]{seon16, narayanan18, shen20}. Specifically, \citet{katz19_sed} use a cosmological radiation hydrodynamics simulation to show that dust preferentially resides in the vicinity of the young stars, thereby increasing the strength of the measured Balmer break. Therefore, we adopt a flexible attenuation law (Section~\ref{subsec:model_sed}) so that we can marginalize over the uncertainty of an unknown attenuation law when constraining the SFHs and stellar ages among other physical properties.

\begin{figure*}
\centering
\includegraphics[width=\textwidth]{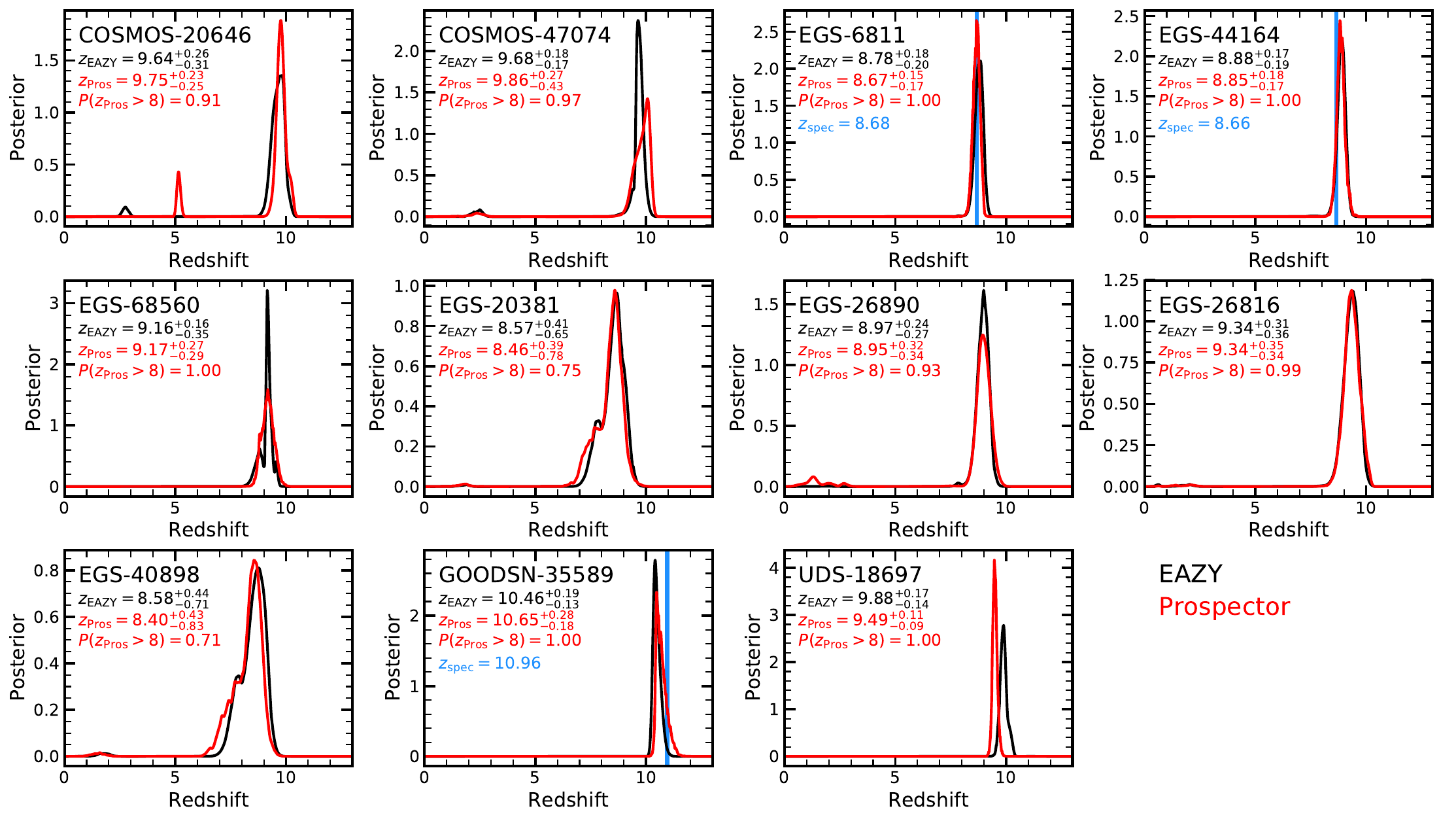}
\caption{Comparison of the photometric redshift ($z_{\rm phot}$) posteriors obtained with \texttt{EAZY} and \texttt{Prospector}. The results from \texttt{Prospector} assume the free-$z$ setup (see Section~\ref{subsec:photo_z}) with a uniform redshift prior between $z=0.1-13$. Each panel shows the \texttt{EAZY} and \texttt{Prospector} $z_{\rm phot}$ posteriors in black and red, respectively. Three galaxies (EGS-6811, EGS-68560 and GOODSN-35589) have spectroscopic redshifts, which are indicated as blue vertical lines. We find good agreement between \texttt{EAZY} and \texttt{Prospector}; see also Figure~\ref{fig:zcomp} for a direct comparison.}
\label{fig:z_phot}
\end{figure*}

Figure~\ref{fig:attenuation_law} shows the resulting posterior distributions of the attenuation as a function of wavelength of all galaxies in our sample. The median and the 16-84th percentiles are shown as solid red lines and red shaded regions, respectively. For comparison, we also plot the \citet{calzetti00} attenuation curve and an SMC-like \citep{pei92} attenuation curve. Furthermore, the blue solid line marks the median of the prior, while the blue shaded region indicates the 16-84th percentile of the prior.

We find that the attenuation laws of our galaxies are consistent with a \citet{calzetti00} law with $R_{\rm V}=3.1$, though significant variations from galaxy to galaxy are present. Interestingly, our obtained attenuation laws are typically shallower than the input prior. Furthermore, the uncertainty for individual galaxies is large, indicating that the attenuation law is not well constrained by our data and that degeneracies with, for example, the stellar age and metallicity exist (see also Figure~\ref{appfig:Auv_age}). Nevertheless, the posteriors of the attenuation laws of individual galaxies look physically sensible, which supports the choice of priors (Table~\ref{tab:parameters}). Importantly, in the remainder of this paper, we marginalize over the uncertainty in the attenuation parameters (i.e. attenuation law).

\subsection{Confirmation of the photometric redshifts}
\label{subsec:photo_z}

\begin{figure}
\centering
\includegraphics[width=0.45\textwidth]{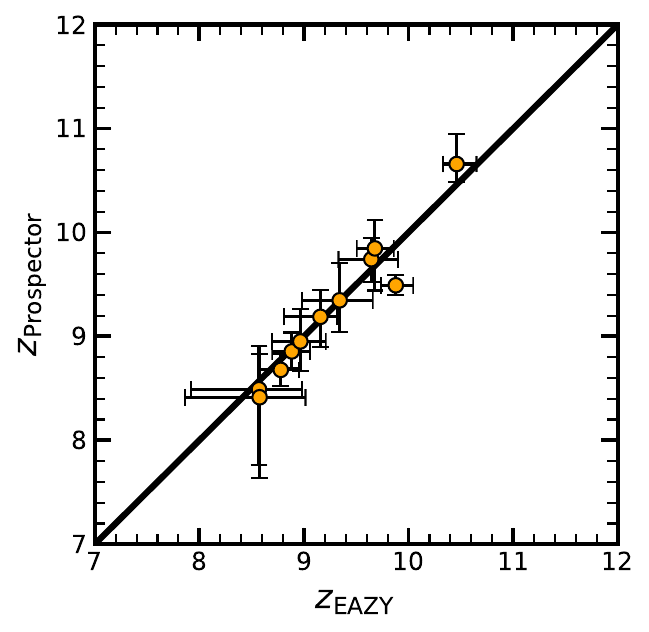}
\caption{A comparison of the \texttt{Prospector}-based photometric redshifts to the \texttt{EAZY}-based photometric redshifts from F21. The points with the errorbars indicate the median and 16-84th percentiles for the redshift posterior after marginalizing over all other SED parameters. The \texttt{Prospector}-based photometric redshifts are obtained with the free-$z$ setup (see Section~\ref{subsec:photo_z}), which assumes a uniform redshift prior between $z=0.1-13$. We find good agreement between the \texttt{Prospector}-based and \texttt{EAZY}-based photometric redshifts.}
\label{fig:zcomp}
\end{figure}

As discussed in Section~\ref{sec:sample} and F21, the photometric-redshift code \texttt{EAZY} has been used to perform our redshift estimation and to select our candidate $z=9-11$ galaxies. Although \texttt{EAZY} allows linear combinations of any number of provided templates, the explored parameter space is limited. In this section, we explore the photometric-redshift constraints that we obtain with \texttt{Prospector} and compare them with the \texttt{EAZY}-based photometric redshifts, finding excellent agreement.

In order to fit for the photometric redshift and also allow low-$z$ solutions, we have to extend the model that we introduced in Section~\ref{subsec:model_sed}. We call this the ``free-$z$ run'', where we let the redshift be free and assume a flat prior between 0.1 and 13. Second, we add dust emission and active galactic nucleus (AGN) emission in order to add flexibility to the SED in the infrared in order to reproduce dusty, lower-$z$ galaxies that have similar SEDs as the high-$z$ dropout candidates. In particular, the dust and AGN emission can dominate the near-IR flux, i.e., the emission in the IRAC bands at low redshifts. At higher redshifts ($z>3$), this emission contributes to bands at longer wavelengths than IRAC covers; hence, we do not have to consider this emission in our fiducial model.

We follow the description of \citet{leja17}, \citet{leja18_AGN} and \citet{tacchella21_quench}, which adds 5 new free parameters to our fiducial model, giving us a model with a total of 19 free parameters. Briefly, the three new parameters for the dust emission are $\gamma_{\rm e}$ (mass fraction of dust in high radiation intensity; log-uniform prior with minimum and maximum of $10^{-4}$ and 0.1), $\mathrm{U}_{\rm min}$ (minimum starlight intensity to which the dust mass is exposed; clipped normal prior with a mean of 2, a standard deviation of 1, minimum and maximum of 0.1 and 15), $\mathrm{q}_{\rm PAH}$ (percent mass fraction of PAHs in dust, uniform prior with minimum and maximum of 0.5 and 7.0). The two new parameters for the AGN emission are $f_{\rm AGN}$ (AGN luminosity as a fraction of the galaxy bolometric luminosity, log-uniform prior with minimum and maximum of $10^{-5}$ and 3) and $\tau_{\rm AGN}$ (optical depth of AGN torus dust, log-uniform prior with minimum of 5 and maximum of 150).

Figure~\ref{fig:z_phot} shows the photometric redshift posteriors obtained from \texttt{EAZY} (black lines) and \texttt{Prospector} (red lines). The results from \texttt{Prospector} assume the aforementioned free-$z$ run with a uniform redshift prior between $z=0.1-13$. Three galaxies (EGS-6811, EGS-68560 and GOODSN-35589) have spectroscopic redshifts, which are indicated in blue. We find overall good agreement between \texttt{EAZY} and \texttt{Prospector}. The two approaches return photometric redshifts that are consistent with each other within the uncertainty. This can also be seen directly in Figure~\ref{fig:zcomp}, which shows a comparison of the \texttt{Prospector}-based and the \texttt{EAZY}-based photometric redshifts. The circles and the errorbars indicate the median and 16-84th percentiles of the redshift posterior, respectively. 

Importantly for this work here, \texttt{Prospector} confirms the high-redshift nature of these galaxies. The probability of lying beyond $z=8$, $P(z>8)$, is larger than 90\% for all galaxies except EGS-20381 ($P(z>8)=0.75$) and EGS-40898 ($P(z>8)=0.71$) for which a tail in the $z_{\rm phot}$ posterior towards $z\sim6$ exists. Furthermore, for the galaxies that show minor peaks at $z\sim1-3$, the posteriors all have $P(z<6) < 0.1$, i.e., $<10\%$ of the posterior volume is at low redshift.

As mentioned above, we assume a flat redshift prior. This might actually not be the ideal prior as a flux-limited survey will contain many more low-$z$ than high-$z$ galaxies. We could therefore think of more complicated priors that, for example, weight according to the luminosity function and consider also the selection function of the survey. A detailed investigation of this is beyond the scope of this work. Nevertheless, since we have probably significantly overestimated the high-$z$ prior volume, we have also performed two additional ``free-$z$ run'', where we split the redshift prior in half by running $z_{\rm phot}$ in the range of $0.1-7.0$ and $7.0-13.0$. Although for some galaxies a viable low-$z$ solution is identified, the high-$z$ solution is preferred for all objects considering both the total $\chi^2_{\rm tot}$ values as well as the Bayes factor (i.e. ratio of the evidences in the Bayesian analysis). Specifically, we find that the high-$z$ run has a lower $\chi^2_{\rm tot}$ by a factor of 2--4 than the low-$z$ run. Furthermore, we obtain for the Bayes factor a median of $\mathcal{Z}_{\rm low-z}/\mathcal{Z}_{\rm high-z}=3\times10^{-2}$ over the whole sample, with all galaxies $\mathcal{Z}_{\rm low-z}/\mathcal{Z}_{\rm high-z}<0.5$. This shows that the high-$z$ model is preferred for all galaxies, which is consistent with our findings above.

Finally, as  mentioned in F21, we have performed the IRAC photometric deblending in two different ways, once with \texttt{TPHOT} (fiducial) and once with \texttt{GALFIT}. We discuss the results of changing from \texttt{TPHOT} to \texttt{GALFIT} IRAC photometry in Appendix~\ref{appsec:galfit_phot} and Figure~\ref{appfig:z_phot_galfit}. In summary, we find consistent photometric redshift estimates for all galaxies except COSMOS-20646 ($z_{\rm phot}=2.63_{-0.16}^{+0.16}$) and EGS-20381 ($z_{\rm phot}=6.79_{-0.13}^{+0.57}$). This is consistent with the \texttt{EAZY}-based results with this photometry from F21 for these two objects. We also find a significant difference for EGS-6811 ($z_{\rm phot}=7.40_{-0.04}^{+0.05}$), where this alternative $z_{\rm phot}$ estimate is inconsistent with the available spectroscopic redshift ($z_{\rm spec}=8.68$).

\section{Chemical Enrichment in Early Bright Galaxies}
\label{sec:enrichment}

\begin{figure}[!t]
\centering
\includegraphics[width=0.49\textwidth]{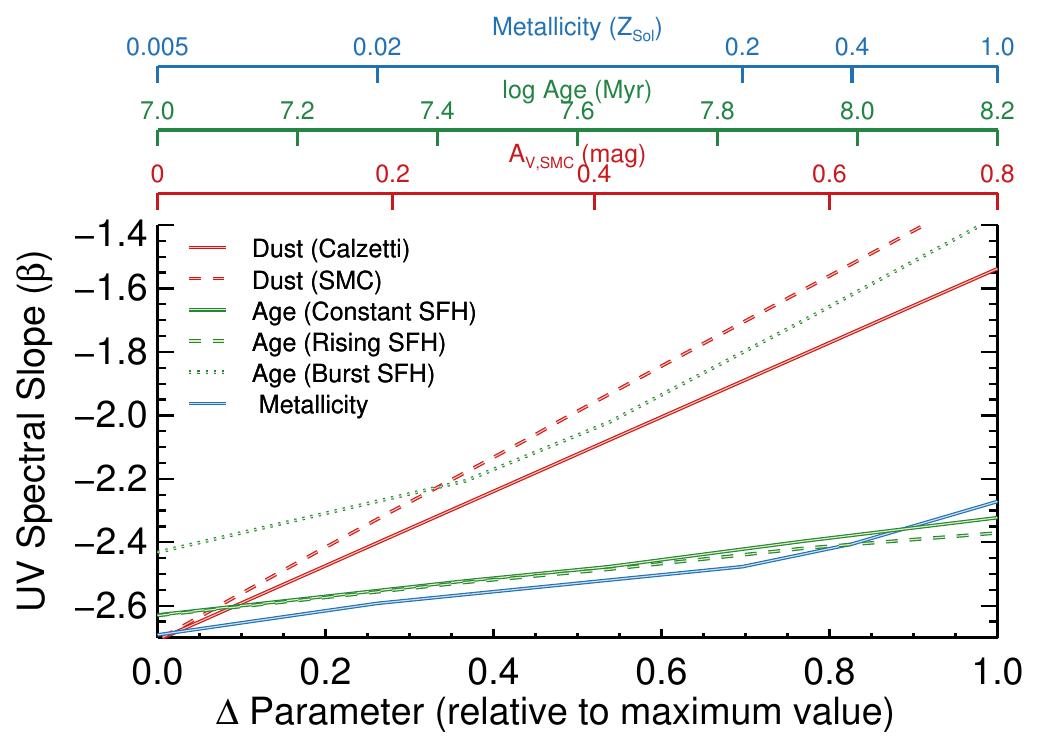}
\caption{Dependence of the UV spectral slope $\beta$ on individual galaxy physical properties. The horizontal axis shows the fractional change (e.g., $\Delta$) in a given parameter relative to the maximal value considered (with the actual values given by the upper horizontal axis). The red, green, and blue curves show the change in $\beta$ for changing dust attenuation, stellar population age, and stellar metallicity, respectively (with the other two parameters held fixed, using fiducial fixed values of A$_{\rm V}=0.2$ mag, age $=$ 50 Myr, and $Z=0.2~Z_{\odot}$, a constant SFH, and a Calzetti attenuation curve). While rising metallicity and age can affect $\beta$, changes in dust attenuation are much more significant, thus using $\beta$ to study dust attenuation is warranted. This is especially true at these early epochs where the maximal stellar age is limited by the short period since the onset of star formation ($\sim200-300$ Myr).}
\label{fig:beta_props}
\end{figure}

\subsection{The UV Spectral Slope}
\label{subsec:beta_pops}

\subsubsection{$\beta$ as a Proxy for Dust Attenuation}

The UV spectral slope $\beta$ (defined as $f_\lambda \propto \lambda^{\beta}$; \citealt{calzetti94}) is often used to quantify stellar populations in the high-redshift universe as it is a straightforward probe of the color of the emergent light from the young, massive stars in these early galaxies. It is also a relatively easy measurement -- $\beta$ is readily measurable if a given galaxy has detections in at least two photometric bands probing the rest-frame UV (free of both the Lyman-$\alpha$ break introduced by the neutral IGM and Balmer/$4000~\mathrm{\AA}$ break). While a number of physical factors can affect the rest-UV color, the observed slope $\beta$ is generally interpreted as a proxy for dust attenuation.

This dust-heavy interpretation of the UV slope is especially true at the highest redshifts we discuss here, as the stellar ages are limited by the very short time since the end of the cosmic dark ages, and metallicities are similarly limited both by the lack of time for significant chemical enrichment and the relative insensitivity of $\beta$ to changing stellar metallicities. In Figure~\ref{fig:beta_props} we use \citet{bruzual03} models to show how the inferred value of the UV spectral slope $\beta$ changes with increasing dust attenuation, stellar population age, and metallicity. For a given curve showing the change in one property, we keep the other two properties fixed, with the fixed values being A$_{\rm V}=0.2$ mag, age $=$ 50 Myr, and $Z=0.2~Z_{\odot}$. For dust attenuation, we consider both a starburst \citep{calzetti00} and an SMC-like \citep{pei92} attenuation curve, while for stellar population age we consider a constant ($\tau=\infty$), rising ($\tau = -300~\mathrm{Myr}$) and extreme burst ($\tau = 0.1~\mathrm{Myr}$) SFH.

\begin{figure*}[!t]
\centering
\includegraphics[width=\textwidth]{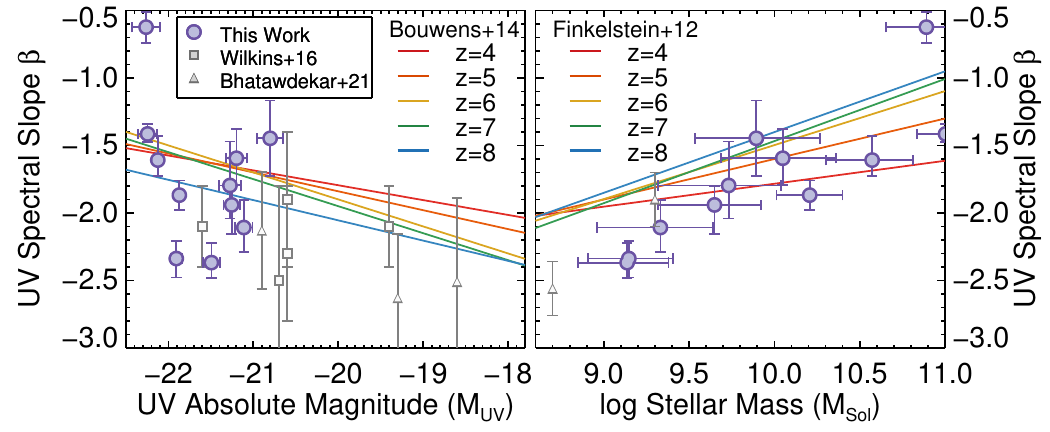}
\caption{The measured UV spectral slope $\beta$ for our sample of 11 $z=9-10$ galaxies (Section~\ref{subsec:beta_measure}) versus their derived UV absolute magnitude (left) and stellar mass (right). We show previously published results for predominantly fainter galaxies as small symbols. In the left panel, the colored lines show the measured correlations between $\beta$ and $M_{\rm UV}$ at $z=4-8$ from \citet{bouwens14}, while with similar lines in the right panel we show the measured correlations between $\beta$ and the stellar mass at $z=4-8$ from \citet{finkelstein12a}, converting from Salpeter to a Chabrier IMF. Our sample of $z=9-10$ galaxies appears to exhibit a strong correlation with the stellar mass (Pearson $r=0.85$) and little-to-no correlation with $M_{\rm UV}$ (Pearson $r=-0.32$).}
\label{fig:beta}
\end{figure*}

Figure~\ref{fig:beta_props} shows that $\beta$ is typically much more sensitive to dust attenuation than it is to age or metallicity, similar to previous analyses \citep[e.g,][]{cortese08, bouwens09, wilkins16, jaacks18b, tacchella18_dust}. The UV slope $\beta$ does get redder with increasing metallicity or stellar population age (at fixed dust attenuation), but the changes are relatively small. The change from $Z =$ 0.005 to 1.0 $Z_{\odot}$ is $\Delta \beta = 0.4$, while the change from $t=10~\mathrm{Myr}$ to $\sim200~\mathrm{Myr}$ (representing a formation redshift of $z=13$ for an observation redshift of $z=9$) is $\Delta \beta = 0.3$. In comparison, a change in SMC-law V-band dust attenuation from 0 to 0.7 mag results in $\Delta \beta = 1.3$. The exception is for the burst SFH, where all of the stellar mass is formed in 0.1 Myr. This population is still fairly blue at t$=$10 Myr ($\beta = -2.4$), but becomes very red ($\beta=-1.4$) by $t=100$ Myr. However, as this galaxy has not formed any stars since its initial burst, its luminosity fades rapidly, dropping three magnitudes from $t=10$ to $t=100$ Myr. Such a galaxy, at $\log(M_{\star}/M_{\odot})=9.5$, would be below the detection limits of our sample. This highlights that precise measures of $\beta$ can be very sensitive to changes in dust attenuation, especially at these high redshifts where changes in the UV slope due to stellar ages are minimized due to the young age of the universe.

However, $\beta$ can still inform on chemical enrichment; Figure~\ref{fig:beta_props} highlights that, for very young and dust-free galaxies, $\beta$ reaches a minimum of $-2.7$ for $Z=0.005~Z_{\odot}$. While this minimum is somewhat model dependent, the search for galaxies with even bluer spectral slopes ($\beta \lesssim -3$) has been an important part of high-redshift studies since the advent of deep near-IR imaging \citep[e.g.,][]{bouwens10b,finkelstein10}. Such blue values would indicate ultra-poor metallicities, potentially even metal-free Population III galaxies. The likelihood for such a discovery is complex, however, as enrichment from the initial burst of Population III stars alone may significantly redden the observed colors of galaxies \citep[e.g.,][]{jaacks18b}. Nonetheless, our sample of well-observed $z\sim9-10$ galaxies presents an excellent opportunity to measure the UV slope $\beta$, and constrain chemical enrichment at some of the highest redshifts yet probed.

\subsubsection{Measurements of $\beta$}
\label{subsec:beta_measure}

While the UV slope $\beta$ can in principle be measured by a single color, additional colors increase the accuracy of the resulting measurements. \citet{finkelstein12a} performed simulations to assess best practices for measuring this quantity, comparing a single color, a power-law fit to multiple colors, and measuring $\beta$ directly from the best-fitting SED model spectrum. They found that, when many colors are available (e.g., at lower redshifts), both the power-law and SED-fitting method outperform the single-color method, while at higher redshifts when information is more limited, the SED-fitting method results in both a smaller scatter and a smaller bias. We therefore elect to use the SED-fitting method here. We note that as our galaxies are fairly bright, we do not expect photometric scatter to result in a bias towards bluer measured UV slopes as found for fainter galaxies \citep{dunlop12}.

This calculation is done by taking the \texttt{Prospector} model spectra (using the fiducial fit with the continuity SFH prior), converting them to $f_{\lambda}$ in the rest frame, and fitting a slope to the spectrum in wavelength windows from \citet{calzetti94} designed to omit spectral emission and absorption features. The \citet{calzetti94} windows span 1268--2580 \AA, however here we omit the three bluest windows to avoid potential contamination from the Ly$\alpha$ break due to the photometric redshift uncertainties, so our bluest window begins at 1407 \AA. We apply this measurement to the spectra from 100 random draws of the \texttt{Prospector} posteriors such that the uncertainty on $\beta$ includes all model uncertainties (including uncertainties on the redshift when relevant). From these 100 draws, we calculate the median value and 68\% confidence range on $\beta$. 

The results for each galaxy are listed in Table~\ref{tab:fiducial_results}. While our measured values span a wide range, interestingly all galaxies have $\beta > -$2.5, implying measurable dust attenuation in every galaxy in our sample. While this may not be unexpected in such relatively massive systems, it implies that significant dust production must be taking place at $z >$ 10 to be observable in this epoch.

The left panel of Figure~\ref{fig:beta} compares these $\beta$ measurements to each galaxy's rest-UV absolute magnitude (taken from F21), compared to previous results at $z \sim$ 9--10 from \citet{wilkins16} and \citet{bhatawdekar21}, as well as to the derived trends between $\beta$ and $M_{\rm UV}$ from \citet{bouwens14}. As our galaxies span a relatively small dynamic range in $M_{\rm UV}$, there is no correlation visible within our small sample (Pearson $r=-0.32$), though the bulk of our galaxies have measured $\beta$ consistent with similarly bright galaxies at lower redshifts ($z\approx4-8$). Our faintest galaxies, at $M_{\rm UV} \sim -$21 have colors that are also consistent with the rest-UV colors measured for a stack of bright $z \sim$ 8 galaxies from \citet{stefanon21_beta}, who measured $J-H$ $\sim$ 0 (for $\beta \sim$ $-$2) at $M_{\rm UV}$ = $-$21.

In the right panel of Figure~\ref{fig:beta} we plot $\beta$ versus our \texttt{Prospector}-derived stellar mass, also including points from \citet{bhatawdekar21} and the derived correlations between $\beta$ and the stellar mass at lower redshifts from \citet{finkelstein12a}. We see that our sample of $z=9-10$ galaxies appears to exhibit a strong correlation between stellar mass and $\beta$ (Pearson $r=0.85$), where more massive galaxies have redder UV spectral slopes, similar to the correlations found by \citet{finkelstein12a}. Furthermore, the measured values of $\beta$ for our sample of galaxies at $\log(M_{\star}/M_{\odot})\gtrsim9.5$ are consistent with those measured for similarly massive galaxies at $z=4-8$. We simulated whether photometric scatter could cause this trend and found that, at the brightness range of our sample, scatter does not appear to input any bias.

\begin{figure*}
\centering
\includegraphics[width=0.85\textwidth]{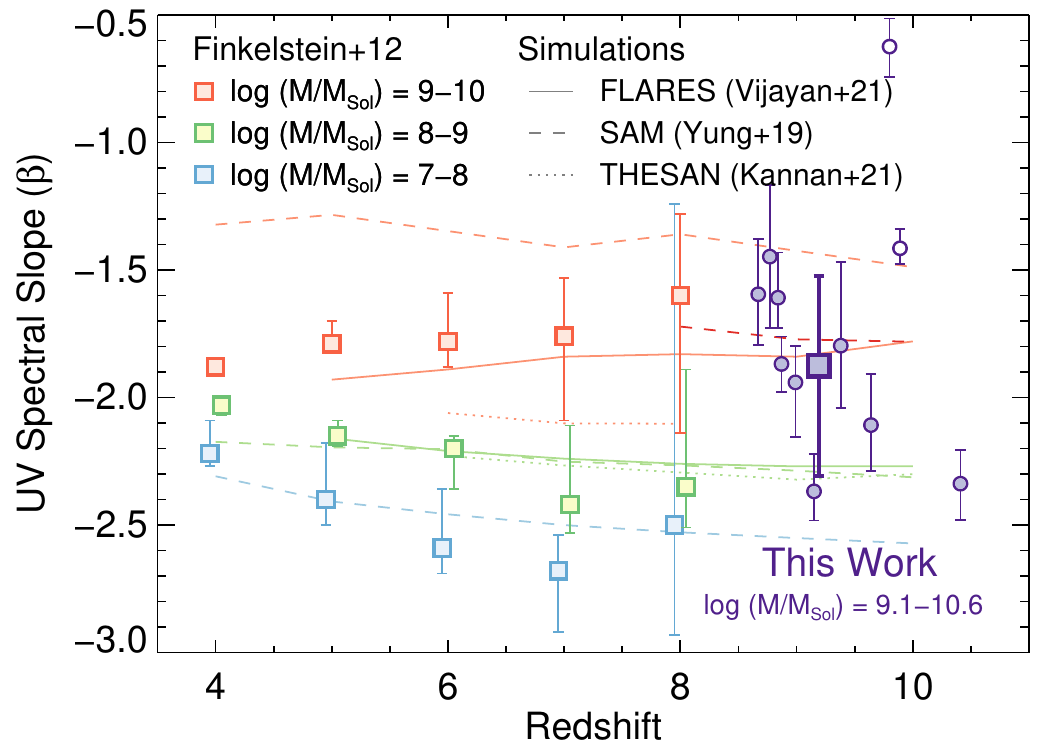}
\caption{The UV spectral slope $\beta$ versus redshift. We show each of our individual galaxies as small purple circles. The large dark purple square shows the median measured $\beta$ from our sample (calculated by stacking the posterior distributions on $\beta)$, excluding the two galaxies with $\log(M_{\star}/M_{\odot})> 10.7$ as their mass may be biased high due to residuals from neighbors in the IRAC imaging (white filled). We compare to results at $z =$ 4--8 from \citet{finkelstein12a}, shown by the lighter-colored squares, using color to denote the stellar mass. We find that our sample of observed modestly massive galaxies (log M/M\sol\ = 9.1--10.6) have measured values of $\beta$ comparable to similarly massive galaxies at $z =$ 4--8. This implies that galaxies of these masses can grow their dust reservoirs in a relatively short period of time, as we are observing many of these galaxies $<$500 Myr from the Big Bang. This is consistent with predictions from multiple simulations (a semi-analytic model \citep{yung19}, the FLARES simulations \citep{lovell21, vijayan21_phot}, and the THESAN radiation-magneto-hydrodynamic simulation \citep{kannan21, garaldi21, smith21}, which predict significant dust reservoirs in these early massive galaxies.}
\label{fig:beta_z}
\end{figure*}

We explore this further in Figure~\ref{fig:beta_z}, where we plot the median values of $\beta$ from $z =$ 4--8 from \citet{finkelstein12a} in mass bins versus redshift alongside the results from our $z=9-11$ galaxy sample. We calculate the median $\beta$ of our sample by combining the measured values of $\beta$ from the 100 posterior draws for each object into a single array and then calculating the median and 68\% confidence range. For our full sample of 11 galaxies, this calculation yields a median of $\beta = -$1.76$^{+0.42}_{-0.49}$. However, we acknowledge that two of our sources appear fairly red and also have the highest derived stellar masses (UDS-18697 and COSMOS-20646). As discussed further in F21, the proximity to bright neighbors makes the IRAC photometry for these sources less reliable; thus it is possible that residual light from the neighbors is contributing to the high stellar mass measurement. If we exclude these two galaxies from this median measurement, we find $\beta = -1.87^{+0.35}_{-0.43}$. While the median is not highly dependent on the inclusion of these two galaxies, we consider this latter value our fiducial value.

Comparing our $z=9-11$ galaxies to the results at lower redshift for similarly massive galaxies in Figure~\ref{fig:beta_z}, we find the surprising result that even though we are probing a few 100 Myr closer to the Big Bang, these relatively massive galaxies appear similarly red to comparable mass galaxies at lower redshifts. This implies that not only does dust build up to significant values very rapidly in modestly massive galaxies, but that this level of attenuation is relatively invariant with redshift at $4<z<10$ at these fixed high stellar masses.

\begin{figure*}[!t]
\centering
\includegraphics[width=\textwidth]{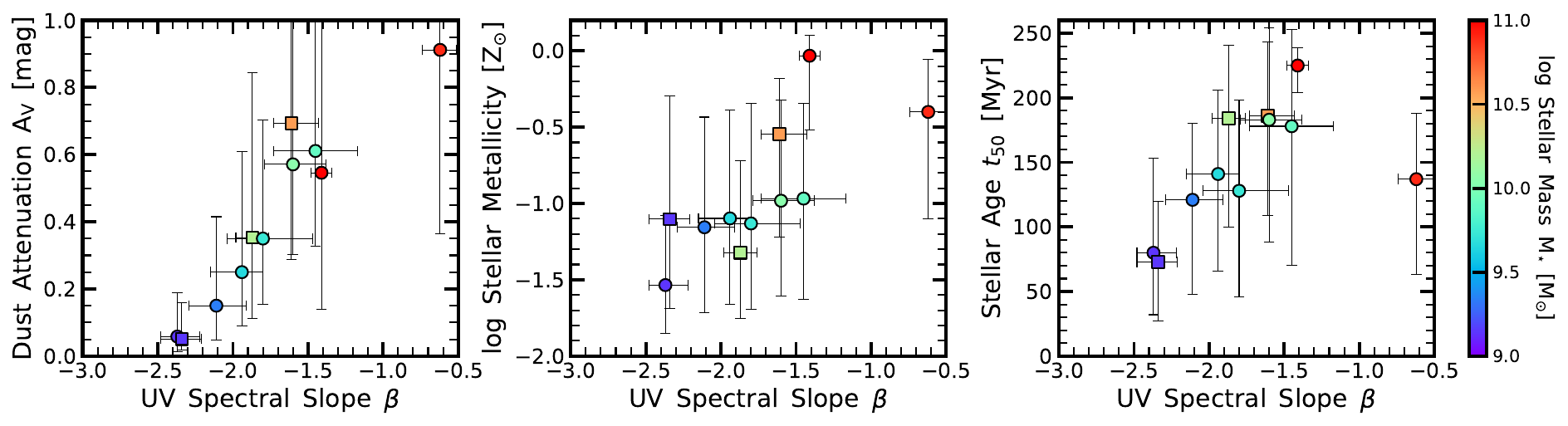}
\caption{The dependence of the \texttt{Prospector}-derived physical quantities on the UV-slope $\beta$, showing V-band dust attenuation (left), log stellar metallicity (middle), and stellar population age $t_{50}$ (right), the lookback time when 50\% stellar mass has been formed. The color-coding denotes increasing log stellar mass as indicated with the colorbar on the right. The squares mark the galaxies with spectroscopic redshifts, while the circles show the galaxies with photometric redshifts. Consistent with our expectation from Figure~\ref{fig:beta_props}, $\beta$ appears to correlate most strongly with dust attenuation. We do not see a strong correlation with stellar metallicity, though the uncertainties on the \texttt{Prospector}-derived metallicities are high, so conclusions from this middle panel are not significant. We do see that the bluest galaxies appear to have the youngest derived stellar population ages, though at $\beta>-2$ there is no visible trend between $\beta$ and age.}
\label{fig:beta_prospector}
\end{figure*}

\subsection{Comparison of $\beta$ to SED-fitting Results}

In Figure~\ref{fig:beta_prospector} we compare our measured values of $\beta$ to the \texttt{Prospector}-derived values of the $V$-band dust attenuation (A$_{\rm V}$), stellar metallicity, and stellar age $t_{50}$\footnote{The stellar age $t_{50}$, i.e. the lookback time when 50\% of the stellar mass has been formed and therefore the median age, is similar -- but not exactly equal -- to the mass-weighted age (a weighted average). We adopt $t_{50}$ throughout this work.}. The stellar age $t_{50}$ is the lookback time at which 50\% of the stellar mass has formed, and it is very similar to the mass-weighted age of the SFH. Starting with dust attenuation, our sample of 11 galaxies exhibits a strong, and nearly monotonic, positive correlation between dust attenuation and $\beta$. This is consistent with what we expected from Figure~\ref{fig:beta_props}, which implied that $\beta$ should inform most strongly about dust attenuation. With the exception of the two bluest galaxies, all galaxies are constrained (at the 1$\sigma$ level) to have non-zero levels of dust attenuation. 

While the lack of a strong observed correlation between $\beta$ and the stellar metallicity in the middle panel is not surprising, as Figure~\ref{fig:beta_props} shows that $\beta$ is not very sensitive to changes in stellar metallicity, we do find that our bluest galaxy (EGS-68560) has the tightest constraints on a low metallicity with $Z < 0.1~Z_{\odot}$ (1$\sigma$). This is consistent with the idea that very blue values of $\beta$ ($<-$3) will imply very low metallicities. While we do not yet see such blue galaxies in our sample, as noted above these are fairly massive systems thus one might expect to see such blue colors (and thus relatively un-enriched galaxies) at lower masses at this same epoch. In the right-hand panel, we see a similar result, where the bluest galaxies have the tightest constraints on a young average age, while at $\beta > -$2, there is little correlation. However, as discussed further in Section 5, these ages are highly dependent on the SFH prior.

The average inferred dust attenuation in our sample of 11 galaxies is A$_{\rm V} =$ 0.4 $\pm$ 0.3 mag. This is larger (though only at $\sim$1$\sigma$ significance) than the average attenuation found in a sample of four fainter galaxies by \citet{wilkins16}, who found an average A$_{1500} =$ 0.5 $\pm$ 0.3, which corresponds to A$_{\rm V}=$0.12 $\pm$ 0.07. This is consistent with the expectation observed at lower redshifts that fainter galaxies have less dust attenuation, though we note that \citet{wilkins16} used the locally derived relation by \citet{meurer99} to convert between $\beta$ and the dust attenuation (while our attenuation is derived from SED fitting using a flexible attenuation curve). However, this is not surprising as these fainter galaxies are presumably less massive, and simulations \citep[e.g.][]{graziani20} predict that the dust mass grows rapidly at higher stellar masses.

\subsection{Implications on Evolution of UV LF}

One of the main conclusions we can make in this section is that the rest-UV colors of $z=9-10$ galaxies at M$_{\rm UV} < -21$ and $\log(M_{\star}/M_{\odot})=9-10$ are similar to those at the same UV luminosities and masses at $z=4-8$ redshift. This has implications for the interpretation of the evolution of the UV luminosity function. Evidence has been growing that the bright end of the rest-UV luminosity function changes little, if at all, from $z =$ 7 to 10. This idea was introduced by \citet{bowler20}, and even more recent luminosity function measurements (including using this same sample here) are continuing to find a higher-than-expected number density of these bright systems \citep[e.g.,][]{rojasruiz20,finkelstein21}. As the bright end of the luminosity function does appear to decline in abundance from $z =$ 4 to 7 \citep[e.g,][]{bouwens15,finkelstein15,bowler15}, this apparent lack in evolution at higher redshift points to a physical change in the galaxies themselves.

The most obvious potential physical change would be in dust attenuation: if more distant galaxies at fixed UV magnitude are less attenuated, then the bright end of the UV luminosity function would evolve more slowly than the faint end, which is exactly what observations suggest. However, our results here cast doubt on this physical interpretation, as we find that the bulk of these bright massive galaxies have similar UV spectral slopes as their $z \sim$ 7--8 counterparts, and thus by extrapolation likely have similar levels of dust attenuation. While our sample is small, if this result holds with larger samples from robust observations with the \textit{James Webb Space Telescope} ({\it JWST}) it implies another physical explanation will be needed, such as changes in the star-formation efficiency \citep[e.g,][]{finkelstein15b, stefanon19}, or time/mass scales for the onset of quenching \citep[e.g.,][]{tacchella18, bowler20}.

\subsection{Implications on Dust Formation in Early Galaxies}

The most surprising result from this analysis is that the UV spectral slope $\beta$ for relatively massive UV-selected systems ($9<\log(M_{\star}/M_{\odot})<10$) changes very little from $z\sim4$ to $z\sim9–10$. This implies that the dust attenuation at this fixed stellar mass is roughly constant with redshift (though, as the effects of reddening due to age and stellar metallicity should be less at higher redshift, it is possible the actual attenuation at the highest redshifts is even higher at fixed $\beta$). Although the most recent constraints from the Atacama Large Millimeter Array imply that dust-obscured star-formation is not dominant in the epoch of reionization \citep[e.g,][]{zavala21}, finding evidence that relatively massive galaxies at these early epochs have significant levels of dust attenuation is not surprising in and of itself, as there are a growing number of direct individual detections of dust emission at $z\gtrsim8$ \citep[e.g.,][]{watson15, laporte17, tamura19, bakx20, schouws21, fudamoto21}. A number of theoretical models have explored these results, finding that with a variety of assumptions the implied dust masses at early times could be formed with our current understanding of dust formation physics \citep[e.g.,][]{mancini15, mancini16,popping17_dust, behrens18, sommovigo20, graziani20}.

An important caveat to highlight is that the connection between dust attenuation and the physical properties of the dust in galaxies (such as the dust mass and the grain properties) is non-trivial. Neglecting the effect of geometry and orientation on attenuation can severely bias the interpretation \citep[e.g.,][]{padilla08}. For example, \citet{chevallard13} show that geometry and orientation effects have a stronger influence on the shape of the attenuation curve than changes in the optical properties of dust grains. Similarly, several studies show that galaxy shape and inclination are the major factors in determining the observed amount of dust attenuation, and not the galaxy dust mass \citep{maller09, kreckel13, zuckerman21}. Although these studies focus on lower-redshift systems ($z<3$), similar effects might drive some of the observed effects we see at $z>4$ regarding $\beta$, $A_{\rm V}$ and the attenuation curve (Section~\ref{subsec:attenuation_curve}). Although parts of this caveat can be alleviated by including far-IR constraints (modulo the assumption regarding energy conservation), this should be kept in mind in the following paragraphs when connecting the attenuation to the physical properties of dust.

The young age of the universe at these observed epochs could in principle constrain the efficiencies of different dust production mechanisms. The formation of dust grains can happen via multiple sources, which have their own timescales, and uncertainties due to various physical assumptions (see, e.g., \citealt{dayal18} and references therein). For example, while dust formation in the ejecta of supernovae (SNe) could lead to the formation of the first dust grains at extremely early times \citep[e.g.,][]{todini01, schneider04, bianchi07, sarangi15, sluder16, marassi19}, the dust destruction timescales are  not well constrained \citep[e.g,.][]{bianchi07, silvia10, bocchio16,micelotta16, Martinez-Gonzalez19, slavin20}, especially in the early universe \citep[e.g.,][]{hu19}. 

Dust can also form in the atmospheres of asymptotic giant branch (AGB) stars \citep[e.g.,][]{gehrz89, ferrarotti06, zhukovska08, ventura12, nanni13, ventura18, dellagli19}, with yields being sensitive to the mass and metallicity of their progenitor stars. Depending on the SFH and on the metallicity of the stellar population, \citet{valiante09} found that AGB stars could plausibly contribute to 30-50\% of the total dust budget in high-redshift galaxies in $\approx$ 300 Myr. Finally, dust can grow in the cold/warm phase of the ISM on seed grains formed by early SNe \citep[e.g.,][]{draine79, draine09, hirashita00, michalowski10, valiante11, asano13, mancini15, lesniewska19}. The timescale for this process may be quite short if dust is formed via the first SNe, thus this formation pathway may be significant at early times. However, we still lack a full understanding of this process at the atomic level, and we equally do not know the phase of the ISM where the process may occur, e.g, molecular \citep{ferrara16, ceccarelli18} versus warm atomic \citep{zhukovska18}. 

As the grain growth timescale is thought to be density dependent \citep{asano13, schneider14, mancini15, popping17_dust}, the expected higher density of star-forming clouds in these early galaxies could lead to this mode of dust production being more efficient. As ISM grain growth also requires initial seed grains, more efficient grain growth at earlier times could point to more efficient dust production by core-collapse SN explosions from low-metallicity massive stars \citep[e.g.,][]{marassi15, marassi19} or pair-instability SN explosions from Population III stars \citep{nozawa03, schneider04}, due to either higher yields \citep[e.g.,][]{schneider04} or earlier Population III star formation times \citep[e.g.,][]{jaacks18b,jaacks19}. However, these seed grains require some chemical enrichment to form, so this entire process is also dependent on the metallicity of the gas.

\citet{graziani20} explored dust formation at high redshift by including dust formation and evolution in a hydrodynamic simulation, accounting for dust formation via both stellar sources (e.g., SNe and AGB stars) and grain growth in the ISM, and several sources of dust destruction. They found that, at $z>10$, dust produced via stellar sources dominates the total cosmic dust mass, with grain growth not playing a significant role until $z<9$. This in principle might predict that massive galaxies at $z>9$ should begin to appear significantly less dusty as they are not yet enriched via grain growth, seemingly at odds with our observation. However, they also found that grain growth becomes dominant for systems with stellar masses of $\log(M_{\star}/M_{\odot})>8.5$, consistent with the mass range of our observed reddened galaxies. It is thus plausible that even at these early epochs, these massive systems have their total dust content enriched via stellar dust production and grain growth, maybe aided by more favorable conditions in their ISM (e.g. higher densities), which is also consistent with the predictions from semi-analytical \citep{popping17_dust, vijayan19} and semi-numerical \citep[e.g.,][]{mancini15, mancini16} models.

We compare our observations to predictions from a semi-analytic model \citep[SAM;][]{yung19}, the FLARES simulations \citep{lovell21, vijayan21_phot}, and the THESAN radiation-magneto-hydrodynamic simulation \citep{kannan21, garaldi21, smith21} to our observations in Figure~\ref{fig:beta_z}. The $\beta$ values for the THESAN simulation are presented in \citet{kannan21_line}. While THESAN does track dust formation and destruction, it does not have any galaxies as massive as those we observe at $z>9$. FLARES and the SAM do not directly track dust, rather both models use the ISM metal abundance to derive a dust attenuation, which should be kept in mind when comparing these models to our observations. It is interesting however that the FLARES predictions agree well with our observations. While the SAM seems to overpredict our observations, we also need to account for our observational bias. If we apply our observational cut of $H <$ 26.6 to the SAM galaxies, we find predicted $\beta$ values well in agreement with our observations (darker red dashed line). This model thus would predict a population of more dusty high-redshift massive systems missed by our UV-selection, similar to those recently discovered by \citet{fudamoto21}. Galaxy selection at redder wavelengths, as will soon be possible with {\it JWST}, will alleviate this potential selection bias.

While our observations cannot alone distinguish between the various competing physical processes, they do point to fairly efficient dust growth in massive galaxies at early times, which can in turn be used to further constrain detailed simulations \citep[e.g.,][]{mckinnon17, aoyama18, graziani20, vogelsberger20}.  The abundance of {\it JWST} Cycle 1 programs targeting the early universe should both allow measures of the rest-UV colors of larger samples of massive galaxies at $z\sim9–10$, as well as push to lower-mass systems at $z>10$ for the first time. Together with radiative transfer simulations \citep[e.g.,][]{behrens18, katz19, shen20, shen21, vijayan21}, this will allow a detailed, more direct comparison between theoretical dust models and observations over a wide range of different galaxies, and thereby constrain the physical processes related to dust growth and chemical enrichment in early galaxies.

\section{Growth of Stellar Mass and Implications on Early Cosmic Star Formation}
\label{sec:mass_growth}

This section presents the key results concerning the early mass growth histories inferred from our SED-modeling analysis. In particular, we present the stellar mass and SFR measurements in Section~\ref{subsec:M_SFR}, followed by an exploration of the inferred SFHs (Section~\ref{subsec:SFH}) and the stellar ages and star-formation timescales (Section~\ref{subsec:age}). Finally, we look into the fraction of mass formed beyond redshift 10 (Section~\ref{subsec:frac_mass}) and the implications for the early evolution of the cosmic SFR density (Section~\ref{subsec:cSFRD}). An important conclusion is that our SFHs depend on the assumed SFH prior, which we introduced in Section~\ref{subsec:prior_sfh}. We then discuss whether our galaxies are overly massive for the $\Lambda$CDM universe (Section~\ref{subsec:too_massive}) and how we can make progress in the future with {\it JWST} (Section~\ref{subsec:jwst}).

\subsection{Stellar masses and star-formation rates (SFRs)}
\label{subsec:M_SFR}

\begin{figure*}
\centering
\includegraphics[width=\textwidth]{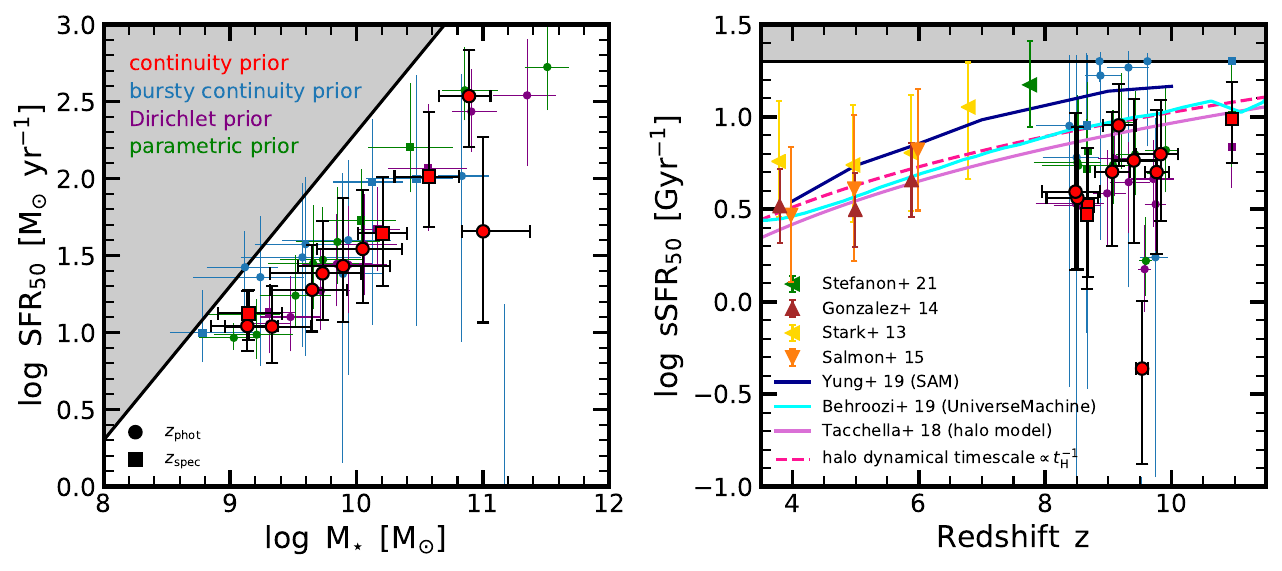}
\caption{Star-formation rate (SFR) and stellar mass ($M_{\star}$) properties of our $z=9-11$ galaxies. The left and right panels show our galaxies in the $\mathrm{SFR}_{50}-M_{\star}$ plane and the $\mathrm{sSFR}_{50}-z$ plane, respectively. The $\mathrm{SFR}_{50}$ and $\mathrm{sSFR}_{50}$ are averaged over the past 50 Myr. The red symbols show the fiducial continuity prior, while the smaller blue, purple and green symbols indicate the results for the bursty continuity prior, the Dirichlet prior, and the parametric prior. The squares mark the galaxies with spectroscopic redshifts, while the circles show the galaxies with photometric redshifts. The gray shaded regions indicate the forbidden parameter spaces, where $\mathrm{SFR}_{50}$ would be too high given the averaging timescale of 50 Myr and the stellar mass $M_{\star}$ (see Eqs.~\ref{eq:max_SFR} and \ref{eq:max_sSFR}). In the right panel, we compare, at fixed stellar mass of $\log(M_{\star}/M_{\odot}) = 9-10$, our measurements to observations by \citet{gonzalez14}, \citet{stark13}, \citet{salmon15} and \citet{stefanon21_beta} and to the models of \citet{yung19}, \citet{behroozi19} and \citet{tacchella18}, where the two latter ones are both consistent with the evolution of the dynamical timescale of halos.}
\label{fig:M_SFR}
\end{figure*}

\begin{figure*}
\centering
\includegraphics[width=\textwidth]{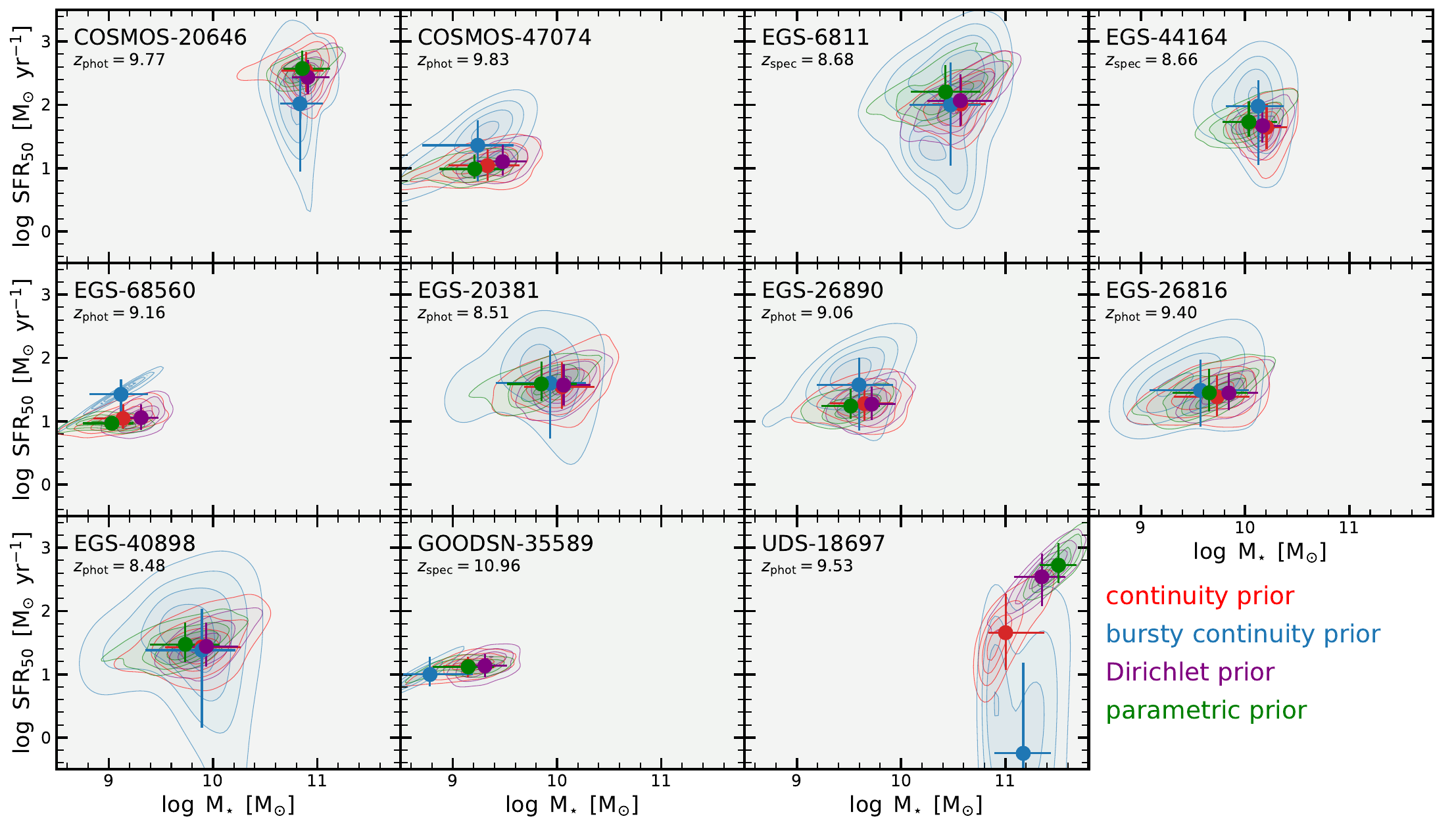}
\caption{Posteriors of the star-formation rate ($\mathrm{SFR}_{50}$) and stellar mass ($M_{\star}$). $\mathrm{SFR}_{50}$ is averaged over the past 50 Myr. The red, blue, purple and green colors indicate the results from assuming the continuity prior, the bursty continuity prior, the Dirichlet prior, and the parametric prior for the SFH. These different priors give rise to similar stellar masses and SFRs (within a factor of 3). An exception is UDS-18697, where $\mathrm{SFR}_{50}$ varies by more than 1 order of magnitude, which stems from the degeneracy with the amount of attenuation.}
\label{fig:SFR_M_posteriors}
\end{figure*}

We present in this section the stellar mass ($M_{\star}$) and SFR measurements of our $z=9-11$ galaxy candidates. If the SFR varies with time, it is important to specify the timescale over which the SFR is measured \citep[e.g.,][]{caplar19}. We choose a timescale of 50 Myr, which is roughly the timescale that the UV light at $1500-3000~\mathrm{\AA}$ probes. We label this SFR as SFR$_{50}$. As the galaxies at these early cosmic times are young, i.e., it is plausible that all the stellar mass of a galaxy has formed within the past 50 Myr, it is useful to consider what this maximal SFR is given the stellar mass $M_{\star}$ \citep{tacchella18}:

\begin{equation}
    \mathrm{SFR}_{\rm max} = \frac{M_{\star}}{t_{\rm SF}},
\label{eq:max_SFR}
\end{equation}
where $t_{\rm SF}=50~\mathrm{Myr}$ is the timescale of the SFR indicator. As an example, for a $M_{\star}=10^{10}~\mathrm{M_{\odot}}$ galaxy, the maximum SFR is $\mathrm{SFR}_{\rm max}=200~M_{\odot}~\mathrm{yr}^{-1}$. Similarly, the maximum specific SFR (sSFR) is given by:

\begin{equation}
    \mathrm{sSFR}_{\rm max} = \frac{\mathrm{SFR}_{\rm max}}{M_{\star}} = \frac{1}{t_{\rm SF}}.
\label{eq:max_sSFR}
\end{equation}
This implies that the maximum sSFR is independent of mass (and cosmic epoch) and only depends on the SFR timescales. In our case, the maximum sSFR is $\mathrm{sSFR}_{\rm max}=20~\mathrm{Gyr}^{-1}$. A corollary is that when considering long SFR timescales (relative to the ages of the galaxies), a perfect correlation between the SFR and $M_{\star}$ is introduced by construction -- important to consider when studying the star-forming main sequence.

After these general considerations, we plot the inferred SFRs and $M_{\star}$ in Figure~\ref{fig:M_SFR}. The left and right panels of Figure~\ref{fig:M_SFR} show the $\mathrm{SFR}_{50}-M_{\star}$ and the $\mathrm{sSFR}_{50}-z$ planes, respectively. The black lines and the gray shaded regions indicate the maximum SFR and sSFR mentioned above. The red datapoints and errorbars show the median and 16-84th percentiles of our fiducial run with the continuity prior for the SFH. The exact values are also given in Table~\ref{tab:fiducial_results}. The smaller blue, purple and green datapoints indicate the results of the bursty continuity prior, the Dirichlet prior and the parametric prior, respectively. The circle symbols show the galaxies with photometric redshift estimates, while the squares mark the objects with a spectroscopic redshift.

Despite the large uncertainty in the individual measurements, we find that more massive galaxies have a higher SFR (left panel of Figure~\ref{fig:M_SFR}). The slope of the $M_{\star}-\mathrm{SFR}$ relation -- estimated with the Orthogonal Distance Regression (ODR) taking into account the uncertainties in $M_{\star}$ and SFR of each galaxy -- is $0.70\pm0.17$ for our fiducial continuity prior, i.e., the higher mass galaxies have typically a lower sSFR than their lower mass counterparts. Although our slope estimates include the propagation of the errors of the inferred $M_{\star}$ and SFR via the ODR, we do not perform a fully hierarchical Bayesian approach to measure the slope, intercept, and scatter of the main sequence \citep{curtis-lake21}. Furthermore, the exact value of this slope depends on the SFH prior: the bursty continuity prior typically leads to a decrease in $M_{\star}$ and an increase in the SFR measurements for the low-mass galaxies, which flattens the $M_{\star}-\mathrm{SFR}$ relation. Specifically, the bursty continuity prior results in a slope of $0.45\pm0.27$, while the Dirichlet prior and the parametric prior lead to a slope of $0.79\pm0.16$ and $0.80\pm0.12$, respectively. 

Measurements of the star-forming main sequence slope have been mainly published at slightly lower redshifts. We focus here on the stellar mass range of $10^9-10^{11}~M_{\odot}$ at $z>4$, where the ``bending'' of the star-forming main sequence plays presumably a minor role. \citet{salmon15} found a rather shallow slope of $0.54\pm0.16$ at $z\sim6$, while \citet{pearson18} and \citet{khusanova21} inferred a slope of $1.00\pm0.22$ and $0.66\pm0.21$ at $z\sim5.5$, respectively. Based on a large literature compilation of $z<7$ studies, \citet{speagle14} inferred a steeper slope of $0.84\pm0.02-(0.026\pm0.003)\times t$, where $t$ is the age of the universe in Gyr (at $z=10$, this inferred slope is $0.83\pm0.02$). All of these estimates are consistent with our estimate when considering the uncertainty. Theoretical models typically produce steeper slopes, closer to 1 \citep{somerville15, tacchella18, behroozi19, yung19}. A more careful comparison between observations and theory of the star-forming main sequence slope (and in particular its scatter) will be useful to shed more light onto the star-formation efficiency in low-mass halos \citep[e.g.,][]{tacchella20} and the underlying assembly of dark matter halos \citep[e.g.,][]{dayal15, khimey21}.

The right panel of Figure~\ref{fig:M_SFR} shows the redshift evolution of the sSFR of galaxies with masses of $M_{\star}\approx10^9-10^{10}~M_{\odot}$. We measure sSFR values in the range of $3-10~\mathrm{Gyr}^{-1}$, which indicate a mass-doubling timescale of $\sim100-300~\mathrm{Myr}$ under the assumption of a constant sSFR, roughly consistent with our age estimate (Section~\ref{subsec:age}). An exception is UDS-18697 for which we measure a low sSFR of $0.4_{-0.3}^{+0.6}~\mathrm{Gyr}^{-1}$ -- an interesting galaxy that seems to have gone through an intense episode of star formation early on and now is showing a declining SFH and a high stellar mass (Section~\ref{subsec:SFH}). An important caveat to this object is that the IRAC deblending uncertainty is large (Appendix~\ref{appsec:galfit_phot}). Despite this uncertainty and even if this object does not have such a low sSFR, it viscerally shows that your prior does not exclude such ``old'' solutions if the data warrant it. For the bursty continuity SFH prior, the sSFR values are typically larger and in some cases reach the maximum sSFR of $20~\mathrm{Gyr}^{-1}$. 

We have also added to the right panel of Figure~\ref{fig:M_SFR} the lower-redshift measurements from \citet{gonzalez14}, \citet{stark13}, \citet{salmon15} and \citet{stefanon21_beta}, and the predictions from the models by \citet{yung19}, \citet{behroozi19} and \citet{tacchella18}. The \citet{behroozi19} and \citet{tacchella18} models both track the evolution of the dynamical timescale of dark matter halos, while the model by \citet{yung19} predicts a steeper increase with redshift. Our fiducial sSFR values are consistent with the lower redshift estimates and lie slightly below the expected, increasing evolution of the theoretical models. This, however, depends on the assumed prior: the burstier SFH prior leads to higher sSFR, slightly higher -- but still consistent -- with theoretical expectations.

\begin{figure*}
\centering
\includegraphics[width=\textwidth]{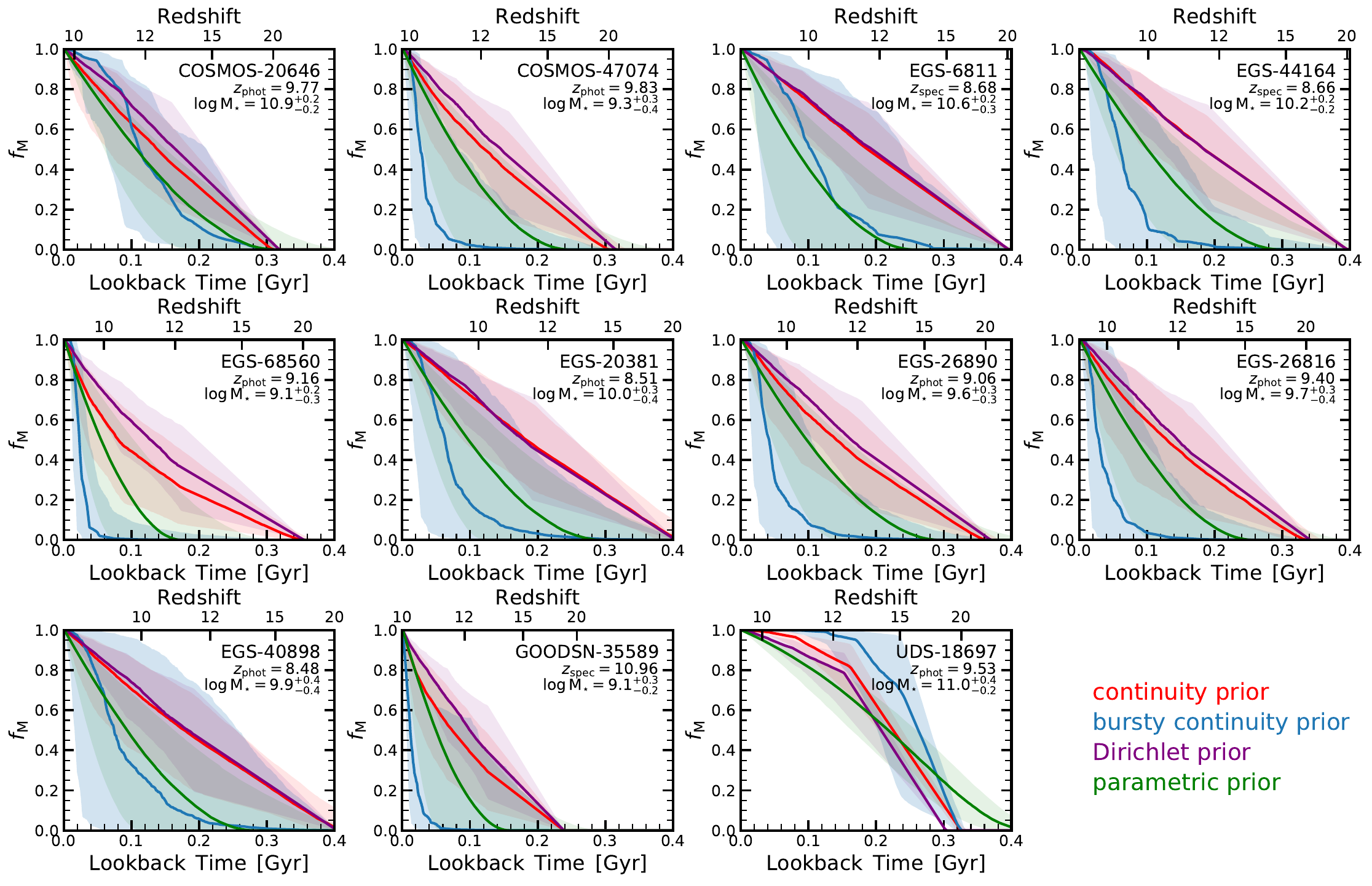}
\caption{Star-formation histories (SFHs) obtained from \texttt{Prospector} assuming different priors. The SFHs are plotted as a fraction of the stellar mass formed. The adopted priors include the continuity prior (red), bursty continuity prior (blue), Dirichlet prior (purple) and parametric prior. These priors are discussed in Section~\ref{subsec:prior_sfh}. The lines and shaded regions show the median and 16-84th percentile of the SFH posterior, respectively. In most cases (exception is the most massive galaxy UDS-18697), the bursty continuity prior leads to more recent star formation than the continuity and Dirichlet priors, which both roughly follow a constant SFH. The parametric prior lies in between those extrema. The data do not prefer any of those priors: the Bayes factor for all models with respect to each other is roughly 1, highlighting that the adopted prior heavily effects the resulting posterior SFH.}
\label{fig:SFH_fm}
\end{figure*}

Figure~\ref{fig:SFR_M_posteriors} shows the detailed posterior distribution for $\mathrm{SFR}_{50}$ and $M_{\star}$ for all the SFH priors. Each panel shows an individual galaxy. It is difficult to draw a single conclusion from this figure, as each galaxy seems to show a different dependence on varying the prior. A common and important feature is that the posteriors of the different priors overlap, i.e., the resulting posteriors are consistent with each other. Furthermore, the typical differences in the median values of the SFR and $M_{\star}$ measurements are of the order of a factor of $2-3$ (except UDS-18697, for which the SFR varies by over 3 orders of magnitude). The parametric SFH prior typically leads to lower masses (as the ages are younger), consistent with findings at lower redshifts \citep{leja19, tacchella21_quench}. The bursty continuity SFH prior shows the largest spread in the posterior space, in particular along the SFR axis.

\subsection{Star-formation histories (SFHs)}
\label{subsec:SFH}

As highlighted in the Introduction and Section~\ref{subsec:beta_pops}, the UV contains plenty of information regarding the age of the stellar population. The key challenge is to differentiate age-related effects from other effects, such as dust attenuation or metallicity variations. Therefore, we have chosen to use a rather flexible SED model (Section~\ref{subsec:model_sed}), which includes a variable dust attenuation law and a range of different SFH priors (Section~\ref{subsec:prior_sfh}). Here we focus on the inferred SFHs and their dependence on the prior, while the next section focuses on the degeneracy between the stellar age and the attenuation.

Figure~\ref{fig:SFH_fm} shows the inferred SFHs by plotting the fraction of the mass formed, $f_{\rm M}$, as a function of lookback time. Each panel shows an individual galaxy, with the red, blue, purple, and green lines showing the median SFHs obtained from the continuity, the bursty continuity, the Dirichlet, and the parametric SFH priors, respectively. The shaded regions show the 16-84th percentiles. The different SFH priors result in different SFH posteriors, i.e., it is important to fully understand how the inferred SFHs are affected by the choice of the prior.

\begin{figure*}
\centering
\includegraphics[width=\textwidth]{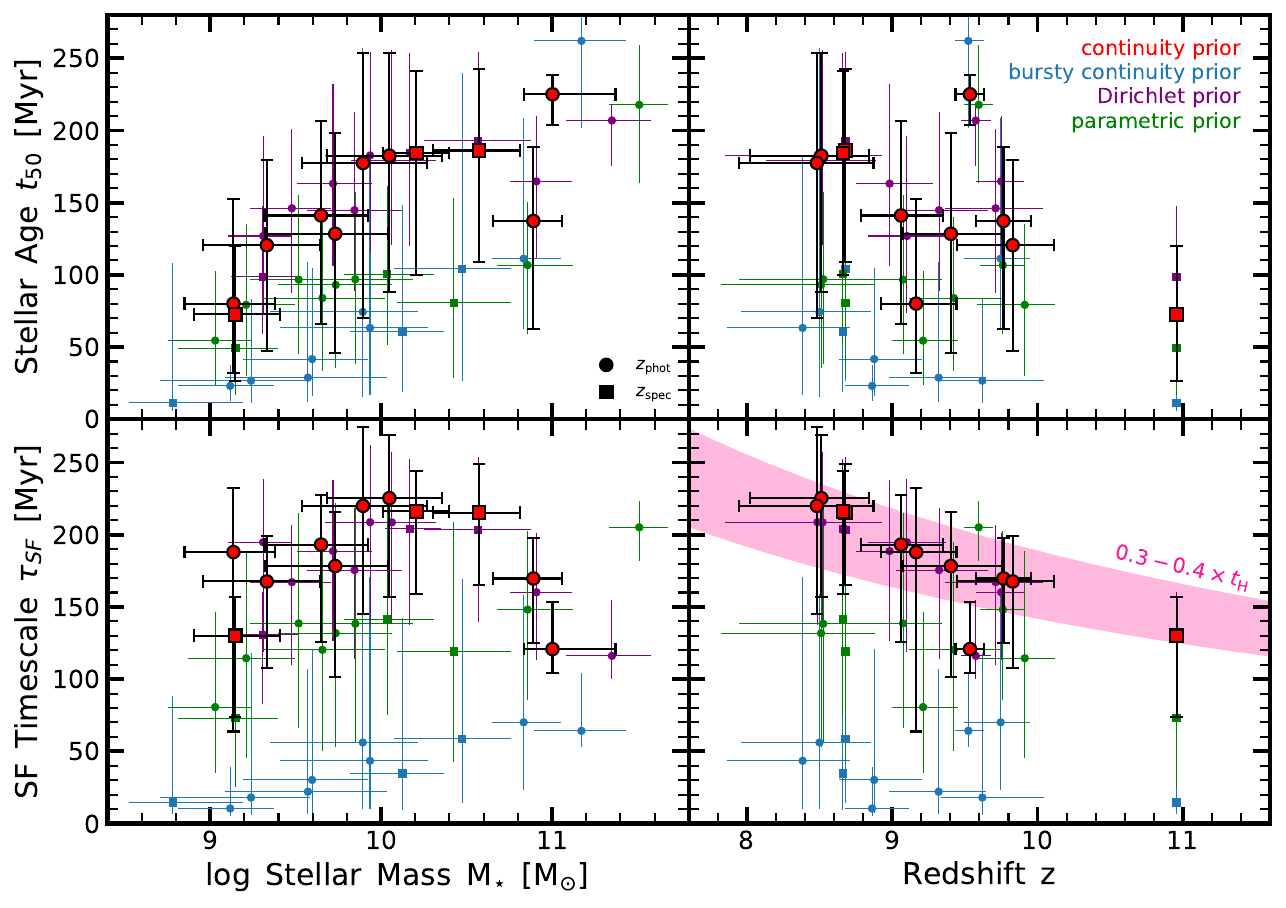}
\caption{Stellar age $t_{\rm 50}$ (top panels) and star-formation timescale $\tau_{\rm SF}$ (bottom panels) as a function of stellar mass $M_{\star}$ (left panels) and redshift $z$ (right panels) of our $z=9-11$ galaxies. The star-formation timescale $\tau_{\rm SF}$ is defined as the time it took to increase the stellar mass from 20\% to 80\%, i.e., it is a measure of how quickly a galaxy formed its stellar mass. The pink shaded region in the bottom right panel indicates $30-40\%$ of the age of the universe at a given redshift $z$. The red symbols show the fiducial continuity prior, while the smaller blue, purple, and green symbols indicate the results for the bursty continuity prior, the Dirichlet prior, and the parametric prior. The squares mark the galaxies with spectroscopic redshifts, while the circles show the galaxies with photometric redshifts. The exact values of $t_{\rm 50}$ and $\tau_{\rm SF}$ depend heavily on the assumed SFH prior: the bursty continuity prior implies younger ages and shorter star-formation timescales, while for the fiducial continuity and Dirichlet prior $\tau_{\rm SF}$ older ages and longer $\tau_{\rm SF}$. Independent of the SFH prior, we find older ages for more massive galaxies and longer star-formation timescales for the lower-redshift galaxies. }
\label{fig:M_age_tauSF}
\end{figure*}

Figure~\ref{fig:SFH_fm} emphasizes when most of the mass has formed, and it stresses that the SFH heavily depends on the assumed SFH prior. The linear increase in $f_{\rm M}$ found when adopting the continuity and the Dirichlet prior is consistent with a constant SFH. Both the parametric and the bursty continuity SFH priors imply a stronger increase in $f_{\rm M}$ in more recent times relative to the fiducial continuity prior. A nominal exception is UDS-18697, where all the priors consistently find an early burst of star formation and little mass growth in the past $\sim100-200$ Myr. In Appendix~\ref{appsec:sfh}, we also show the SFHs plotted as SFR as a function of time (Figure~\ref{appfig:SFH}). There, it is more clearly visible that a few galaxies (i.e., COSMOS-20646, EGS-68560, GOODSN-35589, and UDS-18697) show significant variation in the past $\sim100$ Myr. 

Importantly, the otherwise rather constant behavior for the fiducial continuity and Dirichlet priors is expected, as the expectation value of these two priors are a constant $\mathrm{SFR}(t)$ (Section~\ref{subsec:prior_sfh} and Figure~\ref{fig:prior_sfh}). For the parametric SFH prior, we find for all except one galaxy (UDS-18697) an increasing SFH, something we expect from theoretical models \citep[e.g.,][]{tacchella18}. However, again, this is the expected behavior of the prior. This underscores the worry that the current data provide little constraining power when it comes to the SFH. The different priors, all of which provide equally good fits to the data, are producing rather different SFH posteriors.

\subsection{Inferred ages and star-formation timescales}
\label{subsec:age}

Figure~\ref{fig:M_age_tauSF} plots the mass- and redshift-dependence of the stellar ages ($t_{\rm 50}$) and star-formation timescales ($\tau_{\rm SF}$). The stellar age is the lookback time at which 50\% of the final mass is formed. The star-formation timescale is defined as the time it takes to increase the stellar mass from 20\% to 80\% of the final mass, i.e., it is a measure of how quickly a galaxy formed the bulk of its stellar mass. Figure~\ref{fig:M_age_tauSF} shows $t_{\rm 50}$ on top and $\tau_{\rm SF}$ on the bottom. The colors and symbols are the same as those in Figure~\ref{fig:M_SFR}.

Independent of the SFH prior, we find that more massive galaxies have typically older ages. There is also a hint that galaxies at higher redshifts have younger ages, though the scatter and uncertainties are large. Parts of this trend can probably be explained by an ``envelope effect'', where the galaxy age and the onset of star formation are required to be less than the age of the universe. The star-formation timescales do not show a convincing trend with the stellar mass. However, the trend with the cosmic epoch is more pronounced and roughly follows the age of the universe: these galaxies form their stars on timescales of roughly 30\% to 40\% of the age of the universe, as indicated by the shaded band in the lower-right panel of Figure~\ref{fig:M_age_tauSF}. The exact value of the ages and the star-formation timescales depend on the SFH prior: the bursty continuity prior results in roughly three times younger ages (and shorter star-formation timescales) than the continuity prior. The exact values of the ages and star-formation timescales are given in Table~\ref{tab:ages}. We find a median age $t_{\rm 50}$ over the whole sample of $\sim150$ Myr for the continuity and Dirichlet priors, while the median age is only $\sim50$ Myr for the bursty continuity prior. The parametric SFH prior produces ages that lie in-between with a median age of $\sim90$ Myr. This means we cannot distinguish between median ages of $\sim50-150$ Myr, which corresponds to 50\% formation redshifts of $z_{50}\sim10-12$. These ages are consistent with expectations from the semi-analytical model by \citet{yung19} and the empirical halo model by \citet{tacchella18}, which find stellar ages $t_{50}$ for galaxies at $z\approx10$ of 60 Myr and 40 Myr, respectively.

\begin{deluxetable*}{lccccccccccccccc} \tabletypesize{\scriptsize}
\tablecaption{Stellar age ($t_{\rm 50}$), star-formation timescale ($\tau_{\rm SF}$), and fraction of mass formed before $z=10$ ($f_{\rm M}$) for our galaxy candidates at $z=9-11$. We list the results for the four different SFH priors (Section~\ref{subsec:prior_sfh}). \label{tab:ages}}
\tablehead{
\colhead{} & 
\multicolumn{3}{c}{continuity prior} & \colhead{} & 
\multicolumn{3}{c}{bursty continuity prior} & \colhead{} & 
\multicolumn{3}{c}{Dirichlet prior} & \colhead{} & 
\multicolumn{3}{c}{parametric prior} \\
\cline{2-4}
\cline{6-8}
\cline{10-12}
\cline{14-16}
\\
\colhead{ID} & 
\colhead{$t_{\rm 50}$} & \colhead{$\tau_{\rm SF}$} & \colhead{$f_{\rm M}$} & \colhead{} & 
\colhead{$t_{\rm 50}$} & \colhead{$\tau_{\rm SF}$} & \colhead{$f_{\rm M}$} & \colhead{} & 
\colhead{$t_{\rm 50}$} & \colhead{$\tau_{\rm SF}$} & \colhead{$f_{\rm M}$} & \colhead{} & 
\colhead{$t_{\rm 50}$} & \colhead{$\tau_{\rm SF}$} & \colhead{$f_{\rm M}$} \\
\colhead{} & 
\colhead{[Myr]} & \colhead{[Myr]} & \colhead{[\%]} & \colhead{} & 
\colhead{[Myr]} & \colhead{[Myr]} & \colhead{[\%]} & \colhead{} & 
\colhead{[Myr]} & \colhead{[Myr]} & \colhead{[\%]} & \colhead{} & 
\colhead{[Myr]} & \colhead{[Myr]} & \colhead{[\%]}
}
\startdata
\hline
COSMOS-20646 & $137_{-75}^{+51}$ & $170_{-45}^{+28}$ & $94_{-16}^{+5}$ &  & $111_{-49}^{+98}$ & $70_{-47}^{+88}$ & $99_{-10}^{+1}$ &  & $165_{-53}^{+45}$ & $160_{-47}^{+41}$ & $96_{-4}^{+2}$ &  & $107_{-47}^{+44}$ & $148_{-63}^{+54}$ & $92_{-6}^{+3}$ \\
COSMOS-47074 & $121_{-73}^{+59}$ & $168_{-60}^{+32}$ & $95_{-11}^{+4}$ &  & $27_{-15}^{+57}$ & $18_{-12}^{+100}$ & $53_{-51}^{+45}$ &  & $146_{-58}^{+55}$ & $167_{-58}^{+39}$ & $94_{-6}^{+3}$ &  & $79_{-49}^{+56}$ & $114_{-69}^{+74}$ & $97_{-6}^{+1}$ \\
EGS-6811 & $186_{-77}^{+57}$ & $215_{-50}^{+34}$ & $74_{-21}^{+13}$ &  & $104_{-77}^{+136}$ & $59_{-45}^{+111}$ & $51_{-51}^{+39}$ &  & $193_{-74}^{+61}$ & $203_{-64}^{+50}$ & $75_{-19}^{+12}$ &  & $81_{-52}^{+72}$ & $119_{-77}^{+90}$ & $40_{-40}^{+26}$ \\
EGS-44164 & $184_{-84}^{+57}$ & $216_{-57}^{+28}$ & $73_{-23}^{+13}$ &  & $61_{-42}^{+88}$ & $35_{-25}^{+108}$ & $11_{-10}^{+59}$ &  & $184_{-64}^{+69}$ & $204_{-67}^{+48}$ & $74_{-18}^{+11}$ &  & $100_{-49}^{+61}$ & $141_{-66}^{+79}$ & $49_{-34}^{+18}$ \\
EGS-68560 & $80_{-48}^{+72}$ & $188_{-124}^{+44}$ & $60_{-26}^{+16}$ &  & $23_{-10}^{+15}$ & $10_{-5}^{+28}$ & $1_{-1}^{+12}$ &  & $127_{-53}^{+69}$ & $195_{-76}^{+44}$ & $71_{-16}^{+11}$ &  & $54_{-31}^{+48}$ & $81_{-46}^{+66}$ & $49_{-41}^{+22}$ \\
EGS-20381 & $183_{-95}^{+71}$ & $225_{-68}^{+44}$ & $68_{-25}^{+15}$ &  & $63_{-46}^{+119}$ & $43_{-34}^{+128}$ & $11_{-11}^{+63}$ &  & $179_{-59}^{+76}$ & $209_{-70}^{+49}$ & $69_{-18}^{+13}$ &  & $97_{-59}^{+61}$ & $138_{-82}^{+82}$ & $42_{-42}^{+21}$ \\
EGS-26890 & $141_{-75}^{+65}$ & $193_{-67}^{+34}$ & $74_{-24}^{+15}$ &  & $42_{-26}^{+63}$ & $30_{-23}^{+91}$ & $14_{-14}^{+53}$ &  & $163_{-57}^{+69}$ & $189_{-62}^{+49}$ & $78_{-17}^{+11}$ &  & $97_{-51}^{+58}$ & $139_{-72}^{+76}$ & $64_{-32}^{+13}$ \\
EGS-26816 & $128_{-82}^{+70}$ & $178_{-77}^{+38}$ & $81_{-28}^{+14}$ &  & $29_{-17}^{+80}$ & $22_{-15}^{+85}$ & $24_{-23}^{+73}$ &  & $145_{-56}^{+68}$ & $176_{-62}^{+43}$ & $84_{-11}^{+8}$ &  & $84_{-51}^{+56}$ & $120_{-70}^{+74}$ & $75_{-33}^{+10}$ \\
EGS-40898 & $178_{-107}^{+76}$ & $220_{-75}^{+55}$ & $65_{-29}^{+17}$ &  & $74_{-59}^{+183}$ & $56_{-46}^{+103}$ & $26_{-26}^{+67}$ &  & $183_{-68}^{+72}$ & $209_{-71}^{+53}$ & $68_{-19}^{+14}$ &  & $93_{-57}^{+70}$ & $132_{-79}^{+90}$ & $39_{-39}^{+24}$ \\
GOODSN-35589 & $73_{-46}^{+48}$ & $130_{-56}^{+27}$ & $100_{-0}^{+0}$ &  & $11_{-6}^{+97}$ & $14_{-8}^{+74}$ & $100_{-0}^{+0}$ &  & $99_{-40}^{+49}$ & $131_{-48}^{+30}$ & $100_{-0}^{+0}$ &  & $49_{-32}^{+49}$ & $73_{-48}^{+65}$ & $100_{-0}^{+0}$ \\
UDS-18697 & $225_{-22}^{+13}$ & $121_{-16}^{+32}$ & $99_{-1}^{+0}$ &  & $262_{-60}^{+19}$ & $64_{-11}^{+40}$ & $100_{-0}^{+0}$ &  & $207_{-31}^{+11}$ & $116_{-16}^{+38}$ & $97_{-2}^{+1}$ &  & $218_{-54}^{+41}$ & $205_{-23}^{+18}$ & $96_{-3}^{+2}$ \\
\hline
\enddata
\end{deluxetable*}

In Appendix~\ref{appsec:degeneracy}, Figures~\ref{appfig:mass_age} and \ref{appfig:Auv_age} show the posterior distribution of $M_{\star}$ versus $t_{\rm 50}$ and UV attenuation $A_{\rm UV}$ versus $t_{\rm 50}$, respectively. We show that there is a degeneracy between $M_{\star}$ and $t_{\rm 50}$: a younger age implies a lower stellar mass. This is expected as younger stellar populations are typically brighter at fixed stellar mass. Additionally, we also demonstrate that it is challenging to break the dust-age degeneracy with our current observational data. Specifically, we find an anti-correlation between older ages and more UV attenuation. The UV attenuation is overall not well constrained, i.e., we find rather wide posteriors with uncertainties of more than 1 mag, which can at least in part be explained by the degeneracy with the attenuation law. 

Previously, \citet{stefanon19} measured stellar ages of $22_{-22}^{+69}~\mathrm{Myr}$ (median and 68\% confidence interval over their sample) for 18 bright $z=8$ galaxies ($H_{\rm AB}<25$) with stellar masses of $M_{\star}\sim10^9~M_{\odot}$. They used the FAST code \citep{kriek09} and assumed a fixed  sub-solar $0.2~Z_{\odot}$ metallicity, a constant SFH, and a fixed \citet{calzetti00} attenuation law. Our $10^9~M_{\odot}$ galaxies are slightly older ($50-100$ Myr) when considering the fiducial continuity prior (and therefore an SFH that is similar to a constant as \citealt{stefanon19} assumed), which might imply that fixing the attenuation law leads to this difference as their typical attenuation values are also lower than ours ($A_{\rm V}=0.15_{-0.15}^{+0.30}~\mathrm{mag}$). 

As described in the Introduction, \citet[][see also \citealt{hashimoto18}]{laporte21} study six $z\sim9-10$ galaxies (out of which 3 have a spectroscopic redshift) and perform the SED modeling with \texttt{BAGPIPES} \citep{carnall18}, assuming a fixed \citet{calzetti00} attenuation law and model emission lines self-consistently. They investigate four different SFH prescriptions (delayed, exponential, constant, or burst-like), where the best fit is always obtained for the delayed or constant SFH. They quote ages of $200-500$ Myr, but those ages do not correspond to our $t_{\rm 50}$ (or mass-weighted ages), but to the time since star formation has started. This could partially explain why these ages are significantly (factor of 10) older than what \citet{stefanon19} found and also a factor of $2-3$ older than what we find with our fiducial continuity prior ($\sim150$ Myr). Indeed, when computing $t_{\rm 20}$ (lookback age at which 20\% of the stellar mass was formed), which is a better tracer of the first star-formation episode, we find a median over the whole sample of $260_{-105}^{+59}~\mathrm{Myr}$ and $265_{-90}^{+58}~\mathrm{Myr}$ for the continuity and Dirichlet priors, respectively. These estimates are consistent with the ones inferred by \citet{laporte21}. However, we still infer younger $t_{\rm 20}$ when adopting the bursty continuity ($91_{-72}^{+183}~\mathrm{Myr}$) or parametric ($163_{-99}^{+117}~\mathrm{Myr}$) priors. An exception is UDS-18697, where the results from all priors are consistent with a very early buildup of stellar mass: most of the mass formed around $z\approx15$ and the mass fraction formed before $z>12$ is $>60\%$ (Figure~\ref{appfig:frac_M_z12}). The reason for this is the strong Balmer break (see Figure~\ref{fig:SEDs}, but note that we question the reliability of the IRAC photometry of this object), which can only be explained by older stellar populations. This indicates that these age differences between our work and the work by \citet{laporte21} might not only be caused by the different methodologies and definitions, but also because of sample selection. Indeed, the selection of the sample in different works differs significantly: \citet{laporte21} only selected objects with red IRAC [3.6]$-$[4.5] colors at $z >$ 9 (and detection in both bands) to specifically search for older stellar populations, while we do not employ such a cut. Importantly, we do not claim that we can rule out old ages; we claim that our current data cannot unambiguously confirm old ages in all galaxies in our sample.

\subsection{Fraction of mass formed at $z>10$}
\label{subsec:frac_mass}

\begin{figure*}
\centering
\includegraphics[width=\textwidth]{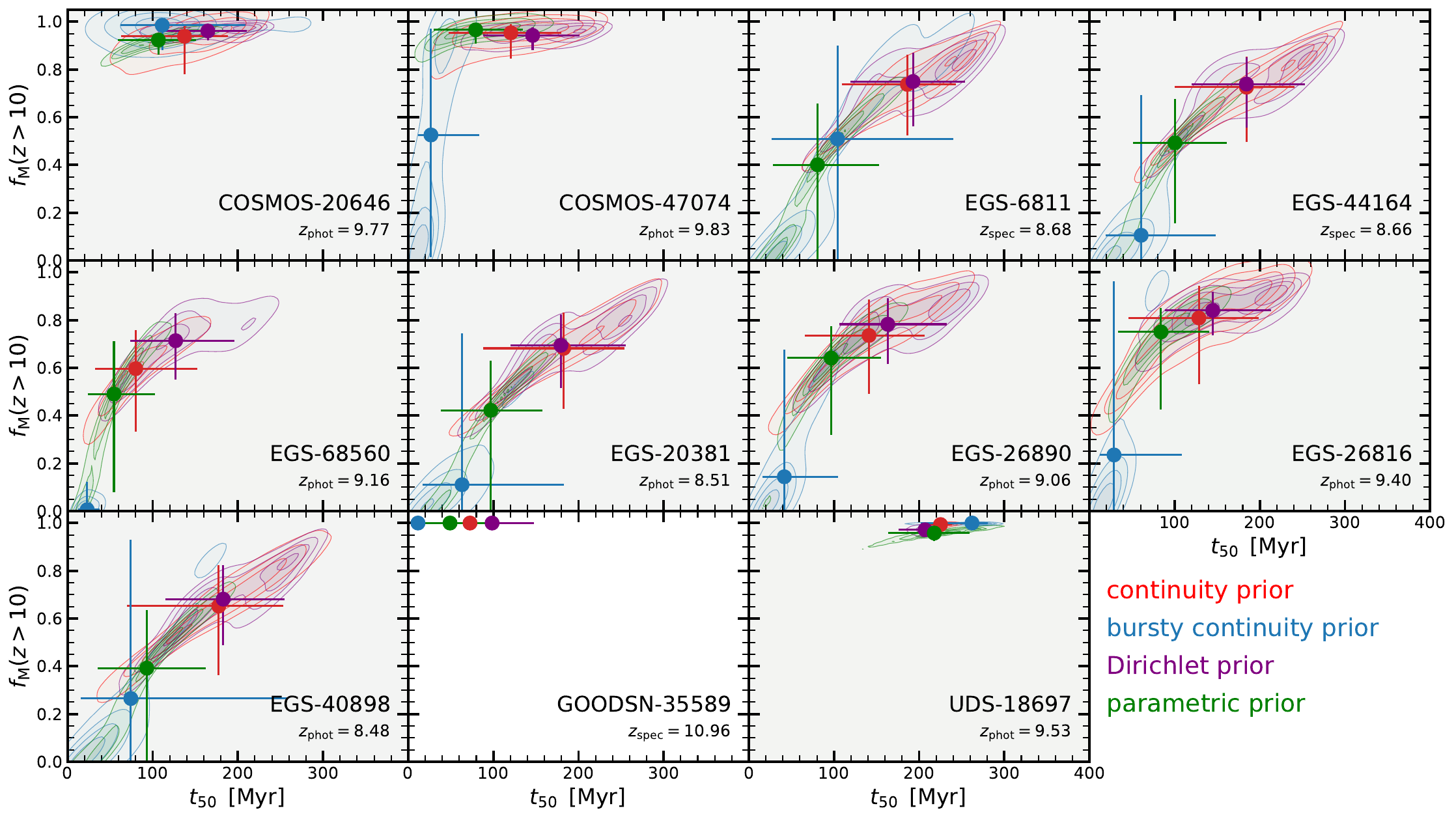}
\caption{Fraction of mass formed previous to redshift $z=10$ ($f_{\rm M}(z>10)$). For each galaxy, we plot the posteriors of $f_{\rm M}(z>10)$ and stellar age ($t_{50}$). As expected, we find a strong correlation between $f_{\rm M}(z>10)$ and age: older galaxies have a higher $f_{\rm M}(z>10)$. More importantly, $f_{\rm M}(z>10)$ (as the age) depends heavily on the assumed prior (shown with the different colors). Since GOODSN-35589 lies at $z_{\rm spec}=10.96$, $f_{\rm M}(z>10)$ is equal to 1.}
\label{fig:frac_M_z10}
\end{figure*}

We now focus on the fraction of stellar mass formed at early times. Figure~\ref{fig:frac_M_z10} shows the posterior distribution of the fraction of mass formed before redshift 10 ($f_{\rm M}(z>10)$) and stellar age $t_{\rm 50}$. The inferred values of $f_{\rm M}(z>10)$ and their uncertainties are also given in Table~\ref{tab:ages}. By definition, $f_{\rm M}(z>10)$ is only a useful number if the redshift of the galaxy is below $z=10$, i.e., $f_{\rm M}(z>10)=100\%$ for galaxy GOODSN-35589 with $z_{\rm spec}=10.96$. Therefore, Figure~\ref{appfig:frac_M_z12} in Appendix~\ref{appsec:frac_M_z12} shows $f_{\rm M}(z>12)$. 

Figure~\ref{fig:frac_M_z10} makes the point that the uncertainty in stellar age directly translates into an uncertainty in $f_{\rm M}(z>10)$. Therefore, different SFH priors can lead to substantially different estimates of $f_{\rm M}(z>10)$. A good example of this is COSMOS-47074, which has $f_{\rm M}(z>10)>90\%$ when considering the continuity prior, the Dirichlet prior, or the parametric prior, while the fraction is only $f_{\rm M}(z>10)=53_{-51}^{+45}\%$, albeit a large uncertainty when adopting the bursty continuity prior. Across our full sample, we find a median (and 68\% percentile) for $f_{\rm M}(z>10)$ of $79_{-28}^{+18}\%$, $39_{-39}^{+61}\%$, $82_{-21}^{+14}\%$, and $65_{-45}^{+31}\%$ when adopting the continuity prior, the bursty continuity prior, the Dirichlet prior, and the parametric prior, respectively. In summary, we conclude that the fraction of stellar mass formed at $z >$ 10 (and also $z >$ 12, see Appendix~\ref{appsec:frac_M_z12}) is not well constrained by our data.

\subsection{Implications for the cosmic SFR density}
\label{subsec:cSFRD}

\begin{figure}
\includegraphics[width=\linewidth]{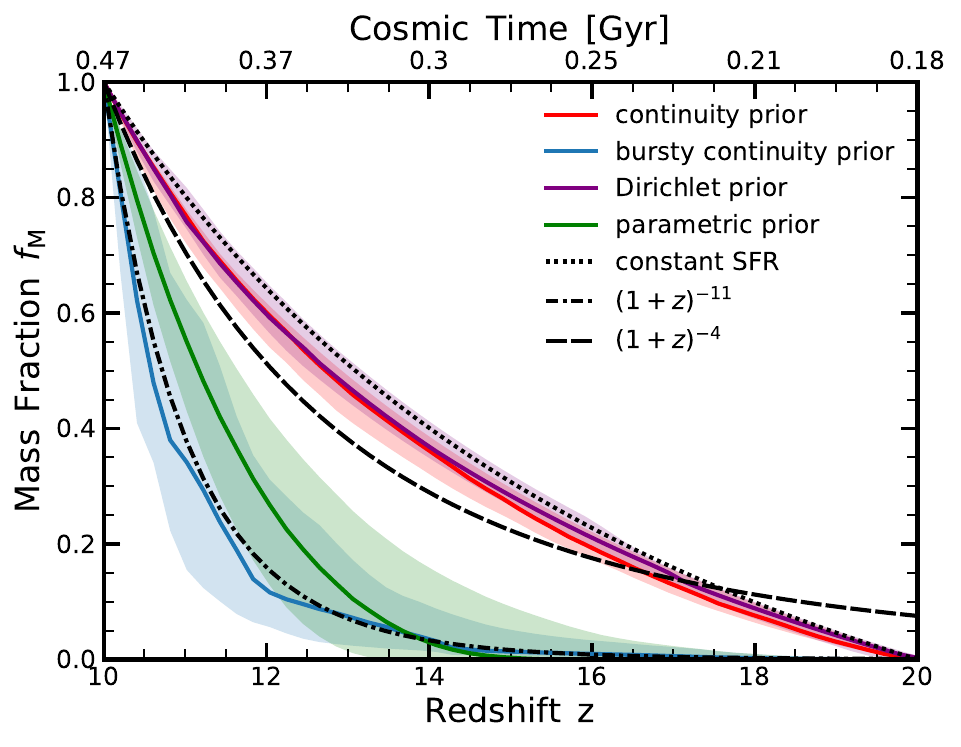}
\caption{Cosmic SFR density implied by our inferred SFHs. We compute the fraction of stellar mass formed from our posterior SFHs averaged over all 10 galaxies that lie at $z<10$ (excluding GOODSN-35589). The red, blue, purple and green lines show the continuity prior, the bursty continuity prior, the Dirichlet prior and the parametric prior for the SFH, respectively. The dotted black line shows the evolution of $f_{\rm M}$ for a constant SFR, while the dashed and dash-dotted lines indicate the rate of decline in the UV luminosity density deduced from large photometric surveys \citep{oesch14, oesch18, bouwens15, mcleod16}. The SFH prior plays a crucial role: assuming the continuity prior or the Dirichlet prior leads to a rather constant SFH, consistent with the rather slow increase with time ($\propto(1+z)^{-4}$; dash-dotted line) as inferred in some UV luminosity studies. On the other hand, assuming the bursty continuity prior or the parametric prior leads to a more rapid and recent increase in $f_{\rm M}$, consistent with the steep increase with time ($\propto(1+z)^{-11}$; dashed line) as inferred in other UV luminosity studies.}
\label{fig:cSFRD}
\end{figure}

An interesting application of the derived SFHs is to study the implications for the early mass assembly of stellar mass as this provides an independent insight into the evolution of the UV luminosity density within the first $\sim500$ Myr. In the literature, it is debated whether the luminosity density over the redshift range $8<z<11$ declines rapidly with $\propto(1+z)^{-11}$ \citep[e.g.,][]{oesch14, oesch18, bouwens15} or slowly with $\propto(1+z)^{-4}$ \citep[e.g.,][]{mcleod16, finkelstein16}. From a theory perspective, a constant star-formation efficiency model together with the buildup of dark matter halos can reproduce the suggested rapid increase in the cosmic SFR density with time \citep{tacchella13, tacchella18, mason15, mashian16, yung19_uvlf}. However, there are also other models that prefer a rather slow increase of the cosmic SFR density at early times \citep{moster18, behroozi20}, consistent with SFHs for the six galaxies studied by \citet{laporte21}.

Following the same approach, we average all our 10 $f_{\rm M}(t)$ SFHs (excluding GOODSN-35589 which lies at $z_{\rm spec}=10.96$) and plot their stacked mass assembly history $f_{\rm M}(z)$ in Figure~\ref{fig:cSFRD}. The important assumption when doing this is that the galaxy sample is representative of the overall galaxy population at this epoch. The red, blue, purple, and green lines show the resulting $f_{\rm M}(z)$ assuming the continuity, the bursty continuity, the Dirichlet, and the parametric SFH priors, respectively. For reference, the dotted, dash-dotted, and dashed black lines indicate $f_{\rm M}(z)$ for a constant SFR, the rapidly increase cosmic SFR density ($\propto(1+z)^{-11}$), and a slowly increase cosmic SFR density ($\propto(1+z)^{-4}$), respectively. 

As expected from the previous sections, our conclusion depends on the adopted SFH prior. If we assume the continuity or the Dirichlet prior, which are overall consistent with a constant SFH, we find they match well with the slowly declining SFR density. This is also compatible with the conclusions of \citet{laporte21}, which is not surprising, as their best fit is a constant or delayed SFH model. On the other hand, if we adopt the bursty continuity prior or the parametric prior, we find $f_{\rm M}(z)$ increases more rapidly at more recent times, which is more consistent with a rapid decline in the cosmic SFR density at $z>10$. We find qualitatively the same results if we include GOODSN-35589 and perform the analysis at $z=12$. In summary, the presently available observational data cannot constrain the SFHs of our ensemble of galaxies well enough to distinguish between a rapid or more smooth decline in the cosmic SFR density at $z>10$, therefore the epoch of first galaxy formation remains to yet be identified. Interestingly, there are indications (even when varying the prior; see in particular UDS-18697) that at least some galaxies have formed significant amounts of stellar mass by $z\approx15$ (Fig.~\ref{fig:SFH_fm}), which is of interest for the interpretation of the 21cm signal and the formation and evolution of black holes in the early universe \citep[e.g.,][]{dayal18}.

\subsection{Overly massive galaxies}
\label{subsec:too_massive}

\begin{figure}
\includegraphics[width=\linewidth]{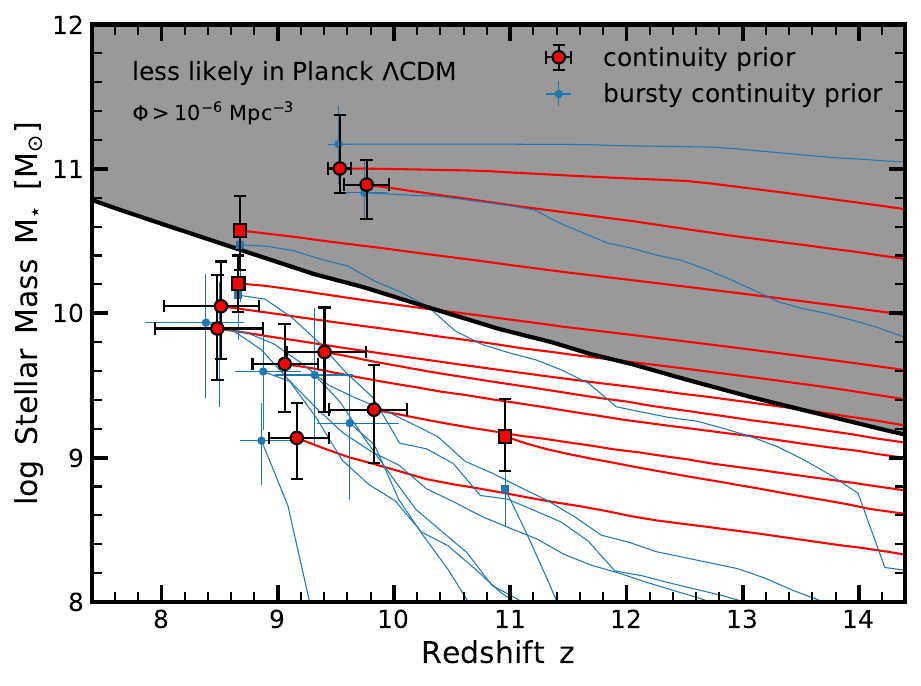}
\caption{Are the $z=9-11$ galaxy candidates overly massive? We plot the stellar mass $M_{\star}$ as a function of redshift $z$ for the galaxies in our sample. The red and blue symbols mark the measurement adopting the continuity and the bursty continuity SFH prior, respectively. The red and blue lines show the inferred mass growth histories from the SFH measurements. The solid black line indicates the threshold stellar mass for a cumulative number density of $\Phi=10^{-6}~\mathrm{Mpc}^{-3}$ \citep[][converted to Chabrier IMF]{behroozi18}, which is roughly our survey volume. The two galaxies with stellar masses larger than the black line (i.e., in the gray shaded region) are in tension with $\Lambda$CDM, though the significance of this tension is low as we discuss in the text.}
\label{fig:overly_massive}
\end{figure}

As discussed above, our stellar mass estimates are overall robust, including the variation of the SFH prior, but some sources might suffer from IRAC systematics as highlighted below. We now investigate whether these stellar masses are above the expectation of the Planck $\Lambda$CDM universe. In particular, given our survey volume of roughly $10^6~\mathrm{Mpc}^{3}$, high-redshift galaxy stellar masses can place interesting limits on number densities of massive halos, which itself constrains cosmology \citep{steinhardt16, behroozi18}. 

Figure~\ref{fig:overly_massive} shows our stellar mass estimates as a function of redshift. The red and blue symbols show the measurements adopting the continuity and the bursty continuity SFH priors, respectively. The red and blue solid lines show the mass growth tracks of individual galaxies as inferred from the SFHs presented in Section~\ref{subsec:SFH}. The black solid line marks the threshold stellar mass for a cumulative number density of $\Phi=10^{-6}~\mathrm{Mpc}^{-3}$, which we adopt from \citet{behroozi18}. The assumption for this stellar mass threshold is a 100\% star-formation efficiency, i.e., the halo mass is related to the stellar mass via $M_{\rm h}=M_{\star}/f_{\rm b}$, where $f_{\rm b}=0.16$ is the cosmic baryon fraction. This can be regarded as the maximal stellar mass since the average SFHs of galaxies inferred from the present-day stellar-to-halo mass relation are much less than the cosmic baryon fraction \citep[e.g.,][]{moster18, behroozi19}.

We find that three galaxies (EGS-6811, COSMOS-20646 and UDS-18697) lie above the black line and in the gray shaded region that is less likely in the Planck $\Lambda$CDM universe. Galaxy EGS-6811 ($\log(M_{\star}/M_{\odot})=10.6_{-0.3}^{+0.2}$ with a spectroscopic redshift) actually lies on the threshold when considering the uncertainty. Therefore, this galaxy is not challenging $\Lambda$CDM. Interestingly, when studying its trajectory in the $M_{\star}-z$ plane, we find that the bursty continuity prior leads to a steep mass growth history, which means that it actually falls below the mass threshold by $z\sim10-11$, while it remains in the less likely $\Lambda$CDM region when adopting the continuity prior. An important note with respect to this statement is that these growth curves assume that there are no previous galaxy mergers involved in the mass growth. Or, equivalently, they represent the summed $M_{\star}$ associated with all galaxies that merge to form the observed objects.

\begin{figure*}
\includegraphics[width=\textwidth]{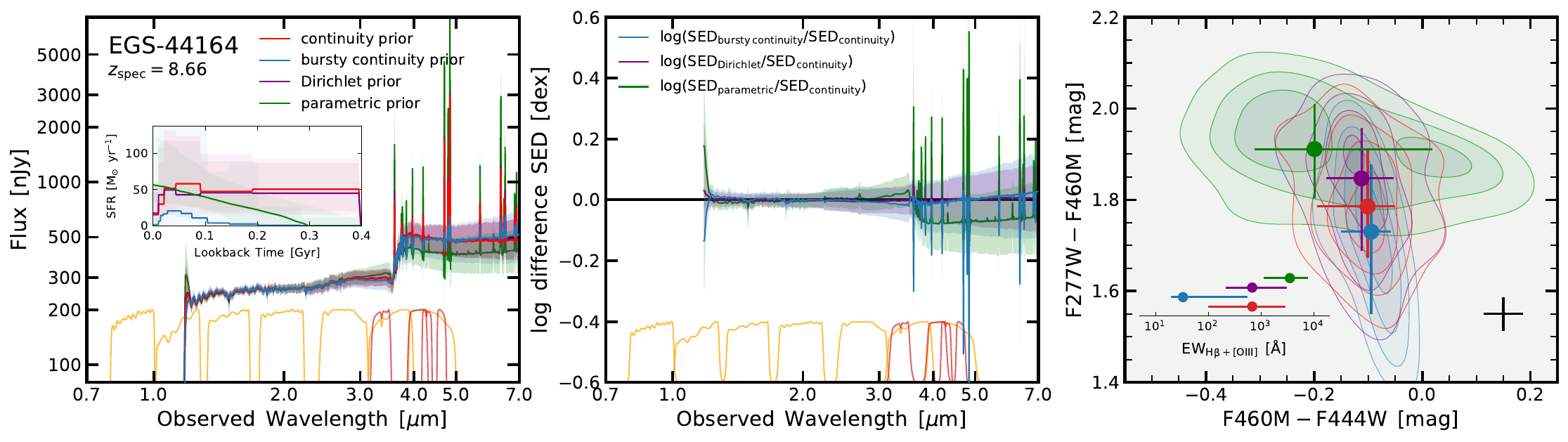}
\caption{Distinguishability of different star-formation histories (SFHs). We plot in the left panel the posterior SED of EGS-44164 resulting from the four different SFH priors: continuity prior (red), bursty continuity prior (blue), Dirichlet prior (purple), and parametric prior (green). The inset show the posteriors of the SFHs. At the bottom, the orange and red lines indicate the widely used {\it JWST}/NIRCam wide (F090W, F115W, F150W, F200W, F277W, F356W, F444W) and medium (F335M, F410M, F430M, F460M) bands, respectively. The middle panel shows the log difference between the bursty continuity prior (blue), Dirichlet prior (purple), and parametric prior (green) relative to the continuity prior. By construction, the SEDs look very similar in the rest-frame UV, while there are some noticeable features around the Balmer break and in the emission line strengths. The right panel shows the {\it JWST} F277W$-$F460M versus F460M$-$F444W color-color diagram and the H$\beta$+[\ion{O}{3}] equivalent width ($\mathrm{EW}_{\rm H\beta+[OIII]}$) distribution. These medium-band colors are sensitive to emission lines as -- for example in this case -- the H$\beta$ emission line straddles the F460M filter. The errorbars at the bottom right illustrate the $5\sigma$ uncertainty for $\sim15$ ksec exposures (see the text for details). }
\label{fig:jwst}
\end{figure*}

We have two galaxies (COSMOS-20646 and UDS-18697) that both lie at $z\sim9-10$ and have stellar masses of $M_{\star}\approx10^{11}~M_{\odot}$ and therefore lie solidly within the less likely $\Lambda$CDM region. However, we acknowledge that these two galaxies (along with EGS-6811) have bright neighbors in their proximity, making the IRAC photometry of these two sources the least reliable of our sample (see F21 and Section~\ref{appsec:galfit_phot}). Thus it is possible that residual light from the neighbors is contributing to the high stellar mass measurement. Nevertheless, taking our fiducial stellar mass and redshift at face value (i.e., ignoring the systematic uncertainty in the IRAC photometry for the moment), we find that COSMOS-20646 and UDS-18697 lie within the less likely $\Lambda$CDM region at $3.0\sigma$ and $4.6\sigma$ significance. 

However, as discussed in detail in \citet[][see their Appendix A]{behroozi18}, the significance of this tension is not as sound as apparent on first sight. Attempts to rule out $\Lambda$CDM are limited by both cosmic variance and observational errors. Considering cosmic variance, as we have selected the galaxies from five different survey fields (Section~\ref{sec:sample}), the chance actually significantly increases that one of the fields will have an ``outlier'' even in a standard $\Lambda$CDM universe \citep[e.g.,][]{trenti08}. The observational errors (considering an uncertainty of $0.2-0.3~\mathrm{dex}$ in $\log~M_{\star}$, but ignoring the larger systematic uncertainty stemming from the IRAC photometry) will inflate the number density of massive halos compared to the underlying true number density \citep[see also the Eddington bias;][]{eddington13} because objects are preferentially scattered to higher masses when drawn from a steep mass function. Taking these effects into account, we calculate that the probability is $\sim20\%$ and $\sim0.4\%$ to find in our survey a galaxy as massive as COSMOS-20646 and UDS-18697, respectively.

Finally, another interesting insight from Figure~\ref{fig:overly_massive} is about the SFH prior, as hinted at above. In the case of the continuity SFH prior, which typically leads to extended SFHs and older ages (Figure~\ref{fig:M_age_tauSF}), the number of galaxies crossing the threshold increases toward higher redshifts. Contrarily, when considering the bursty continuity SFH prior, we find the opposite behavior, and galaxies depart from the less likely $\Lambda$CDM region. This implies that the continuity prior has a larger tendency to violate $\Lambda$CDM than the bursty continuity prior. As the uncertainties in the derived SFHs are large, we however cannot rule out any of the priors at the moment.

\subsection{Towards JWST}
\label{subsec:jwst}

{\it JWST} will transform high-$z$ galaxy evolution studies by providing near-IR (i.e., rest-frame optical) data of unprecedented depth, spatial, and spectral resolution. This will help to better constrain the rest-frame Balmer/$4000~\mathrm{\AA}$-break and therefore get tighter constraints on the SFHs of $z>8$ galaxies. A detailed discussion on the implications for these kinds of measurements is out of the scope of this paper, but see \citet{roberts-borsani21_jwst} for an exploration of the improvement of $z\sim7-11$ galaxy property estimates with {\it JWST}/NIRCam medium-band photometry.

Here we focus on the implications of the different SFH priors regarding the {\it JWST} wavelength coverage. Specifically, we plot in Figure~\ref{fig:jwst} the SED posterior for the four different SFH priors (left panel) and their log differences (middle panel). The orange and red lines on the bottom show the {\it JWST}/NIRCam broad- and medium-band filter curves. We focus here on EGS-44164 as this galaxy has a spectroscopic redshift and is detected with \textit{Spitzer}/IRAC. As we can see from the left panel of Figure~\ref{fig:jwst}, the SEDs from the four different SFH priors are similar in the rest-frame UV wavelength range, while they start diverging in the rest-frame optical. We find strong emission lines but a weak $4000~\mathrm{\AA}$ continuum break for the parametric prior, while the break is stronger but the emission lines are nearly absent for the bursty continuity prior. The continuity and the Dirichlet prior both have a rather strong Balmer/$4000~\mathrm{\AA}$-break and emission lines. These features can be directly understood by looking at the SFHs, which is shown as an inset in the figure. The errorbars at the bottom right illustrate the $5\sigma$ uncertainty for $\sim15$ ksec exposures, estimated from the point source limit with a $0.1\arcsec$ aperture and medium background \citep{williams21}.

The \textit{Spitzer}/IRAC data cannot currently differentiate between these SEDs and SFHs. For {\it JWST}, thanks to its unprecedented sensitivity and its higher spectral resolution, progress can be made, in particular through the inclusion of emission lines. We show the F277W$-$F460M versus F460M$-$F444W color-color diagram in the right panel of Figure~\ref{fig:jwst}. We give the H$\beta$+[\ion{O}{3}] equivalent width ($\mathrm{EW}_{\rm H\beta+[OIII]}$) distribution in the inset of the panel. The significant differences of the very recent ($<20~\mathrm{Myr}$) SFHs between the four priors leads to contrasting EW distributions. For the continuity prior, the bursty continuity prior, the Dirichlet prior and the parametric prior, we find $\mathrm{EW}_{\rm H\beta+[OIII]}=675_{-578}^{+2213}~\mathrm{\AA}$, $\mathrm{EW}_{\rm H\beta+[OIII]}=33_{-14}^{+512}~\mathrm{\AA}$, $\mathrm{EW}_{\rm H\beta+[OIII]}=673_{-463}^{+2361}~\mathrm{\AA}$, and $\mathrm{EW}_{\rm H\beta+[OIII]}=3507_{-2388}^{+4314}~\mathrm{\AA}$, respectively. 

These different EW measurements are then directly reflected in the color-color diagram since the H$\beta$ line straddles the medium-band F460M for this specific case. Hence, these red medium bands will help constrain the recent SFH and will generally provide more stringent constraints on the stellar populations \citep[see also][]{roberts-borsani21_jwst}. Obviously, having higher spectral resolution information (via for example the NIRSpec/Prism or NIRCam/Grism) will further constrain these emission lines. One caveat to this is the rather large uncertainty on the number of ionizing photons that power those emission lines, in particular related to the escape and dust absorption of those photons \citep[e.g.,][]{kimm17, smith17, glatzle19} as well as the production efficiency (stellar binarity and rotation; e.g., \citealt{choi17, eldridge17}).

\section{Summary \& Conclusions}
\label{sec:conclusion}

In this work we carefully assess the current constraints on the stellar populations of galaxies at $z=9-11$. The sample consists of 11 bright ($H <26.6$) galaxies, out of which 3 have spectroscopic redshifts (F21). Given the high-redshift nature of these sources, {\it HST} and \textit{Spitzer}/IRAC data only trace the rest-frame UV and the Balmer/$4000~\mathrm{\AA}$-break of these galaxies. 

We perform a careful inference of the stellar populations by using \texttt{Prospector}, a flexible Bayesian SED fitting code \citep{johnson21}. In particular, we expand upon previous $z>6$ SED investigations by adopting a range of flexible and parametric SFHs, a flexible dust attenuation law, self-consistent modeling for emission lines, and variable IGM absorption. A flexible attenuation law is important because it allows us to assess how degenerate metallicity, age, and SFH constraints behave with the attenuation when extracting this information from a rather limited wavelength coverage at low spectral resolution \citep[e.g.,][]{kriek13, battisti16, salim18_curves, tacchella21_quench}. The self-consistent modeling of emission lines is essential because it allows us to estimate their contribution to the IRAC fluxes and to address the degeneracy with a possible Balmer/$4000~\mathrm{\AA}$ break also present in those bands \citep[e.g.,][]{stark13, hashimoto18}. Finally, the different SFH priors are crucial as we have little knowledge about the ``burstiness'' (i.e., star-formation variability) of these systems \citep[e.g.,][]{smit16, faucher-giguere18, faisst19, iyer20, tacchella20}. 

Our SED-modeling approach (Section~\ref{sec:prospector}) takes into account all of the aforementioned points. The strength of \texttt{Prospector} is its fully Bayesian nature and its flexibility, which allows us to investigate how different assumptions (i.e., priors) affect our results and conclusions. In particular, the choice of the prior for the SFH (Section~\ref{subsec:prior_sfh}) has important consequences on conclusions regarding the early growth of these $z=9-11$ galaxies. We investigate the impact of four priors: the continuity prior, the bursty continuity prior, the Dirichlet prior, and the parametric prior. The behaviors of these different priors are shown in Figure~\ref{fig:prior_sfh}. The continuity prior and Dirichlet prior are both weighted toward a smooth behavior with a constant SFR as the expectation value. The bursty continuity prior allows for bursty star formation, where most of the mass is formed in one or a few time bins. Finally, the parametric prior assumes a delayed-$\tau$ model and -- at these early times -- typically leads to an increasing SFH. We find that all these SFH priors can equally well describe the observational data, and the data do not prefer any of the priors (Bayes factors are about 1).

We measure the rest-frame UV spectral slope ($\beta$) from the \texttt{Prospector} posterior model spectra for our sample of galaxies and find a significant correlation between $\beta$ and the stellar mass, where more massive galaxies are redder. We do not measure a significant correlation between $\beta$ and the UV luminosity in our sample. The measured UV slopes of our massive ($\log(M_{\star}/M_{\odot})=9-10$) $z \sim$ 9--10 galaxies are similar to measured values at the same stellar masses at $z =$ 4--8, indicating that galaxies at these stellar masses rapidly develop a dust reservoir, which then grows more slowly. This is consistent with galaxy formation models that include a significant contribution of dust grain growth. The roughly constant attenuation at these masses across a wide range of redshift implies that the apparent lack of evolution in the number density of UV-bright galaxies at $z >$ 7 is not due to changes in the dust attenuation.

Despite the flexibility in the SED modeling and the investigation of different SFH priors, we find that the stellar masses and the SFRs (averaged over 50 Myr) are rather well constrained with uncertainties of a factor of 2. We find a hint of a star-forming main sequence with a sub-linear slope ($0.7\pm0.2$; Figure~\ref{fig:M_SFR}), i.e., more massive galaxies have a higher SFR. The sSFRs are in the range of $3-10~\mathrm{Gyr}^{-1}$, which indicates a mass-doubling timescale of $\sim100-300~\mathrm{Myr}$ under the assumption of a constant sSFR. 

Stellar population parameters such as the stellar metallicity, the stellar age, and the SFH itself are less well constrained. For example, changing the SFH prior leads to changes in the age of a factor of $\sim3$ (Figure~\ref{fig:M_age_tauSF}): the median measured age over our sample is $147_{-89}^{+78}~\mathrm{Myr}$ and $53_{-40}^{+153}~\mathrm{Myr}$ for the continuity prior and the busty continuity prior, respectively. Even when just selecting one prior, the uncertainty in age for an individual galaxy remains rather large, which can be largely attributed due to the flexibility in the dust attenuation law and the emission line modeling (Figure~\ref{appfig:Auv_age}). More generally, we find that the SFH priors have an important impact on the measured SFHs (Figures~\ref{fig:SFH_fm} and \ref{appfig:SFH}). 

From this, we draw the important conclusion that the current observational data cannot give us tight constraints on how quickly these $z=9-11$ galaxies are building up their mass (Figure~\ref{fig:cSFRD}). In the case of the bursty continuity prior and the parametric prior, which both prefer younger ages, the inferred stellar mass buildup is consistent with a rapidly increasing cosmic SFR density at $z>8$ with time ($\propto(1+z)^{-11}$). In contrast, the continuity prior and the Dirichlet prior, both preferring older ages, are consistent with a slow increase with time in the cosmic SFR density in this epoch ($\propto(1+z)^{-4}$). Therefore, the epoch of first galaxy formation remains to yet be identified. {\it JWST} with its high sensitivity, larger wavelength coverage, and higher spectral resolution (including medium band imaging and spectroscopy) will help solve this mystery. This in turn will help to constrain galaxy formation models and might even shed light onto the nature of dark matter \citep{dayal15, khimey21}.

\section*{Acknowledgments}

We thank the referee for a thorough report that improved and strengthened this work. We are grateful to Ben Johnson and Joel Leja for helpful discussions related to \texttt{Prospector}. We thank Nicolas Laporte for clarifications and useful comments. S.T. is supported by the Smithsonian Astrophysical Observatory through the CfA Fellowship and by the 2021 Research Fund 1.210134.01 of UNIST (Ulsan National Institute of Science \& Technology). S.L.F. acknowledges support from NASA through ADAP award 80NSSC18K0954. L.G. and R.S. acknowledge support from the Amaldi Research Center funded by the MIUR program ``Dipartimento di Eccellenza'' (CUP:B81I18001170001). I.J. acknowledges support from NASA under award number 80GSFC21M0002. Support for {\it Hubble Space Telescope} program \#15862 was provided by NASA through a grant from the Space Telescope Science Institute, which is operated by the Associations of Universities for Research in Astronomy, Incorporated, under NASA contract NAS5-26555.

\bibliographystyle{apj}

\appendix

\section{GALFIT-based IRAC photometry}
\label{appsec:galfit_phot}

\begin{figure*}
\centering
\includegraphics[width=\textwidth]{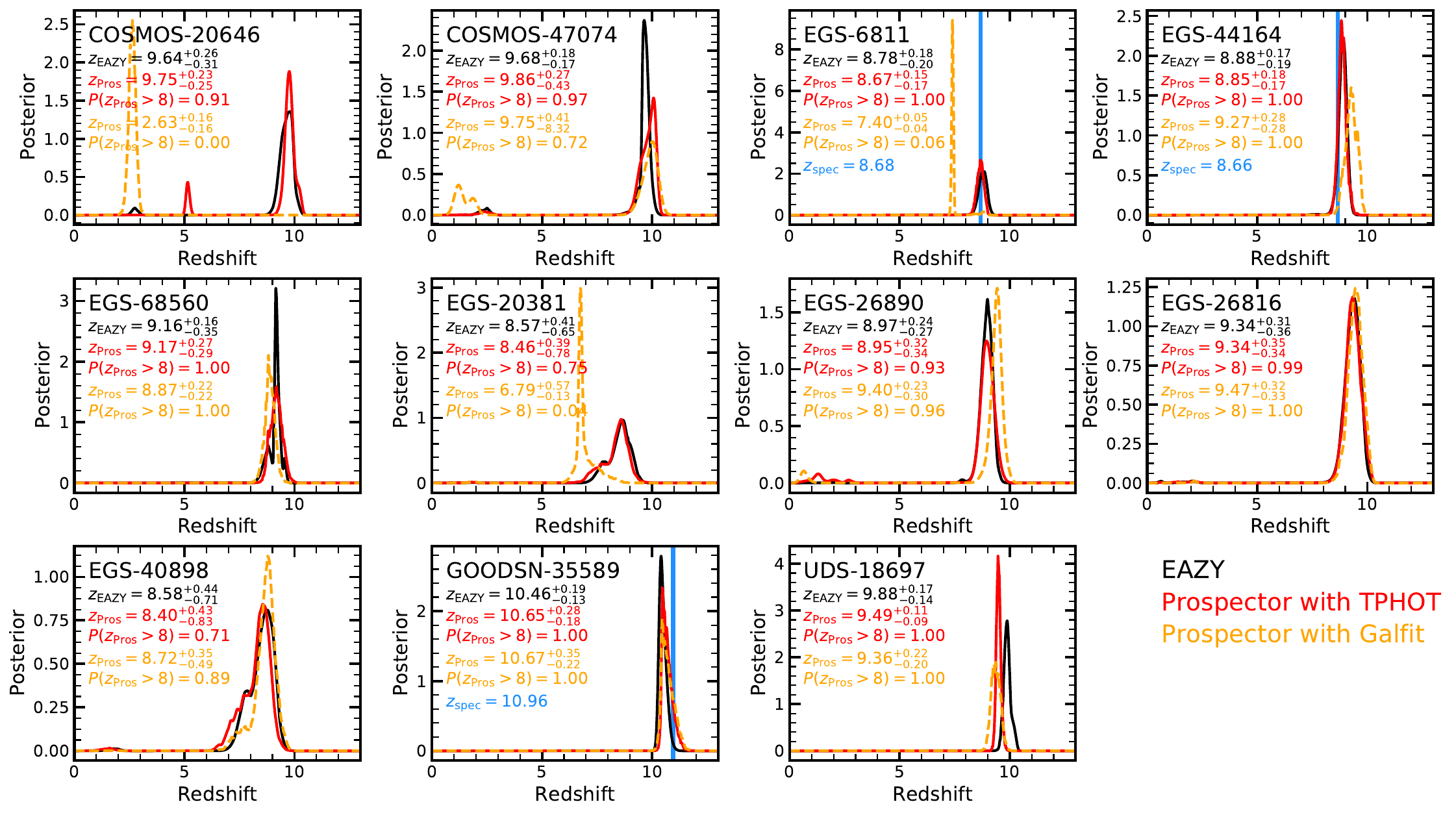}
\caption{Impact of IRAC photometry on resulting photometric redshift ($z_{\rm phot}$) posteriors. Following Figure~\ref{fig:z_phot}, each panel shows the $z_{\rm phot}$ posteriors obtained by \texttt{EAZY}, by \texttt{Prospector} with \texttt{TPHOT} (fiducial) IRAC photometry, and by \texttt{Prospector} with GALFIT IRAC photometry in black, red, and orange, respectively. Three galaxies (EGS-6811, EGS-68560 and GOODSN-35589) have spectroscopic redshifts, which are indicated in blue. We find a significantly different $z_{\rm phot}$ in 3 out of the 11 galaxies (COSMOS-206464, EGS-6811, and EGS-20381), while the other galaxies have a consistent $z_{\rm phot}$ within the uncertainty. Importantly, the $z_{\rm phot}$ of EGS-6811 with GALFIT-based photometry is inconsistent with its $z_{\rm spec}$.}
\label{appfig:z_phot_galfit}
\end{figure*}

\begin{figure*}
\centering
\includegraphics[width=\textwidth]{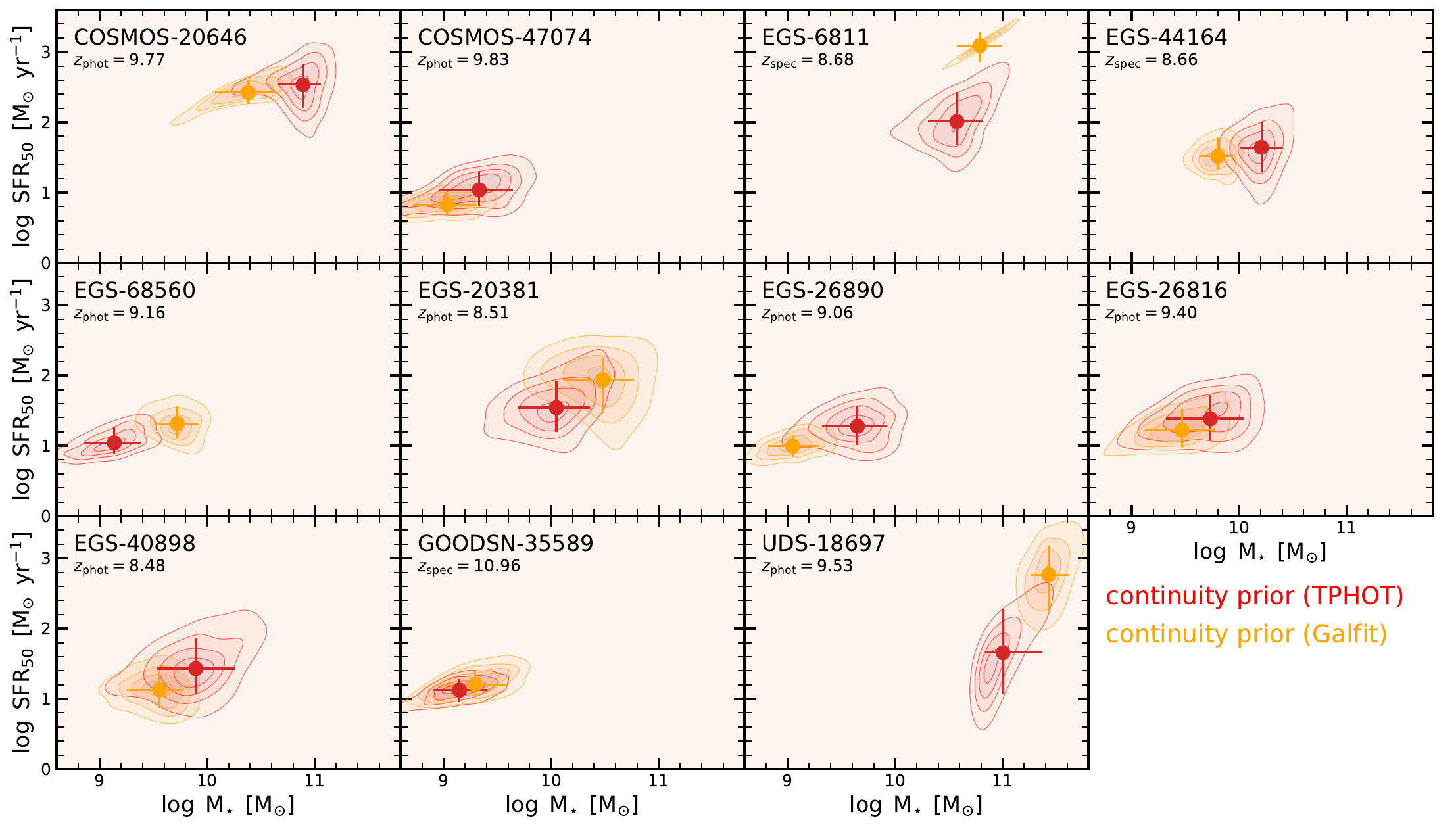}
\caption{Posteriors of the star-formation rate ($\mathrm{SFR}_{50}$) and stellar mass ($M_{\star}$) as in Figure~\ref{fig:SFR_M_posteriors}, but for the \texttt{TPHOT} (fiducial) and \texttt{GALFIT} IRAC photometry in red and orange, respectively. We find that changes in the IRAC photometry lead to differences in the stellar mass and SFR that can be larger than the uncertainties. }
\label{appfig:M_SFR_galfit}
\end{figure*}

We discuss extensively the uncertainty of the IRAC photometry in F21. In that work, we have performed the deblending and flux measurements of the IRAC photometry with both \texttt{TPHOT} (our fiducial approach) and \texttt{GALFIT} (see Table 6 in F21). We discuss in this section how the systematic uncertainty of the IRAC photometry impacts our results.

Figure~\ref{appfig:z_phot_galfit} shows the photometric redshift posterior when adopting \texttt{TPHOT} (solid red lines) or \texttt{GALFIT} (dashed orange lines) IRAC photometry. Similar to Figure~\ref{fig:z_phot}, we also plot the redshift posterior of \texttt{EAZY} for comparison, and the vertical blue lines indicate the spectroscopic redshift when available. This figure assumes the ``$z$-free model'' introduced in Section~\ref{subsec:photo_z}. 

We find that COSMOS-20646 with \texttt{GALFIT}-based IRAC photometry strongly prefers a $z\sim2.5$ solution, consistent with the finding by F21. Furthermore, EGS-6811 and EGS-20381 shift mildly but significantly toward lower redshifts ($z\sim7$). However, in the case of EGS-6811, the \texttt{GALFIT}-based photometric redshift is inconsistent with the spectroscopic redshift. For all other galaxies, the \texttt{GALFIT}-based photometric redshifts are consistent with the \texttt{TPHOT}-based ones within the uncertainties. 

Figure~\ref{appfig:M_SFR_galfit} investigates the SFRs and stellar masses obtained when swapping between \texttt{TPHOT}- and \texttt{GALFIT}-based IRAC photometry. Here we assume our standard model, i.e., the continuity SFH prior and the \texttt{EAZY} redshift posterior as the redshift prior. For most galaxies, the SFR and $M_{\star}$ posteriors are consistent with each other, but exceptions are EGS-6811 and UDS-18697. EGS-68560 and UDS-18697 are more massive when using the \texttt{GALFIT} IRAC photometry, while COSMOS-20646, EGS-44164 and EGS-26890 have a tendency to be less massive. The SFRs are overall similar, but we find a significant increase for EGS-6811 and UDS-18697.

This also has important implications for the discussion whether our galaxies are too massive at these early cosmic times (Section~\ref{subsec:too_massive}). Both COSMOS-20646 and UDS-18697 have stellar masses of $M_{\star}\approx10^{11}~M_{\odot}$ and therefore lie solidly within the less likely $\Lambda$CDM region in Figure~\ref{fig:overly_massive}. Figure~\ref{appfig:M_SFR_galfit} shows that the stellar mass of COSMOS-20646 reduces by roughly a factor of 3 when adopting the \texttt{GALFIT} photometry, making it consistent with the $\Lambda$CDM boundary when considering the uncertainty. On the other hand, UDS-18697 seems to get even more massive (and star forming) when adopting the \texttt{GALFIT} photometry. 

In summary, these differences depending on which photometric method is used highlight a systematic uncertainty when making use of deblended photometry from low-resolution imaging, something which will be alleviated soon with {\it JWST}. We adopt throughout this work the \texttt{TPHOT}-based IRAC photometry as it does overall a better job of removing the bright neighboring sources (F21). The object COSMOS-20646 has the largest uncertainty regarding which method is more accurate; we therefore caution the reader that the nature of this galaxy is still somewhat uncertain.

\section{Star-formation histories: SFR versus time}
\label{appsec:sfh}

\begin{figure*}
\centering
\includegraphics[width=\textwidth]{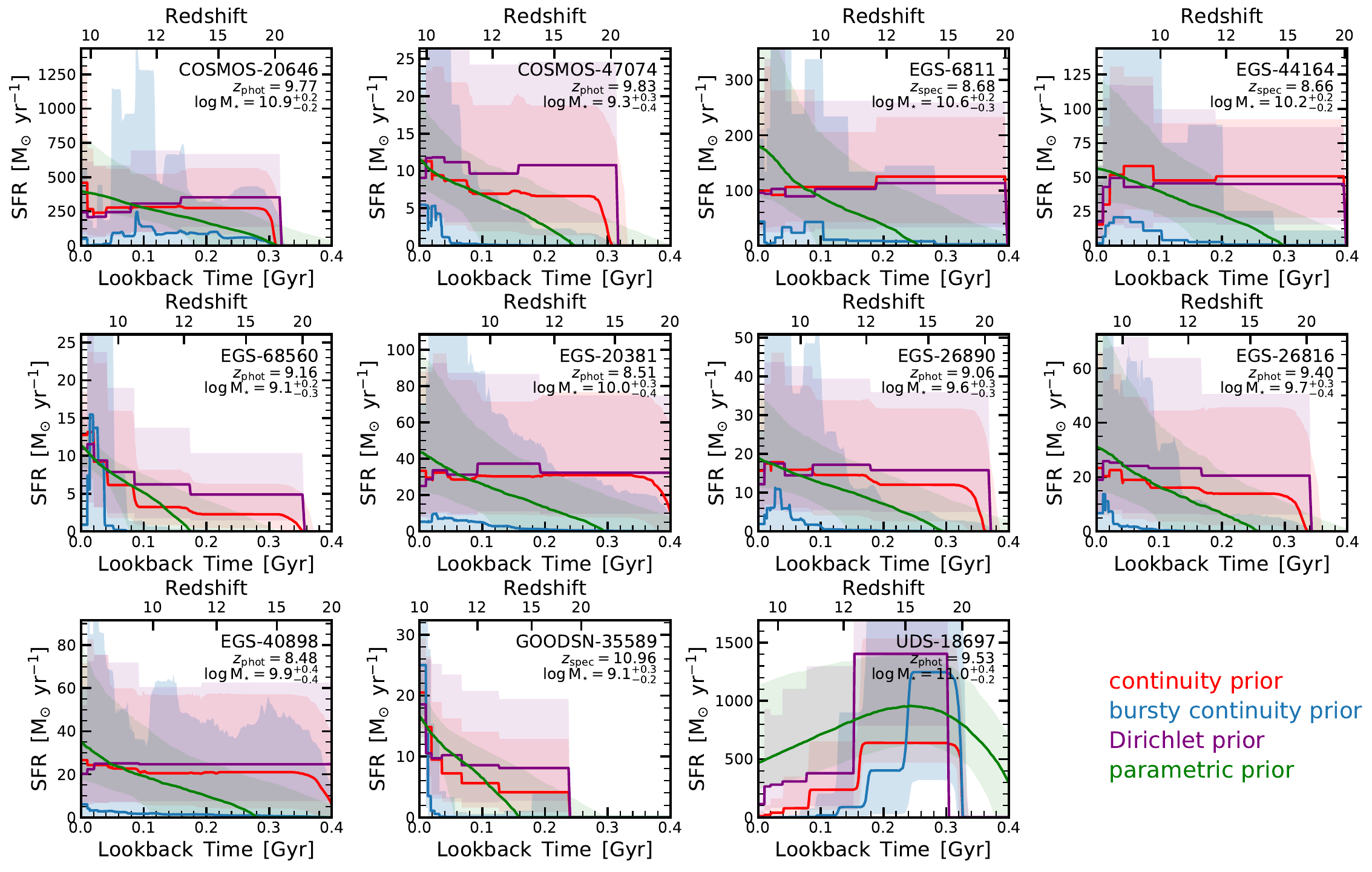}
\caption{Star-formation histories (SFHs) obtained from \texttt{Prospector} assuming different priors. The adopted priors include the continuity prior (red), bursty continuity prior (blue), Dirichlet prior (purple), and parametric prior. These priors are discussed in Section~\ref{subsec:prior_sfh}. The lines and shaded regions show the median and 16-84th percentile of the SFH posterior, respectively. The continuity and Dirichlet prior typically lead to a rather constant SFH, the parametric prior (delayed-tau parameterization) to an increasing SFH, and the bursty continuity prior to a bursty SFH, which biases the median SFH low (see text). The key point of this figure is that the adopted prior heavily effects the resulting posterior SFH. }
\label{appfig:SFH}
\end{figure*}

For completeness, Figure~\ref{appfig:SFH} plots the inferred SFHs as the SFR as a function of time, while Figure~\ref{fig:SFH_fm} in the main text plots the SFHs as fraction of mass formed as a function of time. Each panel shows an individual galaxy, with the red, blue, purple, and green lines showing the median SFHs obtained from the continuity, the bursty continuity, the Dirichlet, and the parametric SFH priors, respectively. The shaded regions show the 16-84th percentiles. The different SFH priors result in different SFH posteriors, i.e., it is important to fully understand how the inferred SFHs are affected by the choice of the prior.

For the fiducial continuity and Dirichlet priors, the resulting SFHs are similar and roughly constant with time. For a few galaxies (i.e., COSMOS-20646, EGS-68560, GOODSN-35589, and UDS-18697), there is significant variation in the past $\sim100$ Myr. The otherwise rather constant behavior is expected, as the expectation value of this prior is a constant $\mathrm{SFR}(t)$ (Section~\ref{subsec:prior_sfh} and Figure~\ref{fig:prior_sfh}). For the parametric SFH prior, we find for all except one galaxy (UDS-18697) an increasing SFH, something we expect from theoretical models \citep[e.g.,][]{tacchella18}. However, again, this is the expected behavior of the prior. This underscores the worry that the current data only provide little constraining power when it comes to the SFH.

The median SFH of the bursty continuity prior seems to lie significantly below the other SFHs, which -- at first glimpse -- implies a lower stellar mass. However, this is not the case as we have showed in Figure~\ref{fig:SFR_M_posteriors}. The explanation is that this prior tends to form the stars in one or a few time bins, which leads to a peaked SFH. When computing the median (as a function of time), this leads to a bias low, but peaky behavior can still be seen in the 16-84th percentiles (shaded region). In order to circumvent this problem, we can normalize first by mass before computing the median. This is done in Figure~\ref{fig:SFH_fm}.

\section{Degeneracies between age and attenuation}
\label{appsec:degeneracy}

\begin{figure*}
\centering
\includegraphics[width=\textwidth]{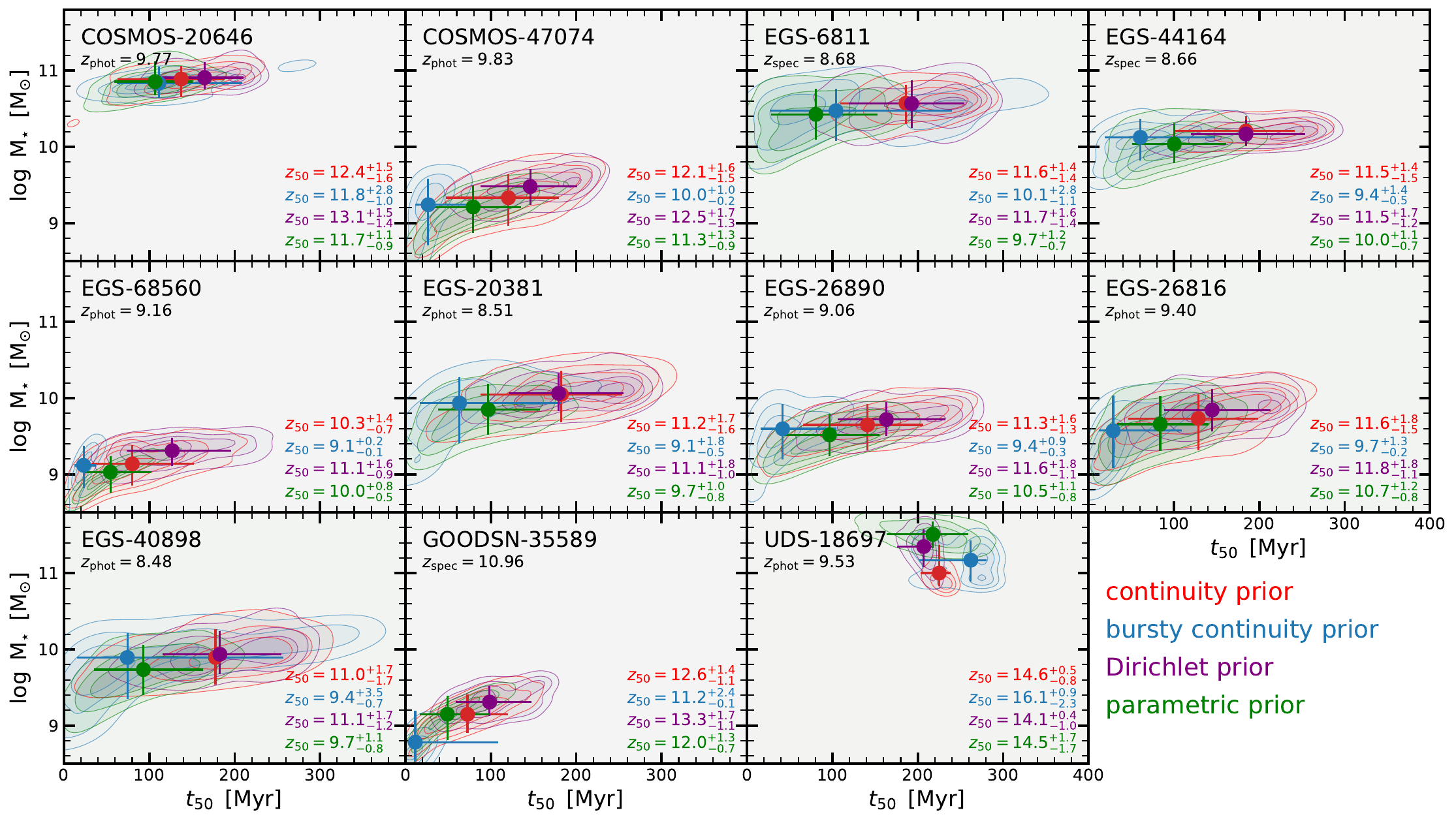}
\caption{Posteriors of stellar mass ($M_{\star}$) and stellar age ($t_{50}$). The red, blue, purple, and green colors indicate the continuity prior, the bursty continuity prior, the Dirichlet prior, and the parametric prior for the SFH. We also indicate the formation redshift $z_{50}$ at which 50\% of the stellar mass has been formed. Although both the stellar mass and the age are consistent within the uncertainty with each other, assuming different priors, the absolute values for the ages vary significantly. Consistent with Figure~\ref{fig:SFH_fm}, the bursty continuity prior and the parametric prior lead to younger ages than the continuity prior and the Dirichlet prior.}
\label{appfig:mass_age}
\end{figure*}

\begin{figure*}
\centering
\includegraphics[width=\textwidth]{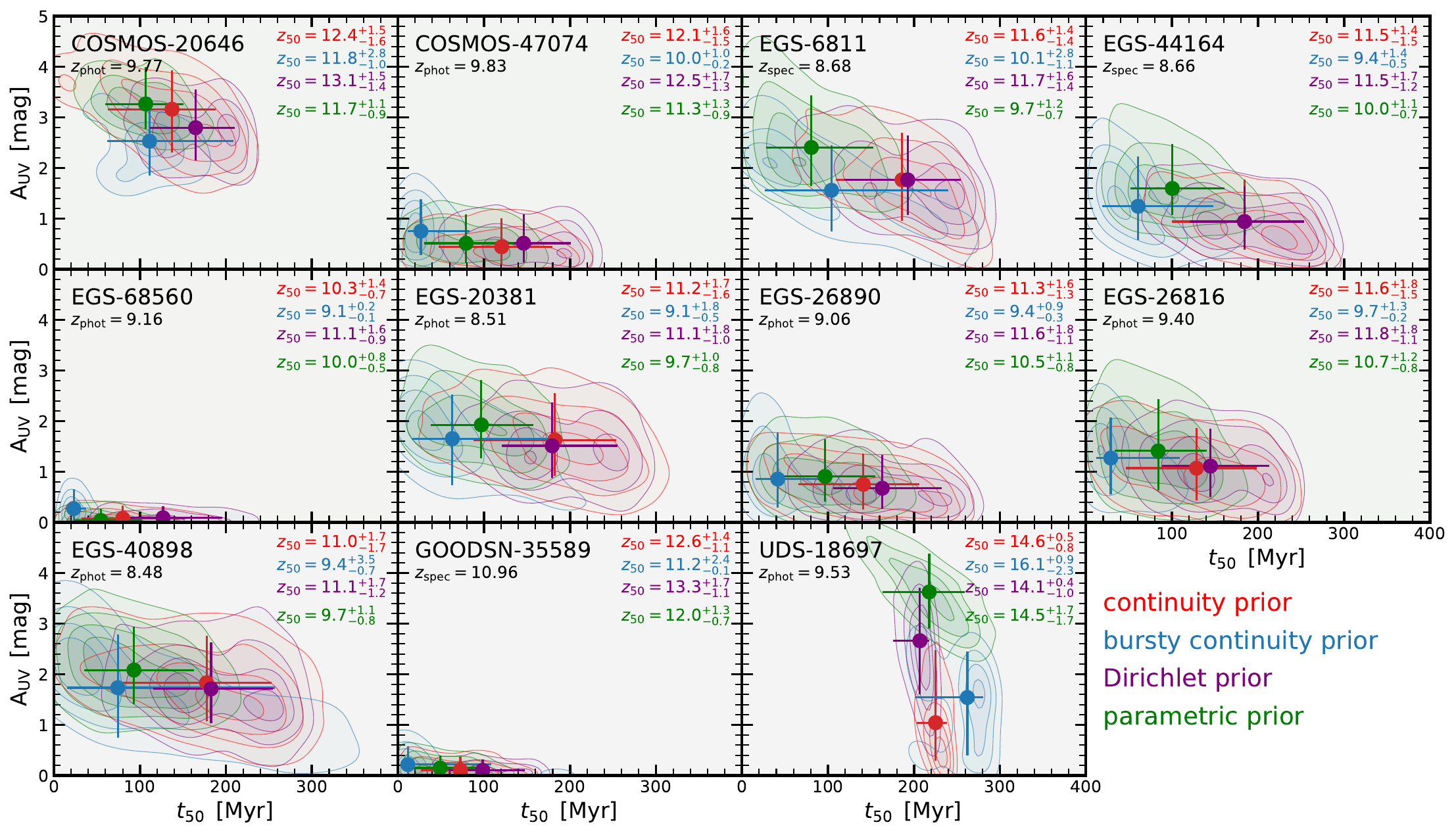}
\caption{Posteriors of UV attenuation ($A_{\rm UV}$) and stellar age ($t_{50}$). The red, blue, purple, and green colors indicate the continuity prior, the bursty continuity prior, the Dirichlet prior, and the parametric prior for the SFH. Although both the UV attenuation and the age are consistent within the uncertainty with each other, assuming different priors, the absolute values for both quantities vary significantly. Furthermore, there is a clear degeneracy between the $A_{\rm UV}$ and age, which we are not able to break with the current observational data.}
\label{appfig:Auv_age}
\end{figure*}

Figures~\ref{appfig:mass_age} and \ref{appfig:Auv_age} show the posterior distribution of $M_{\star}$ versus $t_{\rm 50}$ and UV attenuation ($A_{\rm UV}$) versus $t_{\rm 50}$, respectively. These two figures follow the same layout as Figure~\ref{fig:SFR_M_posteriors} in the main text. Figure~\ref{appfig:mass_age} shows that there is a degeneracy between $M_{\star}$ and $t_{\rm 50}$: a younger age implies a lower stellar mass. This is expected as younger stellar populations are typically brighter at fixed stellar mass. There is also a clear rank ordering of the SFH prior: the bursty continuity prior produces younger ages, followed by the parametric prior, while the Dirichlet and the continuity priors produce the oldest galaxies. As these galaxies lie within a redshift range of $z\approx9-11$, we also give the formation redshift $z_{50}$, i.e., the redshift by which 50\% of the mass of the galaxy has formed. We find that these galaxies typically form around $z_{\rm f}\approx11-13$, with UDS-18697 forming the earliest at $z_{\rm f}\sim15$. Importantly, $z_{\rm f}$ should not be confused with the epoch when the star-formation initially started. As we can see in Figure~\ref{appfig:SFH}, star formation in most galaxies start around $z\approx20$, which is basically by construction as constraining this is out of reach for the current data (Section~\ref{subsec:prior_sfh}). 

Figure~\ref{appfig:Auv_age} shows that it is challenging to break the dust-age degeneracy with our current observational data. Specifically, we find an anti-correlation between older ages and more UV attenuation. The UV attenuation is overall not well constrained, i.e., we find rather wide posteriors with uncertainties of more than 1 mag. This can at least in part be explained by the degeneracy with the attenuation law, over which we marginalize here. Furthermore, we do not find any trend with the SFH prior. 

\section{Fraction of mass formed at $z>12$}
\label{appsec:frac_M_z12}

\begin{figure*}
\centering
\includegraphics[width=\textwidth]{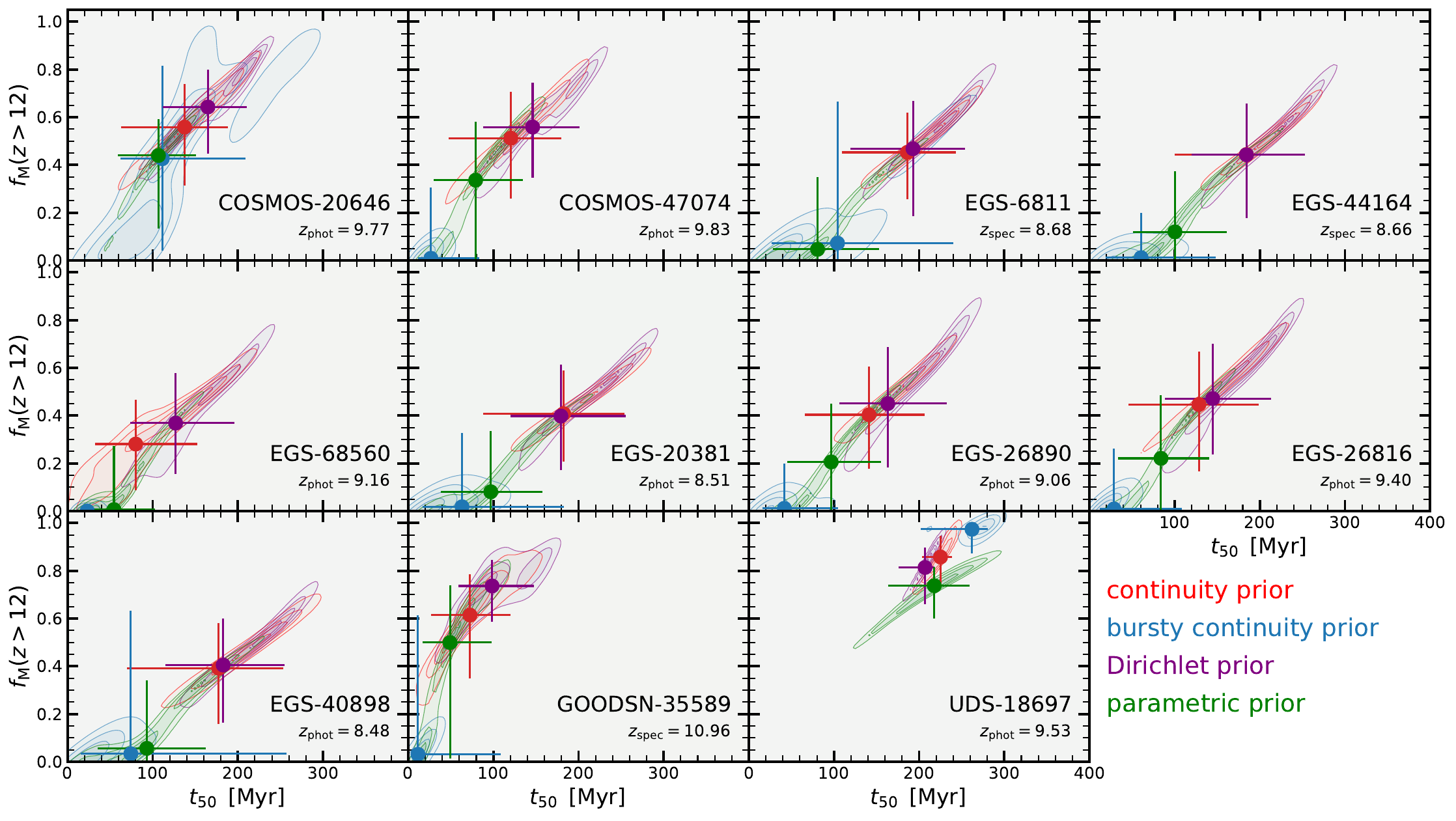}
\caption{Same as Figure~\ref{fig:frac_M_z10}, but considering the fraction of mass formed previous to redshift $z=12$. The fraction of mass formed before $z=12$ can be significant, but it depends on the assumed SFH prior.}
\label{appfig:frac_M_z12}
\end{figure*}

We present in Section~\ref{subsec:frac_mass} the results on the fraction of mass formed before $z=10$ ($f_{\rm M}(z>10)$; see also Table~\ref{tab:ages} and Figure~\ref{fig:frac_M_z10}). We find that all of our $z=9-11$ galaxies have formed at least 50\% of their mass before $z=10$, though the exact number depends on the assumed prior. In particular, for the bursty continuity prior, for some galaxies $f_{\rm M}(z>10)$ drops to less than 10\%. 

By definition, for galaxies with redshifts of about or larger than 10 $f_{\rm M}(z>10)$ will be close to or exactly 1. In order to quote a more meaningful number of these objects, we quantify in this appendix the fraction of mass formed previous to redshift $z=12$, i.e., $f_{\rm M}(z>12)$. 

Figure~\ref{appfig:frac_M_z12}, following the same layout as Figure~\ref{fig:frac_M_z10}, shows the posterior distribution of $f_{\rm M}(z>12)$ and stellar age $t_{\rm 50}$. We find again by construction the large degeneracy between $t_{\rm 50}$ and $f_{\rm M}(z>12)$. Furthermore, the continuity and the Dirichlet priors lead to fractions around 50\%, while the bursty continuity prior typically leads to significantly lower fractions, many times consistent with 0\%. This highlights again that the fraction of stellar mass formed before $z=12$ is not well constrained by our current data and that it is heavily depends on the assumed SFH prior.

\end{document}